\documentclass[a4paper,11pt]{article}
\pdfoutput=1 

\usepackage{jcappub} 
                     
\usepackage{aas_macros}
\usepackage[utf8]{inputenc}
\usepackage[english]{babel}
\usepackage[T1]{fontenc}
\usepackage{hyperref}
\usepackage{amsmath}
\usepackage{amsfonts}
\usepackage{dsfont} 
\usepackage{color} 
\usepackage{caption}
\usepackage{bbold}
\usepackage{comment}
\usepackage[export]{adjustbox}
\usepackage{dcolumn}

\usepackage[english]{babel}    
\usepackage{xspace}
\usepackage{microtype}    
\usepackage[utf8]{inputenc}    

\usepackage[T1]{fontenc}
\usepackage[extramarks]{titleps}
\usepackage{lipsum}

\usepackage{amsmath,amssymb,amsfonts,dsfont,bm}     
\allowdisplaybreaks     
\usepackage{mathrsfs}
\usepackage[usenames,dvipsnames]{xcolor}
\usepackage{slashed}
\usepackage{youngtab}
\usepackage{physics}
\usepackage{tensor}

\usepackage{natbib}
\usepackage{graphicx}
\graphicspath{{./Figures/}}

\newcommand{\bea}{\begin{eqnarray}}
\newcommand{\eea}{\end{eqnarray}}
\newcommand{\be}{\begin{equation}}
\newcommand{\ee}{\end{equation}}

\title{
Universal signatures of singularity-resolving physics in photon rings of black holes and horizonless objects
}

\author[a]{Astrid Eichhorn,}
\author[b,c,d]{Aaron Held,} 
\author[a]{Philipp-Vincent Johannsen}

\affiliation[a]{CP3-Origins, University of Southern Denmark,
\\
Campusvej 55, DK-5230 Odense M, Denmark}
\affiliation[b]{Theory Group, Blackett Laboratory,
\\
Imperial College London, SW7 2AZ, London, UK}
\affiliation[c]{Theoretisch-Physikalisches Institut, Friedrich-Schiller-Universit\"at Jena,\\Max-Wien-Platz 1, 07743 Jena, Germany}
\affiliation[d]{The Princeton Gravity Initiative, Jadwin Hall, Princeton University,\\Princeton, New Jersey 08544, U.S.}

\emailAdd{eichhorn@cp3.sdu.dk}
\emailAdd{aaron.held@uni-jena.de}

\abstract{
Within quantum-gravity approaches and beyond, different mechanisms for singularity resolution in black holes exist. Under a set of assumptions that we spell out in detail, these mechanisms leave their imprint in shadow images of spherically symmetric black holes. 
We find that even current EHT accuracy is sufficient to place nontrivial constraints on the scale of new physics within one modified spacetime, if the EHT measurement of M87* is combined with an independent measurement of the black-hole mass.
In other spacetimes, increased accuracy is required that the next-generation EHT may deliver.\\
We show how the combination of $n=1$ and $n=2$ photon rings is a powerful probe of the spacetime geometry of regular black holes, even when considering astrophysical uncertainties in accretion disks. Further, we generate images containing a localized emission region, inspired by the idea of hotspots in accretion flows.
Finally, we investigate the photon-ring structure of a horizonless object, which is characterized by either two or no photon spheres. We show how photon rings annihilate each other, when there is no photon sphere in the spacetime.
}

\begin{document}
\maketitle


%
\section{Motivation}
Black holes are fascinating objects that are currently under intense observational scrutiny in the strong-field regime 
\cite{Abbott:2016blz,paper1,paper2,paper3,paper4,paper5,paper6,paper7}. 
These observations test the predictions of General Relativity (GR) which has so far passed all observational tests (within the systematic uncertainties) \cite{TheLIGOScientific:2016src,Do:2019txf,LIGOScientific:2019fpa,Cardoso:2019rvt,LIGOScientific:2020tif,Volkel:2020xlc,LIGOScientific:2021sio}. Nevertheless, there is no doubt that GR does not completely describe black holes, because GR predicts unphysical curvature singularities \cite{Penrose:1964wq}. 
Thus, the question is not whether the Kerr paradigm will fail, the question is, when and how it will do so.

There is consensus that trans-Planckian curvature scales are affected by quantum effects that are not included in GR.  However, there is no consensus as to whether quantum-gravity effects may already set in at sub-Planckian curvature scales. The restriction of quantum effects to trans-Planckian scales follows a very simple dimensional analysis, that may very well be too naive to properly account for quantum-gravity effects. For instance, i) quantum gravity might feature a dynamically generated scale that is different from the Planck scale;  ii) quantum gravity might feature more than one scale (e.g., Starobinsky inflation~\cite{Starobinsky:1980te}, with a very large value of the curvature-squared operator, is an example for a gravity model with two very different scales)
iii) black holes, being defined through a non-local object, namely a horizon, may not obey estimates that rely on a comparison of the local curvature with the Planck scale. 
Thus, \cite{Almheiri:2012rt,Rovelli:2014cta,Haggard:2016ibp,Giddings:2016btb,Giddings:2019jwy,Held:2019xde,Bacchini:2021fig} even postulate or search for quantum gravity effects on horizon scales. Further, \cite{Carballo-Rubio:2019nel,Carballo-Rubio:2021ayp} argue that time-dependence is essential for black holes from quantum gravity, leading to large deviations from GR at some time during the evolution of black holes, where the time scale itself is a matter of discussion \cite{Bonanno:2020fgp,Barcelo:2020mjw}.

Moreover, quantum gravity may not be the only source of beyond GR effects; GR may be modified at the classical level at significantly sub-Planckian curvature scales.
For instance, spherically symmetric black-hole solutions in curvature-squared gravity have been investigated in \cite{Lu:2015psa,Lu:2017kzi,Pravda:2016fue,Podolsky:2019gro} and can indeed differ from solutions of GR; steps towards studying the stability of such theories can be found, e.g., in \cite{Brito:2013wya,Cayuso:2020lca,Held:2021pht}.

Thus, the theoretical possibilities are wide open:
the breakdown of the Kerr paradigm may or may not be accessible with current or near-future observations and may or may not be a quantum-gravity effect. 
The novel observational opportunities in gravitational waves and radio-very-long-baseline-interferometry (radio VLBI) enable us to remain agnostic with respect to the scale(s) at which beyond-GR effects set in. Instead, we can use observations to constrain the scale(s). \\

In this paper, we extend the \emph{principled-parameterized} approach to black holes beyond GR \cite{Eichhorn:2021etc,Eichhorn:2021iwq}. This approach treads a middle ground between a principled approach based on specific theories beyond GR on the one side and a parameterized approach based on general parameterizations \cite{1979GReGr..10...79B,Cardoso:2014rha,Johannsen:2015pca,Konoplya:2016jvv,Delaporte:2022acp} on the other side. It combines physical input from the principled approach with the comprehensiveness of the parameterized approach. Specifically, in this paper, we base our investigations on four principles (locality, regularity, simplicity, Newtonian limit). These principles may encode fundamental physics from several theories at an effective level. At the same time, they result in a parameterization of black-hole spacetimes that focuses on physically well-motivated deviations from GR. We aim to understand whether these principles give rise to universal, observable features of black-hole spacetimes. The underlying motivation is that instead of constraints on deviation parameters from the Kerr spacetime, as, e.g., obtained in \cite{Cardenas-Avendano:2019zxd,Shashank:2021giy}, we can obtain information on physical principles realized in theories beyond~GR.\\

In our work, we focus on spherical symmetry. This is motivated by technical simplicity and the observation that M87* is observed at near-vanishing inclination, resulting in a nearly spherical shadow. Implications of similar principles in spacetimes with finite spin have been investigated in~\cite{Eichhorn:2021etc,Eichhorn:2021iwq}. Beyond spherical symmetry, the above four principles result in non-circular spacetimes \cite{Delaporte:2022acp}, with corresponding technical complications.
\\

This paper is structured as follows: In Sec.~\ref{sec:principles} we spell out, how the four above principles together imply that black-hole horizons and photon spheres are more compact than in GR. We connect to the previous literature by highlighting that regular black holes of the Hayward~\cite{Hayward:2005gi}, Dymnikova~\cite{Dymnikova:1992ux} and Simpson-Visser~\cite{Simpson:2019mud} type, among others, satisfy these principles.

In Sec.~\ref{sec:Observationalconstraints}, we explore observational constraints on these regular black holes which in turn constrain the single free parameter in spacetimes of the principled-parameterized approach, namely the scale of new physics.
Current EHT data, combined with post-Newtonian information on stellar dynamics can already constrain the scale of new physics for Simpson-Visser spacetimes. 
Further, we explore how the simplicity, regularity and locality principle together leave an imprint in image features which may be accessible to future EHT observations, without the reliance on data from other observational campaigns. 
We investigate two promising approaches: (i) the relative separation of photon rings and (ii) spacetime tomography~\cite{Tiede:2020jgo}, for instance, via time-dependent localized emission sources (hotspots~\cite{Broderick:2005my,Broderick:2005jj}).
Regarding photon rings, we find that their relative separation increases in spacetimes constructed according to the four principles. Remarkably, for the largest choices of new-physics scale in the Simpson-Visser spacetime, the separation may approach 10 $\mu\rm as$ for M87*, which is the nominal resolution achievable with a ground-based array at 345 GHz. Therefore, a future ngEHT, assuming that it achieves this resolution and also a high enough dynamic range to detect the $n=1$ and $n=2$ photon ring \cite{Pesce:2021adg}, might be able to constrain the new-physics scale based solely on photon-ring observations. 
Finally, as a first step towards spacetime tomography, we explore inhomogeneous disk models and in particular localized patches of emission (see Sec.~\ref{sec:localized-emission}).

We  turn to horizonless objects in Sec.~\ref{sec:horless}, where we review how additional photon rings appear inside the former shadow boundary. We explore how these image features evolve under increases of the new-physics scale which leads to a loss of the two photon spheres that characterize these spacetimes.

We conclude in Sec.~\ref{sec:conclusions}. A technical understanding of simplicity in terms of series expansions is given in App.~\ref{app:simplicity-expansions}. A discussion of Post-Newtonian constraints is given in App.~\ref{app:PN}.

\section{Simplicity, regularity and locality imply a more compact event horizon and photon sphere}
\label{sec:principles}
We consider a general spherically symmetric and static spacetime with line-element
	\begin{align}
	ds^2 = -A(r)dt^2 + \frac{dr^2}{B(r)} + r^2(d\theta^2 + \sin^2\theta\,d\phi^2).\label{eq:generalsph}
	\end{align}	
This general form contains the Schwarzschild metric as a special case with
\bea
B_{\rm Schw}(r) &=& A_{\rm Schw}(r),\quad A_{\rm Schw}(r) = 1- \frac{2M}{r}. \label{eq:ABSchw}
\eea
To go beyond the Schwarzschild metric, we allow deviations from Eq.~\eqref{eq:ABSchw} that respect the symmetries of the spacetime and are constrained by the following four principles:
\begin{enumerate}
\item[(i)] \emph{(Locality principle)} Modifications of the metric from the Schwarzschild form depend on the local value of curvature invariants of the Schwarzschild spacetime.
\item[(ii)] \emph{(Regularity principle)} All (non-derivative) curvature invariants of the resulting line-element have to  be regular at all spacetime points.
\item[(iii)] \emph{(Simplicity principle)} Modifications of the metric from the Schwarzschild form are characterized by a single new-physics scale, not by several distinct scales. 
\item[(iv)]  \emph{(Newtonian limit)} The spacetime is asymptotically flat. Near asymptotic infinity, the Newtonian limit is recovered.
\end{enumerate}
The locality principle follows from an effective-field-theory perspective: GR holds in regions of spacetime with low curvature, but breaks down in regions of high curvature, with GR becoming a worse approximation, the larger the local curvature becomes. Thus, the local spacetime curvature (evaluated for the Schwarzschild metric) determines the size of the deviations.

The regularity principle enforces  that the new physics beyond GR produces a non-singular spacetime. It does not necessarily result in geodesic completeness, which would be an additional requirement to impose.\footnote{For a general overview of geodesically complete regular black holes, see e.g.~\cite{Carballo-Rubio:2019fnb}.}\\
The simplicity principle derives from the assumption that a theory beyond GR contains only one new-physics scale which is relevant for black holes.

The Newtonian limit can be understood as an observational constraint.

We will see that the combination of the four principles results in a powerful constraint on the metric and in observational imprints which do not follow from the first two principles alone. This constitutes an example how a focus on physics principles can provide a map (not necessarily one-to-one) between observational features and fundamental principles. In the following, we will refer to black holes which satisfy all four principles as simple, regular black holes.\\

To implement the locality principle, we start from a polynomially complete basis of non-derivative curvature invariants~\cite{1991JMP....32.3135C, 1997GReGr..29..539Z} for the Schwarzschild spacetime. Due to the high degree of symmetry and Ricci-flatness of the Schwarzschild solution, there is just one independent invariant, the Kretschmann scalar.\footnote{For the general metric in Eq.~\eqref{eq:generalsph}, additional invariants are nonzero and polynomially independent of the Kretschmann scalar, with the Ricci scalar being the lowest-order one.} For the general spherically symmetric metric Eq.~\eqref{eq:generalsph}, it is given by	
\bea
	K &=& R_{\mu\nu\kappa\lambda}R^{\mu\nu\kappa\lambda}\nonumber\\
	&=&\frac{1}{4r^4A(r)^4}\Bigl[16A(r)^4\left(B(r) - 1\right)^2 + 8r^2A(r)^2B(r)^2A'(r)^2 + 8r^2A(r)^4B'(r)^2 \nonumber\\
&{}&+r^4\left(A(r)A'(r)B'(r) - B(r)\left(A'(r)^2 - 2A(r)A''(r)\right)\right)^2\Bigr].\label{eq:Kretschmann}
\eea
For $A(r) = B(r)$, $K$ reduces to
	\begin{align}
	K = \frac{4\left[\left(A(r) - 1\right)^2 + r^2A'(r)^2\right]}{r^4} + A''(r)^2,\label{eq:KretschmannAequalB}
	\end{align}
and for the Schwarzschild spacetime, $K$ becomes
\be
K_{\rm Schw}=\frac{48 M^2}{r^6}.
\ee
To implement the first three principles, we follow \cite{Eichhorn:2021etc,Eichhorn:2021iwq} and promote $M$ in Eq.~\eqref{eq:ABSchw} to functions $M_{A/B}(K) = M_{A/B}(r)$. These functions and their first two derivatives must be non-singular everywhere and are constrained at small $r$ to ensure \emph{regularity}. Additionally, $M_{A/B}(r)$ are constrained at large $r$ to ensure the \emph{Newtonian limit}. (Post-Newtonian constraints from weak-field tests will be considered in App.~\ref{sec:PN}.)
This still leaves a lot of freedom in the choice of $M_{A/B}(r)$. In the literature, simple mass profiles $M_{A/B}(r)= M_{\rm eff}(r)$ have been considered, such as the Bardeen black hole \cite{Bardeen}, the Dymnikova black hole \cite{Dymnikova:1992ux}, the Hayward black hole \cite{Hayward:2005gi}, the Simpson-Visser black hole \cite{Simpson:2019mud}. Here, we show that they can all be rewritten in terms of $M_{\rm eff}(K_{\rm Schw})$ (thus satisfying the locality principle) and we spell out the underlying simplicity principle in detail to make the underlying physics assumption clear: The mass functions in the examples transition between the two asymptotic behaviors (at large and small $r$, respectively) without extra intermediate features, i.e., they are monotonically increasing functions of $r$. This implies simplicity in terms of the underlying physics, because they depend on a single scale, namely the transition scale between the two asymptotic behaviors.

\subsection{Implementing simplicity}
\label{sec:AneqB}
The deviations of $A(r)$ and $B(r)$ in Eq.~\eqref{eq:generalsph} from their classical form can be parameterized by two functions $f_{1,2}$. To implement locality, $f_{1,2}$ are functions of $K_{\rm Schw} r_{\rm NP}^4$, where $r_{\rm NP}$ is the new-physics scale. Here, we have  already implemented a prerequisite of simplicity, by allowing only a single new-physics scale.
\bea
\label{eq:f1}
A(r) &=& 1-\frac{2M}{r}f_1\left(
K_{\rm Schw}r_{\rm NP}^4
\right),\\
\label{eq:f2}
B(r)&=&1-\frac{2M}{r}f_2\left(
K_{\rm Schw}r_{\rm NP}^4
\right).
\eea
Before we can fully implement \emph{simplicity}, we first need to fix $f_{1,2}$ at small and large $K_{\rm Schw}$ (corresponding to large and small $r$, respectively) by the other principles.
To ensure the \emph{Newtonian limit}, we require that  
\be
f_{1,2}(K_{\rm Schw} \rightarrow 0) \rightarrow 1.
\ee
 To ensure the \emph{regularity principle},  we demand that curvature invariants are everywhere regular. In particular, we focus on the limit  $K_{\rm Schw} \rightarrow \infty$ ($r \rightarrow 0$), where we assume that $f_1\left(K_{\rm Schw}r_{\rm NP}^4\right)$ and $f_2\left(K_{\rm Schw}r_{\rm NP}^4\right)$ have a Taylor expansion in $\frac{1}{K_{\rm Schw}r_{\rm NP}^4}$. Plugging these expansions into the expression for the Kretschmann scalar yields non-vanishing coefficients $\sim r^{-n}$, $n \leq 6$, in the expansion of the Kretschmann scalar of the metric in Eq.~\eqref{eq:generalsph} about $r=0$.
These can be set to zero by demanding that 
 \be
 f_{1,2}\left(K_{\rm Schw}r_{\rm NP}^4 \rightarrow \infty \right) 
 \simeq \mathcal{O}\left(\frac{1}{\sqrt{K_{\rm Schw}r_{\rm NP}^4}}\right)= \mathcal{O}(r^3).
 \ee
Thus $f_{1,2}$ interpolate between $f_{1,2} \rightarrow 0$ at $K_{\rm Schw}\rightarrow \infty$ and $f_{1,2}\rightarrow 1$ at $K_{\rm Schw} \rightarrow 0$. The intermediate behavior is determined by the \emph{simplicity} principle: This principle states that $f_{1/2}(K_{\rm Schw} r_{\rm NP}^4)$ must be monotonically decreasing functions of $K_{\rm Schw} r_{\rm NP}^4$ (i.e., monotonically increasing functions of $r$), with the transition between the two limits determined by the new-physics scale $r_{\rm NP}$.
More complicated functions, e.g., with a (local) extremum of $f_{1/2}$, would be associated to a second scale and thus contradict the single-scale assumption of the \emph{simplicity} principle, see also App.~\ref{sec:series} for the connection between simplicity and series expansions.  
Simplicity thus restricts $f_{1,2}$ to satisfy $f_{1,2}\left(K_{\rm Schw}r_{\rm NP}^4\right)<1$ for $0<K_{\rm Schw}<\infty$, i.e., for the interval $r \in \left[0, \infty\right)$. This will be important for the observational imprints.

\subsubsection{Examples for black-hole spacetimes satisfying all four principles}
The literature contains several examples of black-hole spacetimes that satisfy all four principles. They correspond to the line-element Eq.~\eqref{eq:generalsph} with $B(r)=A(r)$ and with the mass $M$ upgraded to a function $M_{\rm eff}(r)$. The mass function can be written such that the dependence on $K_{\rm Schw}$, required to satisfy \emph{locality}, becomes obvious, by using that $r = \left(48 M^2/K_{\rm Schw} \right)^{1/6}$. Here, we keep the notation in terms of $A(r)$ and $f(r)$.

\begin{figure}[!t]
\begin{center}
\includegraphics[width=0.5\linewidth]{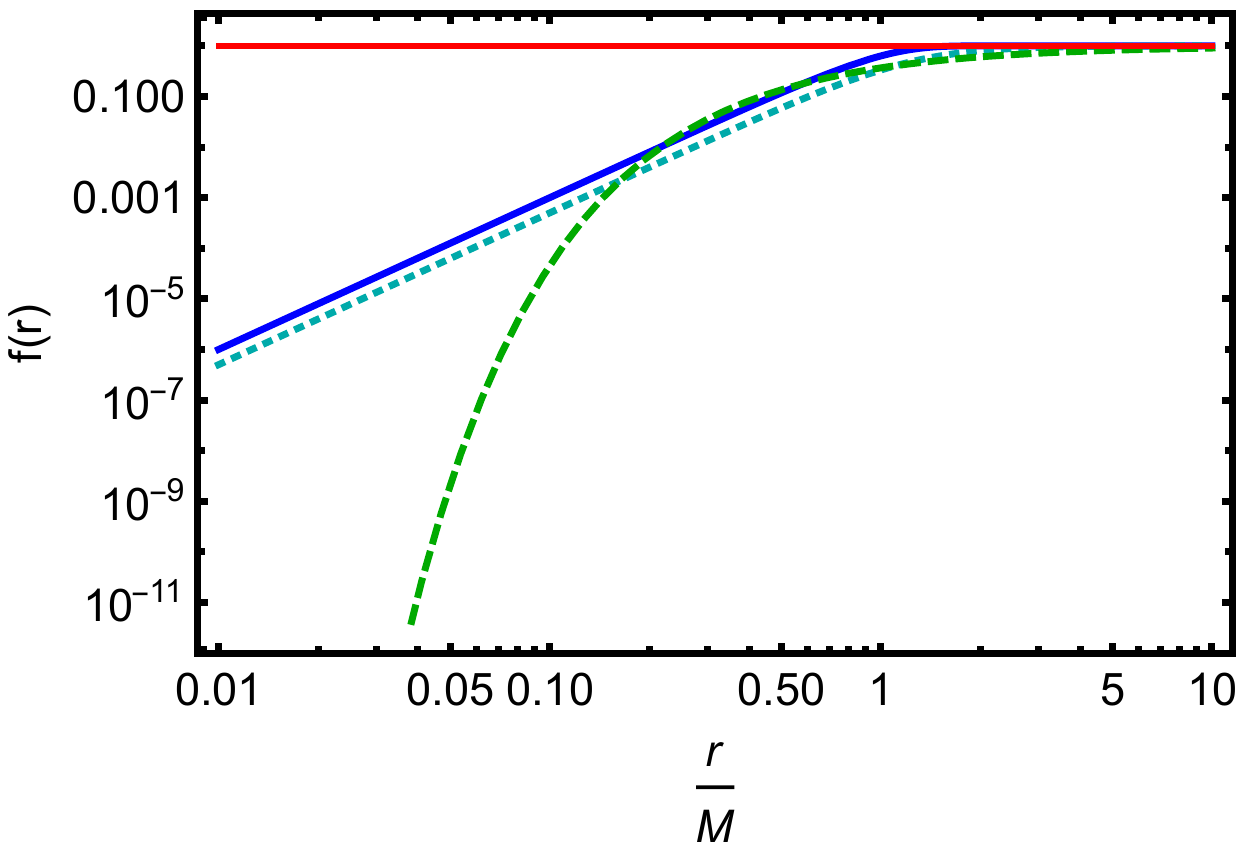}
\caption{\label{fig:frsimplicity} We show $f(r)$ for Dymnikova (blue, continuous), Hayward (cyan, dotted) and Simpson-Visser (greed, dashed), with $r_{\rm NP}= M$. All three functions increase from 0 to 1 (indicated by the red line) monotonically, as $r$ increases and do not exhibit a local extremum.
}
\end{center}
\end{figure}

Specifically, the metric functions are given by
\begin{enumerate}
\item[i)] Dymnikova \cite{Dymnikova:1992ux}: $A_{\rm D}(r) = 1- \frac{2M}{r}\left(1-\exp\left(-\frac{r^3}{M r_0^2}\right)\right) = 1- \frac{2M}{r} \left( 1-\exp\left(-\frac{\sqrt{48}}{\sqrt{K_{\rm Schw}}\,r_{\rm NP}^2} \right)\right)$,
\item[ii)] Hayward \cite{Hayward:2005gi}: $A_{\rm H}(r) = 
1- \frac{2M}{r}\frac{r^3}{r^3+2\,M\,l^2}
= 1 - \frac{2M}{r}\left(\frac{1}{1+ 2\,r_{\rm NP}^2\sqrt{\frac{K_{\rm Schw}}{48}}} \right)$,
\item[iii)] Simpson-Visser \cite{Simpson:2019mud}: $A_\text{SV}(r)=1-\frac{2M}{r}\exp\left(-\frac{a}{r}\right) = 1- \frac{2 M}{r}\exp\left(-\left(\frac{r_\text{NP}^4 K_{\rm Schw}}{48}\right)^{1/6}\right)$.
\end{enumerate}
In all three cases, we have chosen to identify the new-physics parameter of the respective spacetime, i.e., $r_0=r_{\rm NP}$ for Dymnikova, $l=r_{\rm NP}$ for Hayward, and $a=(r_\text{NP}^2M)^{1/3}$ for Simpson-Visser, such that $f(r/r_\text{NP})$ depends on $K$ and $r_{\rm NP}$ only.
To show that these three cases satisfy the \emph{simplicity} criterion, we plot the corresponding functions, see Fig.~\ref{fig:frsimplicity}, 
\bea
f_{\rm D} &=& \left( 1-\exp\left(-\frac{\sqrt{48}}{\sqrt{K_{\rm Schw}}\,r_{\rm NP}^2} \right)\right),\label{eq:fDym}\\
f_{\rm H}&=& \left(\frac{1}{1+ 2\,r_{\rm NP}^2\sqrt{\frac{K_{\rm Schw}}{48}}} \right),\label{eq:fHay}\\
f_{\rm SV}&=& \exp\left(-\left(\frac{r_\text{NP}^4 K_{\rm Schw}}{48}\right)^{1/6}\right)\label{eq:fSV}.
\eea
Simplicity is obvious from Fig.~\ref{fig:frsimplicity}, which shows a monotonic increase of all three functions without local extrema. 

Several mass functions have been discussed in a quantum-gravitational context, e.g., in \cite{Bonanno:2000ep,Nicolini:2008aj,Platania:2019kyx,Nicolini:2019irw}, where $r_{\rm NP}$ is typically equated to the Planck scale. However, both within quantum gravity as well as beyond, a more agnostic perspective is to keep $r_{\rm NP}$ as a free parameter which is to be constrained by observations, both gravitational as well as non-gravitational ones, e.g., \cite{Zhou:2020eth}. Under the assumption of simplicity, observations at all distances depend on $r_{\rm NP}$, albeit typically through the ratio $r_{\rm NP}/r$, meaning that observations at large distances are less sensitive to $r_{\rm NP}$- effects. 

In contrast, the asymptotic-safety-inspired black hole in \cite{Bonanno:2000ep} is an example for which simplicity does not hold (and a different new-physics-scale determines the large-distance behavior), because it is described by
\be
\label{eq:fBR}
f_{\rm BR}= \frac{1}{1+ \frac{\tilde{\omega}}{r^2}+ \frac{\gamma\, \tilde{\omega} M}{r^3}},
\ee
which contains two scales, namely $\tilde{\omega}$ and $\gamma\, \tilde{\omega}$. Indeed, this choice encodes asymptotic-safety inspired corrections near the center of the black hole, as well as the leading-order quantum-gravity correction from effective field theory \cite{Donoghue:1993eb} at large distances from the core.

\subsection{Increased compactness of simple, regular black holes}

The family of simple, regular black-hole spacetimes, parameterized by different functions $f_{1,2}$ (cf.~Eqs.~\eqref{eq:f1} and~\eqref{eq:f2}) and a single, dimensionful free parameter $r_{\rm NP}$  share a common characteristic feature: simple, regular black holes are more compact than their classical counterparts. This holds for the horizon, as well as other closed surfaces such as the photon sphere.

Herein, the simplicity principle is crucial. We demonstrate this by constructing a non-simple regular black hole with $f_\text{BR}$ in Eq.~\eqref{eq:fBR} -- although only for an unphysical choice of new-physics parameters. The quantum effects in \cite{Bonanno:2000ep} result in $\gamma>0$ and $\tilde{\omega}>0$ and thus in a monotonically increasing function $f_\text{BR}(r)$, even in the non-simple case. The opposite case of mathematically allowed but physically unmotivated choice $\gamma<0$ and $\tilde{\omega}<0$ results in a non-simple regular black hole for which $f(r)$ is non-monotonic and the conclusions below do not apply, so that, e.g., a more compact horizon comes with a less compact photonsphere, cf.~App.~\ref{app:simplicity-expansions}.

\subsubsection{Increased compactness of the event horizon}

We provide both a mathematical as well as a physical explanation for the increased compactness of the horizon.

Mathematically, in our choice of coordinates, the location of the horizon is determined by 
\be
g^{rr} = B(r) =  1- 2 \frac{M}{r}f_2 \left(K_{\rm Schw} r_{\rm NP}^4 \right) =0.
\ee
Since $f_2 \left(K_{\rm Schw} r_{\rm NP}^4 \right)<1$ for $K_{\rm Schw} r_{\rm NP}^4>0$, and $K_{\rm Schw} = \frac{3}{4 M^4}$ at the classical horizon ($r=2M$), $g^{rr}>0$ at $r=2M$. Thus, the event horizon does not lie at $r=2M$.
Instead, $g^{rr}=0$ is realized at $r_{\rm H}<2 M$, where the value $r_{\rm H}$ depends on the profile of the mass function. Thus, the regular black hole is more compact.\\
To circumvent the connection between regularity and increased compactness, simplicity has to be given up: If $f_{2}>1$ in the region around $r \approx 2 M$, then $r_{\rm H}>2 M$ is achievable. In turn, $f_2>1$ is not compatible with simplicity, as discussed above, because it requires the presence of a global extremum in $f_2$.

Physically, the location of the horizon is determined by the causal properties of the spacetime and corresponds to the outermost location that does not permit outward-traveling null geodesics. To regularize a black-hole spacetime, gravity must weaken.  Therefore, at the location of the classical horizon, the gravitational pull is weaker than classically and outward-traveling null geodesics exist. Thus, the regular black hole is more compact.

In physical terms, simplicity states that the weakening of gravity, required in the core of a regular black hole,  is only shielded completely at asymptotic infinity, and cannot be overcompensated by a strengthening of gravity at finite geodesic distance to the center. Such an overcompensation would be required to shift  the horizon outwards, i.e., to decrease compactness. To achieve such an overcompensation, the weakening of gravity must be shielded by the presence of a second scale, which is incompatible with simplicity.

Thus, simple, regular, spherically symmetric black holes are more compact than their classical counterparts.\\

Regular black-hole spacetimes that satisfy the simplicity criterion come as a family of spacetimes that is parameterized by one new-physics scale $r_\text{NP}$ or, equivalently, one dimensionless coefficient $\gamma$. For this coefficient, there is a critical value $\gamma_{\rm crit}$ beyond which the spacetime describes a horizonless compact object. This follows because $B(r=0)=1 = B(r \rightarrow \infty)$ together with $B(0<r<\infty)<1$ can only be realized with an even number of zeros of $B(r)$.
Our simplicity criterion precludes the occurrence of more than two zeros, but does not exclude that there are no zeros. At $\gamma= \gamma_{\rm crit}$, the two real zeros become complex, such that a horizonless object is left behind, cf.~Sec.~\ref{sec:horless}.
\\

\subsubsection{Increased compactness of the photon sphere}\label{sec:morecompactgamma}

The same increase in compactness affects other closed surfaces in the spacetime, not just the horizon. In particular, the photon sphere is also more compact. Thus, the singularity-resolving new-physics effect is expected to be visible in the black-hole shadow. Further, just like for the event horizon, there is a critical value $r_{\rm NP,\, crit,\,2}$, or equivalently $\gamma_{\rm crit,\, 2}$ for which there is no longer a photon sphere. 
As we will see in Sec.~\ref{sec:horless}, $r_{\rm NP,\, crit,\,2}>r_{\rm NP,\, crit}$.
Just as for the event horizon, the loss of photon sphere occurs, because a second (inner) photon sphere appears for $r_{\rm NP}>0$. At $r_{\rm NP,\, crit,\,2}$, the inner and outer photon spheres merge and become complex-valued for $r_{\rm NP}>r_{\rm NP,\, crit,\,2}$.
\\

Next, we show explicitly that the photon sphere is more compact in simple, regular black holes.
In static spherically symmetric spacetimes, geodesic motion is separable and can without loss of generality be confined to the equatorial plane ($\cos(\theta) = 0$). As a result, null geodesics obey the following equations as a function of the affine parameter $\lambda$, see, e.g., \cite{Cardoso:2008bp},
\begin{align}
	\left(\frac{dr}{d\lambda}\right)^2 = - V_r(r),
	\quad\quad\quad
	\frac{d\phi}{d\lambda} = \frac{L}{r^2},
	\quad\quad\quad
	\frac{dt}{d\lambda} = \frac{E}{A(r)},
\end{align}
where the energy $E$ and the angular momentum $L$ are constants of motion and the effective radial potential is
\begin{align}
	\label{eq:Vr}
	V_r(r) = - B(r)\left[\frac{E^2}{A(r)} - \frac{L^2}{r^2}\right].
\end{align}
A “generic” null geodesic moves between different radii. There can, however, be special radii in the spacetime, at which null geodesics can follow closed circular orbits, i.e., $r= \rm const$. For black holes in GR, there is one such radius, and the corresponding closed circular orbits make up the photon sphere. To find the location of the photon sphere, $r = \rm const$, the following conditions have to hold \cite{Bardeen:1972fi}
\begin{align}
	\label{eq:circular-orbit-cond}
	V_r(r =r_{\gamma}) = 0 = V_r'(r=r_{\gamma}).
\end{align}
These ensure that $r = \rm const$ not just momentarily, but that $r(\lambda)=\rm const$ for all $\lambda$.

For the Schwarzschild spacetime, there is a single photon sphere which is also the only closed photon orbit. For static spherically symmetric spacetimes beyond GR, multiple photon spheres as well as other closed photon orbits with varying $r$ can occur, cf.~Sec.~\ref{sec:horless}. The second derivative of the radial potential $V_r''(r)$ at $r=r_\gamma$ determines whether the photon sphere at $r_\gamma$ is stable or unstable. The two conditions $V(r=r_{\gamma})=0$ and $V_r'(r= r_{\gamma})=0$ become
\bea
0&=& \frac{E}{L} - \sqrt{\frac{A(r)}{r^2}},\label{eq:relnEandL}\\
0 &=& A(r= r_{\gamma}) - \frac{r}{2}A'(r= r_{\gamma}).\label{eq:photonsphere}
\eea
To arrive at Eq.~\eqref{eq:photonsphere}, the relation Eq.~\eqref{eq:relnEandL} has to be used.
The first condition, $V(r= r_{\gamma})=0$, translates into a condition that relates $E$ and $L$. This sets the initial conditions for null geodesics in the circular bound orbit.

For simple, regular black holes, $r_{\gamma} < r_{\gamma\, \rm Schw}$, i.e., the photon sphere is more compact than for a Schwarzschild black hole with the same mass parameter. To show this, we write
\be
A(r) = 1- \frac{2M}{r}f_1(r).
\ee
Thus, Eq.~\eqref{eq:photonsphere} becomes
\be
0 = 1- \frac{3M}{r}f_1(r=r_\gamma) + M f_1'(r=r_\gamma). \label{eq:photonsphereintermediate}
\ee
For the Schwarzschild case, this results in $r_{\gamma\, \rm Schw}=3M$. For simple, regular black holes, it holds that $f_1(r)<1$ and $f_1'(r)>0$. Therefore Eq.~\eqref{eq:photonsphereintermediate} can be rewritten in the form
\be
r_\gamma = \frac{3M\, f_1(r=r_\gamma)}{1+M\, f_1'(r=r_\gamma)}< \frac{3M}{1+M\, f_1'(r=r_\gamma)}< 3M.
\ee
Thus, the photon sphere of a simple, regular black hole, if it exists, i.e., if Eq.~\eqref{eq:photonsphereintermediate} has a real solution, is necessarily more compact than for a Schwarzschild black hole with the same mass parameter.

Because these effects occur universally for all spacetimes that obey regularity, locality, simplicity and the Newtonian limit, we focus on selected examples, such as the Simpson-Visser spacetime, in the following Sec.~\ref{sec:Observationalconstraints} and~\ref{sec:horless}. The only difference between mass functions is a quantitative one: The critical values of $r_{\rm NP}$ differ, as do the relative sizes of the new-physics effect.
\\

\section{Towards observational constraints on simple, regular black holes}\label{sec:Observationalconstraints}
Three types of seminal observations of black holes open up the possibility to test deviations from GR black holes: First, post-Newtonian constraints are available for Sag-A*~\cite{Ghez:2008ms,2015MNRAS.447..948C,Do:2019txf} and M87*~\cite{Gebhardt:2011yw}.
Such constraints can be imposed in the weak-field regime, where higher-order coefficients in the post-Newtonian expansion are subleading; deriving constraints on the post-Newtonian expansion in the strong-field regime is more subtle \cite{Volkel:2020xlc}.
Second, constraints in the strong-field regime, mostly from the spacetime region around the photon sphere are made possible by EHT observations \cite{paper1,paper6}. Third, strong-field constraints from the spacetime region closer to the horizon arise from LIGO/Virgo observations of mergers of compact objects \cite{Abbott:2016blz,TheLIGOScientific:2016src,LIGOScientific:2019fpa,LIGOScientific:2020tif} and will in the future become stronger, e.g., from the ringdown phase. Reviews of the field can be found in \cite{Berti:2015itd,Cardoso:2019rvt}.

In the previous sections, we have shown that the horizon of a spherically symmetric, simple, regular black holes is more compact  than in a GR black hole of the same mass. This increased compactness results in a more compact photon sphere and thus a more compact black-hole shadow, see, e.g., \cite{Abdujabbarov:2016hnw,Held:2019xde,Kumar:2019ohr,Liu:2020ola,Stuchlik:2019uvf,Eichhorn:2021etc,Eichhorn:2021iwq,Wielgus:2021peu}. The shadow boundary itself does not provide an observational handle to test the underlying effects, because the spherically symmetric shadow boundary of a regular black hole is degenerate with the shadow boundary of a GR black hole of smaller mass. In addition, the shadow boundary is not an actual observable, because it contributes a negligible part to the total image intensity \cite{Johnson:2019ljv}.

In the following, we investigate three ways to constrain simple, regular black holes with (ng)EHT observations. The first is a combination of the $n=1$ photon ring with post-Newtonian observations, cf.~\cite{Held:2019xde,EventHorizonTelescope:2021dqv}. The second measures the relative distance between the $n=1$ and the $n=2$ photon ring. The third is spacetime tomography via, for instance, time-dependent localized emission.

\subsection{Combination of post-Newtonian with EHT observations}
\label{sec:PN+EHT}
As has been discussed in \cite{Held:2019xde}, $r_{\rm NP}$ can be constrained based on a combination of post-Newtonian observations, e.g., of stellar dynamics, with EHT observations. As highlighted in \cite{EventHorizonTelescope:2021dqv}, the mass reconstructed from stellar dynamics agrees with the mass reconstructed from EHT observations, within the respective systematic uncertainties and under the assumption of GR. Going beyond GR, the reconstructed mass as well as the systematic uncertainty of the mass would change. \\

From stellar dynamics, one reconstructs a mass of $M_{\text{PN}}=\left(6.6 \pm 0.4\right)\cdot 10^9\, M_{\odot}$  \cite{Gebhardt:2011yw} for M87*, while the EHT provides $M_{\text{EHT}}=\left(6.5 \pm 0.2_{\rm stat}\pm 0.7_{\rm syst}\right) \cdot 10^9\, M_{\odot}$ \cite{paper5}. \\
To infer whether this gives rise to constraints on $r_{\rm NP}$, we proceed as follows: We work with the effective mass function $M_{\rm eff}$ for the Hayward and Simpson-Visser case,
\be
M_{\rm eff}(r) = M \cdot f_{\rm H/SV},
\ee
where $f_{H}$ and $f_{\rm SV}$ are defined in Eq.~\eqref{eq:fHay} and \eqref{eq:fSV}, respectively. From measurements at large distances, $r \gg M_{\text{PN}}$, one extracts $M_\text{PN}=M_\text{eff}(r\geqslant M) \approx M$. This holds for the mass of M87* reconstructed from stellar dynamics. In contrast, mass measurements at $r \simeq M_{\text{EHT}}$, that is, reconstructed from EHT data, give rise to a smaller reconstructed mass.

To properly determine the exact value of $r_{\rm NP}$ which is excluded for different mass functions, the EHT mass extraction should be performed with GRMHD simulations (or simulations within a modified dynamics) for the appropriate spacetime. Instead, we proceed differently to obtain an estimate of the excluded range of $r_{\rm NP}$. 
We assume that the mass reconstructed from the EHT measurement corresponds to the effective mass at the photon sphere $r_{\gamma}$, which is obtained by solving Eq.~\eqref{eq:photonsphere}.

The calculated $M_\text{EHT}=M_{\rm eff} (r = r_{\gamma})$ can be compared to the mass $M_{\rm PN}$ extracted from stellar dynamics at large $r$. To determine up to which value of $r_{\rm NP}$ the two masses are compatible with each other, we require an estimate for the error size of $M_{\rm eff} (r = r_{\gamma})$.
We assume that the relative error is the same as the relative error on the mass that the EHT collaboration provides under the assumption of GR. Such an assumption has been argued for in \cite{EventHorizonTelescope:2021dqv}. 

\begin{figure}[!t]
\includegraphics[width=0.495\linewidth]{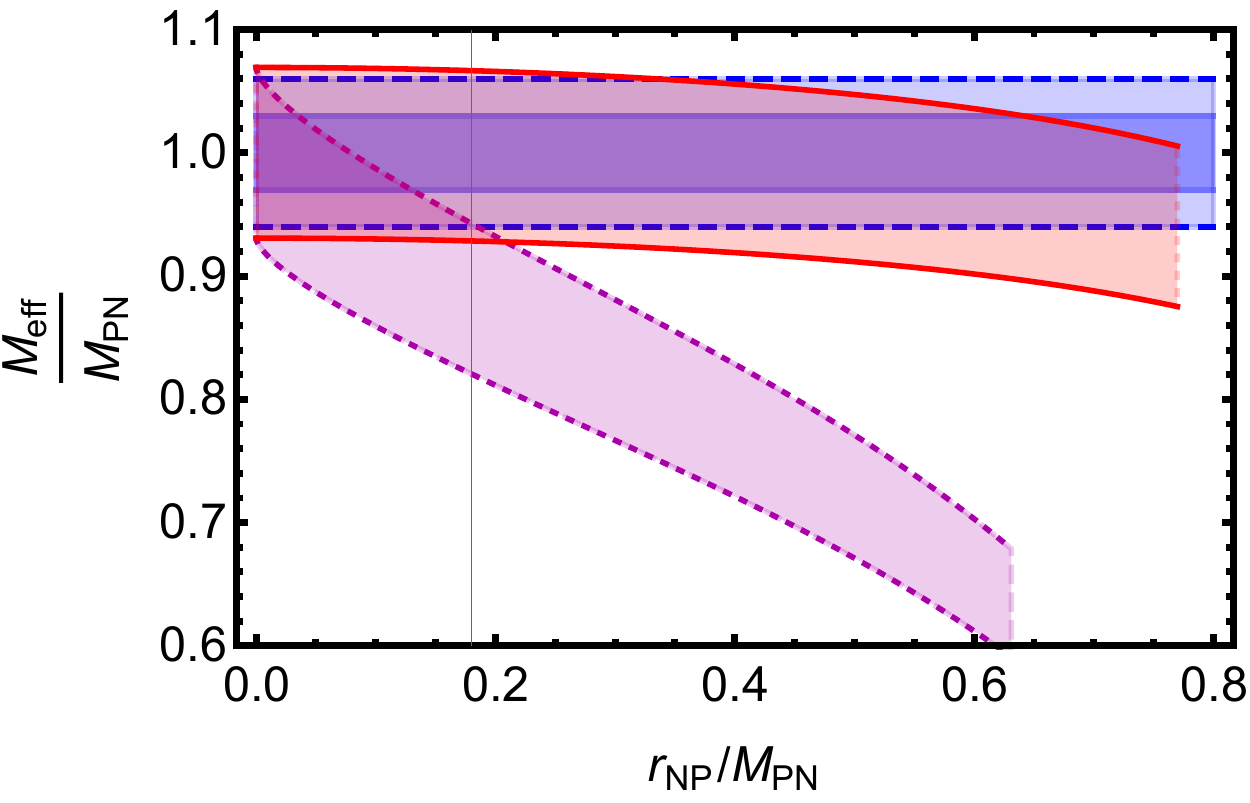}
\hfill
\includegraphics[width=0.465\linewidth]{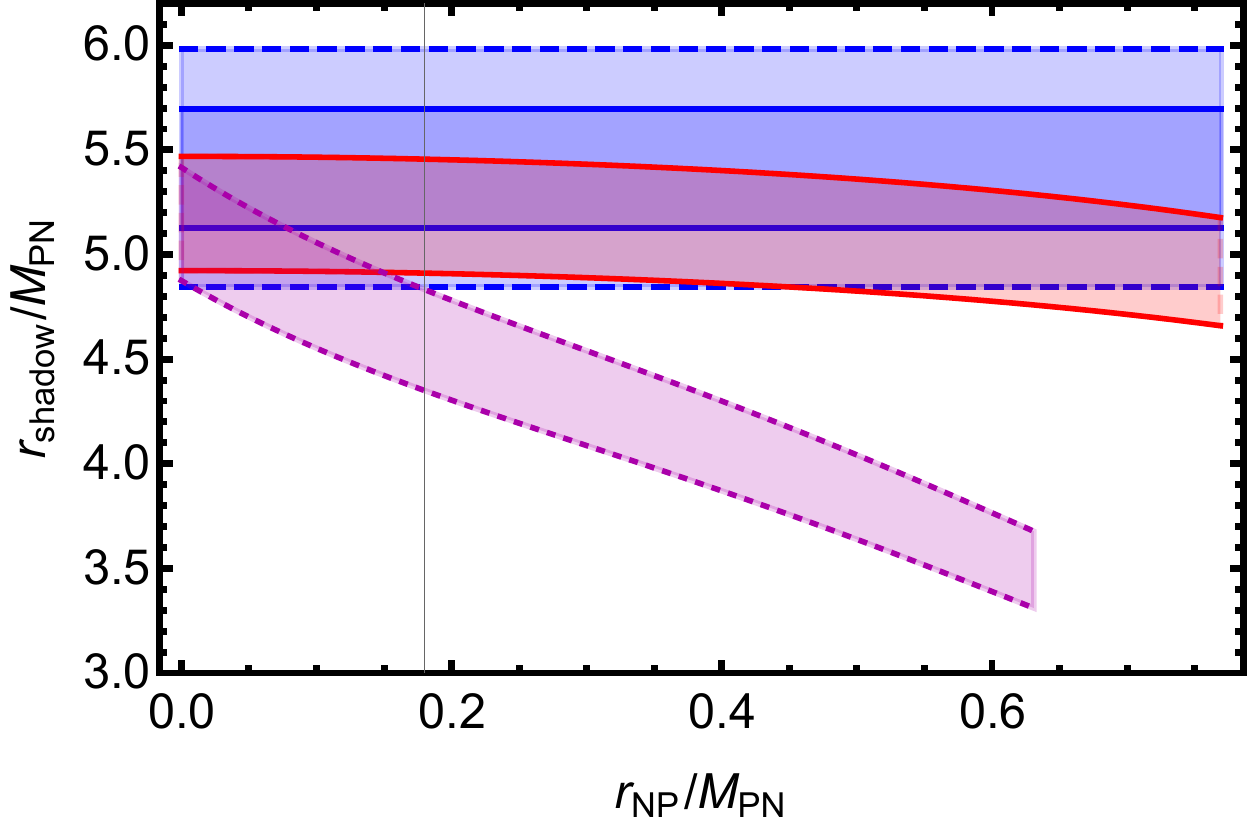}
\caption{\label{fig:exclusionSV} Left panel: We show the effective mass as a function of $r_{\rm NP}$ for the Hayward (continuous red lines) and Simpson-Visser (magenta dotted lines) spacetime. The errors on both effective masses correspond to the relative error of the M87* mass measurement by the EHT.
We also indicate the 1-$\sigma$ and 2-$\sigma$ errors (darker and lighter blue regions) on the mass extracted from stellar dynamics.\\
Right panel: We show the shadow radius in units of $M_{\rm PN}$ that is reconstructed from the EHT measurement, together with 1-$\sigma$ (blue continuous line) and 2-$\sigma$ (blue dashed line) contours. We compare with the shadow radius for the Hayward spacetime (red lines) and Simpson-Visser spacetime (magenta dotted lines). The errors on both shadow radii correspond to the same relative error as that of the EHT measurement.}
\end{figure}

The results are shown in the left panel of Fig.~\ref{fig:exclusionSV}, where for the Hayward spacetime, no value of $r_{\rm NP}$ in the range $0\leq \frac{r_{\rm NP}}{M_{\rm PN}} \leq \sqrt{\frac{32}{52}}=\frac{r_{\rm NP,\, crit}}{M_{\rm PN}}$ is excluded. This is consistent with the previous results in \cite{Held:2019xde} and \cite{EventHorizonTelescope:2021dqv}, where values of $r_{\rm NP}$ that result in black-hole spacetimes with a horizon cannot be excluded. In contrast, for the Simpson-Visser spacetime, a significant range of values for $r_{\rm NP,\,crit}>r_{\rm NP}>0.18 M_{\rm PN}$ can be excluded, because it falls outside the $2\sigma$ region of $M_{\rm PN}$.
The stricter the requirement, how significantly a certain value of $r_\text{NP}$ is excluced, the smaller the range of excluded $r_\text{NP}$. Here, we make a less conservative choice by using $2\sigma$, because we expect that future data may have smaller error budgets and thus lead to similar exclusions at higher significance.
\\

To cross-check our results, we proceed in a second way. This way does not use the error on the mass-reconstruction by the EHT. Instead, we calculate the absolute shadow size from the EHT measurement of shadow size in $\mu$as and the known distance $D=16.8^{+0.8}_{-0.7}\;\text{Mpc}$ to M87*~\cite{paper6, 2010A&A...524A..71B, 2009ApJ...694..556B, 2018ApJ...856..126C}. To convert this into a dimensionless number, we use the mass measurement $M_{\rm PN}$ extracted from stellar dynamics. The error on this quantity is thus the combined error on the shadow-size from the EHT reconstruction, the distance to M87* and the error on the mass from stellar dynamics. We then calculate the radii of the idealized shadow boundary in the Hayward and Simpson-Visser spacetime, i.e., the critical impact parameters $\Lambda_{\rm c}$. These follow from Eqs.~\eqref{eq:relnEandL} and~\eqref{eq:photonsphere} with $\Lambda=E/L$ as the impact parameter of null geodesics perpendicular to the observer's screen. The respective values of $\Lambda_{\rm c}$ need to be determined numerically.
In the region of $r_{\rm NP}$ of interest to us, they can be fit by the following polynomials
\bea
\frac{\Lambda_{{\rm c,}\,\rm Hayward} (r_{\rm NP})}{M} &=& 3 \sqrt{3} + 0.012 \frac{r_{\rm NP}}{M} - 0.498 \frac{r_{\rm NP}^2}{M^2} +0.267 \frac{r_{\rm NP}^3}{M^3}-0.356 \frac{r_{\rm NP}^4}{M^4},\label{eq:shadowHayward}\\
\frac{\Lambda_{{\rm c,}\,\rm SV}(r_{\rm NP})}{M} &=& 5.15 - 4.002 \frac{r_{\rm NP}}{M} + 6.983 \frac{r_{\rm NP}^2}{M^2} -11.496 \frac{r_{\rm NP}^3}{M^3}-6.189 \frac{r_{\rm NP}^4}{M^4}. \label{eq:shadowSV}
\eea
We assume that the EHT measurement of the shadow diameter indeed corresponds to the idealized shadow boundary. For the actual EHT image reconstruction, this is not the case, see \cite{paper5}, where the conversion factor between the shadow diameter and the mass is not $3 \sqrt{3}$, as it would be in GR if the shadow boundary was indeed observable. A similar shift would likely occur, if the EHT image reconstruction and mass extraction was performed within a modified spacetime, e.g., the Hayward or Simpson-Visser spacetimes.\\
To estimate an error for the shadow size in these modified spacetimes, we assume that the relative error is the same as it is for the EHT extraction of the shadow size. 

The results are shown in the right panel of Fig.~\ref{fig:exclusionSV}. Within the respective errors, the results are well compatible with our results from the other method to constrain $r_{\rm NP}$, and yield no constraint for the Hayward spacetime and a constraint of $r_{\rm NP}<0.18\, M_{\rm PN}$ for the Simpson-Visser spacetime. We also see that the two methods would yield different results, if the error on the mass from stellar dynamics, and the error on the shadow size in $\mu$as from the EHT were smaller. Thus we conclude that future constraints on $r_{\rm NP}$, e.g., from the ngEHT, require a more sophisticated analysis, where GRMHD simulations are performed in modified spacetimes.

In summary, combining mass measurements from different sources is a first way of constraining $r_{\rm NP}$. To constrain $r_{\rm NP}$ quantitatively, a specific mass function $M_{\rm eff}(r)$ has to be assumed. However, a universal constraint on all regular, simple black holes can be obtained, because $M_{\rm eff}(r \approx M) < M_{\rm PN}$ holds universally. Thus, if mass measurements from future EHT observations and stellar dynamics remain in agreement within errors, a universal, qualitative constraint on simple, regular black holes arises that limits their parameter space.

\subsection{Photon rings}
\label{sec:rings}
In the spherically symmetric case, the shadow boundary is insufficient to constrain $r_{\rm NP}$ due to a degeneracy between $M$ and $r_\text{NP}$. As emphasized in \cite{Lima:2021las}, even if the shadow boundary is degenerate, the gravitational lensing is not. Thus, observing photon  rings allows us to distinguish different spacetimes even in the spherically symmetric case.
Photon rings are caused by strong gravitational lensing \cite{Johnson:2019ljv} and arise as follows in the spherically symmetric case: The image of a black hole consists of multiple (in principle infinitely many) images of the light source, e.g., an accretion disk. The diffuse emission from the disk constitutes the $n=0$ component of the image intensity. Emission that is picked up by geodesics that orbit the black hole  by half a turn constitutes the $n=1$ component. The more often geodesics orbit the black hole, the closer to the shadow boundary they arrive on the image plane. The black hole image therefore features an exponentially stacked set of photon rings. In the idealized case in which the accretion disk is transparent, i.e., its absorptivity is set to zero, the peak intensity of the $n^{th}$ photon ring increases with $n$. Simultaneously, the width  of the $n^{th}$ photon ring decreases with $n$. At finite absorptivity, the increase of peak intensity with $n$ is prevented. The $n \rightarrow \infty$ ring is therefore irrelevant for practical observations, as it contributes negligibly  to the total image intensity. In practise, $n=1$ may be observable with ground-based VLBI \cite{Blackburn:2019bly}, while, for Kerr black holes, $n=2$ might be observable with a station on the moon and $n=3$ with a station in the second Lagrange point, see \cite{Johnson:2019ljv}.
\\
The photon rings are images of the accretion disk and hence depend on astrophysics. Nevertheless, spin and mass measurements are possible from the photon rings \cite{Broderick:2021ohx}, providing tests of GR. Given these results, two questions are crucial to answer: First, how strong can constraints on $r_{\rm NP}$ based on the observation of $n=1$ alone be? Second, because photon rings are more separated in beyond-Kerr-spacetimes, can $n>1$ photon rings be observed at lower resolution than in GR?
\\

Photon rings for regular black holes have first been considered in \cite{Eichhorn:2021iwq}; a comparison between Kerr photon rings and photon rings from several exotic compact objects can also be found in \cite{Wielgus:2021peu}; and a photon-ring study in the parameterized approach has been put forward in \cite{Ayzenberg:2022twz}.
We will discuss how the relative distance between the first and second ring depends on the new-physics parameter. This has been done, e.g., for Hayward-like mass functions in \cite{Eichhorn:2021iwq,Wielgus:2021peu}, here we review those results and add results for the Simpson-Visser mass function.
Measuring two ratios of photon ring radii thus may provide a powerful test of GR. At fixed disk model, such a test would be conclusive. However, in practise the parameters of the astrophysical accretion disk are not known exactly, and variations in disk parameters might alter or even mimick the signatures of physics beyond GR. Thus, we will test the impact of variations in the density profile of a simple disk model on image features and the relative distances of photon rings.\\

Further, emission structures depend on time, since material is being accreted. The time-dependence of such localized emission provides additional information, even from the $n=0$ diffuse emission and $n=1$ ring alone. For instance, one can model localized emission by a narrowly peaked radial profile (i.e., effectively a ring) that moves from higher to lower radii as a function of time. In such an idealized model, the observation of emission from two radii suffices to measure mass and spin uniquely~\cite{Broderick:2021ohx}. In the presence of $r_{\rm NP}$, the mass extracted from emission at a third radius is expected to not agree with the mass extracted from the previous two measurements, similar to the case of a charge in a Kerr-Newman black hole investigated in \cite{Broderick:2021ohx}. This may provide a constraint on $r_{\rm NP}$. 
\\

We consider a disk model described by the following number density, similar to \cite{Broderick:2021ohx}:
\bea
\label{eq:disk-model}
n(r,\theta)&=& n_0
\begin{cases}
 r^{-\alpha}\, \exp[-\frac{\cos(\theta)^2}{2h^2}],\quad\quad\quad\quad\quad\quad\quad\;\; \mbox{for } r>r_{\rm cut}\\
 r^{-\alpha}\, \exp[-\frac{\cos(\theta)^2}{2h^2}]  \exp[- \frac{(r-r_{\rm cut})^2}{w^2}] \quad \mbox{for } r<r_{\rm cut}
\end{cases},
\eea
where $h$ is is a dimensionless parameter determining the disk height, $\alpha$ determines the power-law falloff at large $r$, $r_{\rm cut}$ sets the inner cutoff, $w$ sets the width of the inner cutoff region and $n_0$ is a normalization.

We model a disk with nonzero emissivity, but vanishing absorbtivity. This choice is motivated by observations that indicate that the accretion disks of supermassive black holes are optically thin \cite{Johnson:2015iwg}. We also neglect the frequency dependence of the emission profile and do not account for redshift effects here. This approximation is motivated by the fact that EHT observations are essentially monochromatic  \cite{paper1} and that redshift only plays an important role for emission very close to the horizon. 

Under these approximations, the radiative transfer equation (Boltzmann equation) for the intensity $I_{\nu}$ along a null geodesic, parameterized by the affine parameter $\lambda$, reduces to
\begin{align}
\label{eq:radiative-transfer-eq-simplified}
	\frac{d}{d\lambda}\left(\frac{I_\nu}{\nu^3}\right) = C\,n\left(x^\mu(\lambda)\right),
\end{align}
where $C$ is a normalization. The combination $C\, n_0$ drops out, once the calculated intensity profiles are normalized.

\subsubsection{Low-order photon rings}
\begin{figure}
\centering
\includegraphics[height=0.32\linewidth]{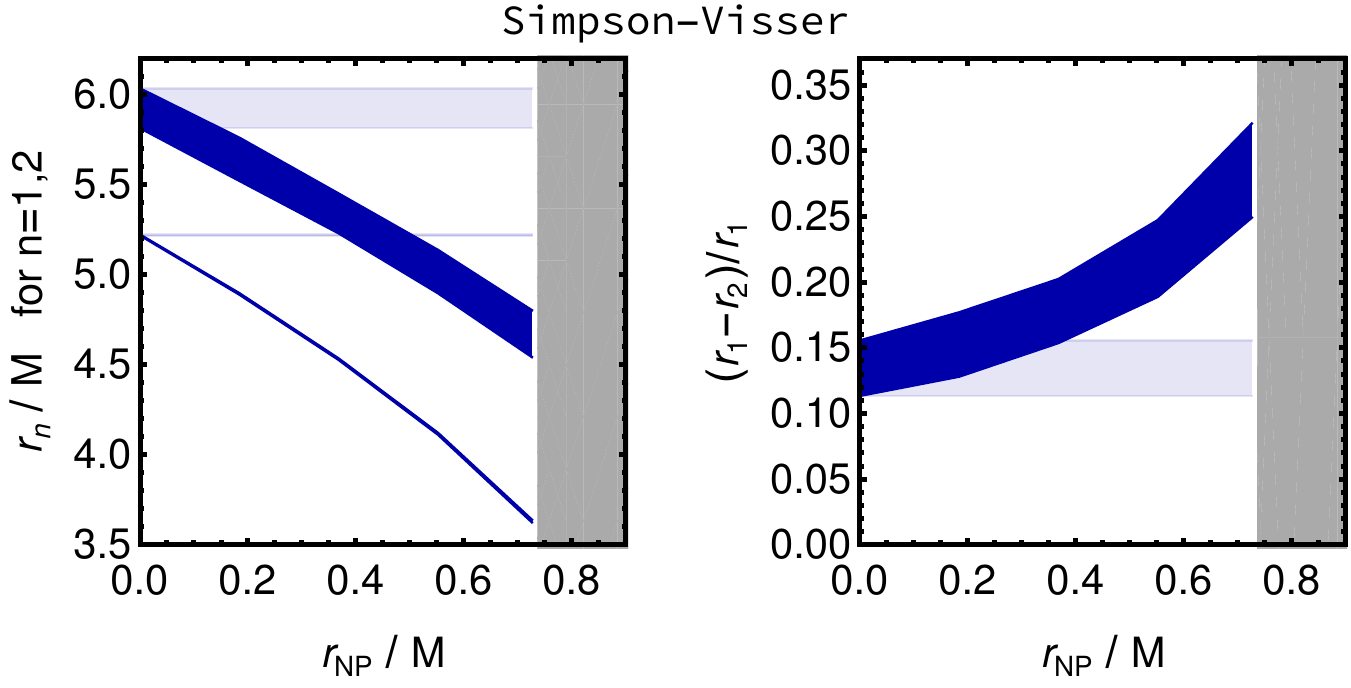}
\hfill\vline\hfill
\includegraphics[height=0.32\linewidth]{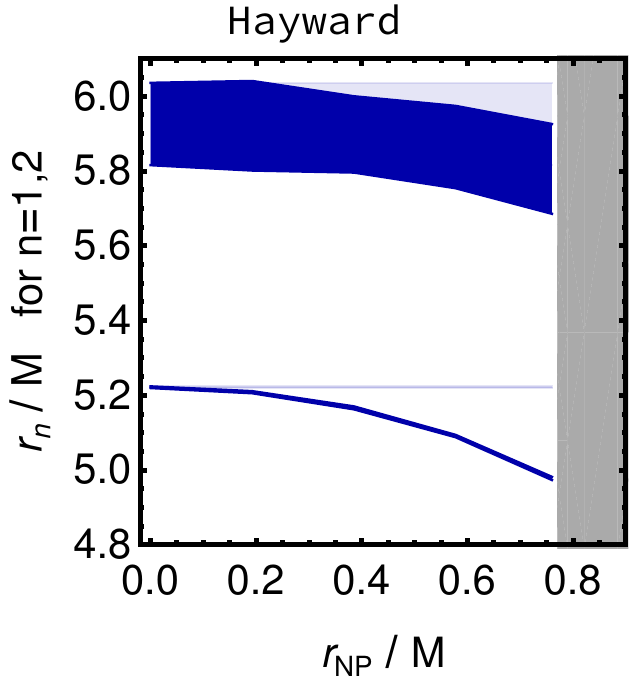}
\caption{\label{fig:rings_avg_disk}
Radial location $r_{\rm n}$ of the $n=1$ (upper band) and $n=2$ (lower band) photon rings in the image plane. 
The dark blue bands indicate the range from the minimal to the maximal possible location (within the investigated set of disk models).
The left- and right-hand panel show the behavior with varying new-physics parameter $r_\text{NP}$ of Simpson-Visser and Hayward mass functions. For comparison, we also continue the range of the respective Schwarzschild value (light blue). The gray region (on the right side of each plot) indicates parameter values which result in horizonless objects.
\newline
For the Simpson-Visser spacetime, we also show the relative separation between intensity peaks, i.e., $(r_1 - r_2)/r_1$. For the Hayward spacetime, the relative separation stays constant within the systematic uncertainties and is therefore not shown. 
}
\end{figure}
Pbservationally resolving $n \geq 1$ rings is challenging, and the challenge grows with increasing $n$, see \cite{Johnson:2019ljv}. Thus, it is important to understand how much information about new physics can be extracted from low $n$, given astrophysical uncertainties in the disk properties. To provide a tentative answer, we vary the parameters of the disk model and extract the image location of the $n=1$ and $n=2$ rings. 
We vary the disk parameters in the following intervals
\begin{align}
	\alpha \in \left[\frac{1}{5},\,1\right],
	\quad
	r_{\rm cut} \in \left[2M,\,12M\right],
	\quad
	w \in \left[1,\,3\right],
	\quad
	h \in \left[\frac{1}{10},\,\frac{1}{2}\right].
\end{align}
The minimum and maximum values for $h$ correspond to a typical geometrically thin-disk and thick-disk model, respectively. The inner cutoff is varied in a wide region around the location of the ISCO; we thus expect it to cover all physically relevant ranges.
For the resulting 108 disk models, we obtain image cross sections from which we extract the peak location of the photon rings. 

Simple, regular black holes have a universal behavior of photon rings, in which the specific choice of mass function $M_{\rm eff}(r)$ only enters quantitatively, but not qualitatively: because $M_{\rm eff}(r)$ decreases with decreasing $r$, high-$n$ photon rings are shifted inwards more strongly than low-$n$ photon rings. Thus, as a universal property of simple, regular black holes, the relative photon ring separation increases.

In Fig.~\ref{fig:rings_avg_disk}, we show the resulting behavior as a function of  the new-physics scale $r_\text{NP}$ for Simpson-Visser and Hayward spacetimes, respectively. Assuming (i) an independent mass measurement with a negligible error (cf.~Sec.~\ref{sec:PN+EHT}), (ii) no prior knowledge about disk physics, and (iii) that the present set of disk models is representative of astrophysical uncertainty, we make the following observations:
\begin{itemize}
	\item
	Resolving the $n=1$ intensity peak is sufficient to constrain a Simpson-Visser-type falloff in the mass function for $r_{\rm NP} \approx 0.2M$.
	\item
	Resolving the $n=1$ intensity peak is insufficient to constrain a Hayward-type polynomial falloff in the mass function. Constraints are even weaker for a Dymnikova-type mass function.
	\item
	Resolving the $n=2$ intensity peak is sufficient to significantly constrain both mass functions.
\end{itemize}
This analysis demonstrates that astrophysics impacts the $n=1$ photon ring (let alone $n=0$) considerably. To the contrary, our analysis indicates that astrophysics does not significantly impact the $n=2$ photon ring which constitutes a powerful probe of the underlying geometry, also~cf.~\cite{Gralla:2020srx,Broderick:2021ohx,Wielgus:2021peu}.

Fig.~\ref{fig:rings_avg_disk} also shows that, as expected, the relative separation between $n=1$ and $n=2$ grows as a function of $r_{\rm NP}$. The growth of relative separation is an intriguing feature, because, at fixed shadow radius, less resolution is required to resolve the difference between the $n=1$ and $n=2$ photon rings than is required for black holes in GR. To explore the implications,
we assume an angular resolution of $10\, \mu \rm as$, which is the nominal resolution achievable with a ground-based VLBI array at a frequency of 345 GHz \cite{Blackburn:2019bly}.
We compare this resolution to  the separation between $n=1$ and $n=2$ (defined as the separation between the center of the bands in Fig.~\ref{fig:rings_avg_disk}), assuming that the radius associated to $n=2$ corresponds to 42 $\mu \rm as$, as in M87*~\cite{paper1,paper6}. This results in an expected separation of about 6 $\mu \rm as$ for $r_{\rm NP} = r_{\rm NP,\, crit}$ for the Simpson-Visser case; while the difference in $n=2$ diameter to $n=1$ diameter amounts to  12 $\mu\rm as$.
For the Hayward case, the expected difference in diameters at $r_{\rm NP} = r_{\rm NP,\, crit}$ would be about 7 $\mu \rm as$, i.e., the ring separation would be about 3.5 $\mu\rm as$.
Thus, a ground-based ngEHT array could be close to achieving a constraint on the Simpson-Visser metric which is independent of mass measurements by other observational missions. 

Of course it should be kept in mind that here, we have proceeded in a  simplified manner and a more thorough analysis that performs an image reconstruction (based on potential ngEHT configurations) of simulated images in the Simpson-Visser spacetime is necessary to make a more robust statement.

\subsubsection{Robustness of $n\geq2$ rings as probes of geometry}
In \cite{Broderick:2021ohx}, an emission region corresponding to a thin ring located relatively close to the horizon led to a reversal of the ordering of the rings: The “standard” ordering, associated to emission regions far enough away from the event horizon, consists in $n=0$, direct emission, arriving at larger distance from the image center than $n=1$, which in turn arrives at larger distance from the center than $n=2$, and so on. This ordering can be reversed for the low-order rings for emission regions located close to the horizon of the Kerr spacetime, such that the image exhibits a “ring flip”. However, the image location of $n \geq 2$ rings is, for practical purposes, independent of the location of the emission region. This result further strengthens the case that geometric information can be separated from astrophysical information and thus black-hole images can be used to test GR.
\begin{figure}
\centering
	\includegraphics[height=0.3\linewidth]{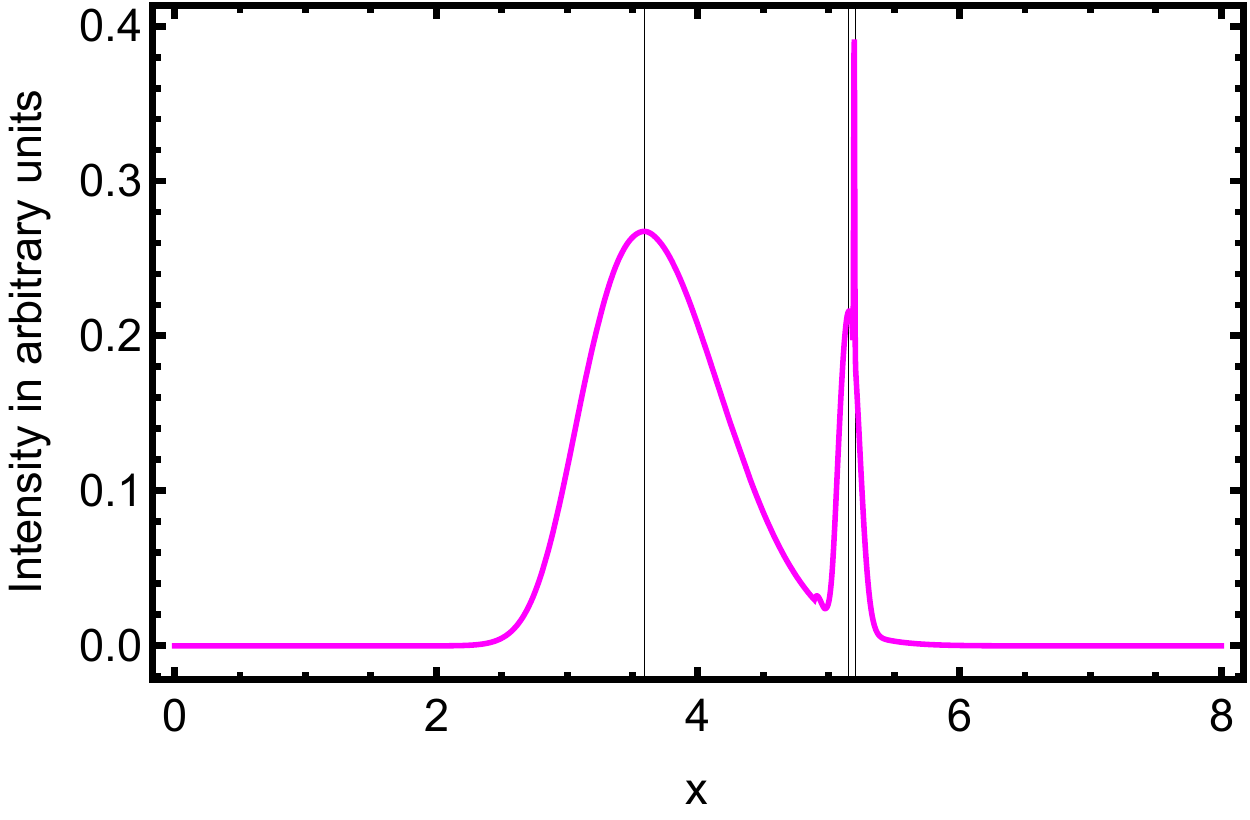}
	\hspace*{1em}
	\includegraphics[height=0.3\linewidth]{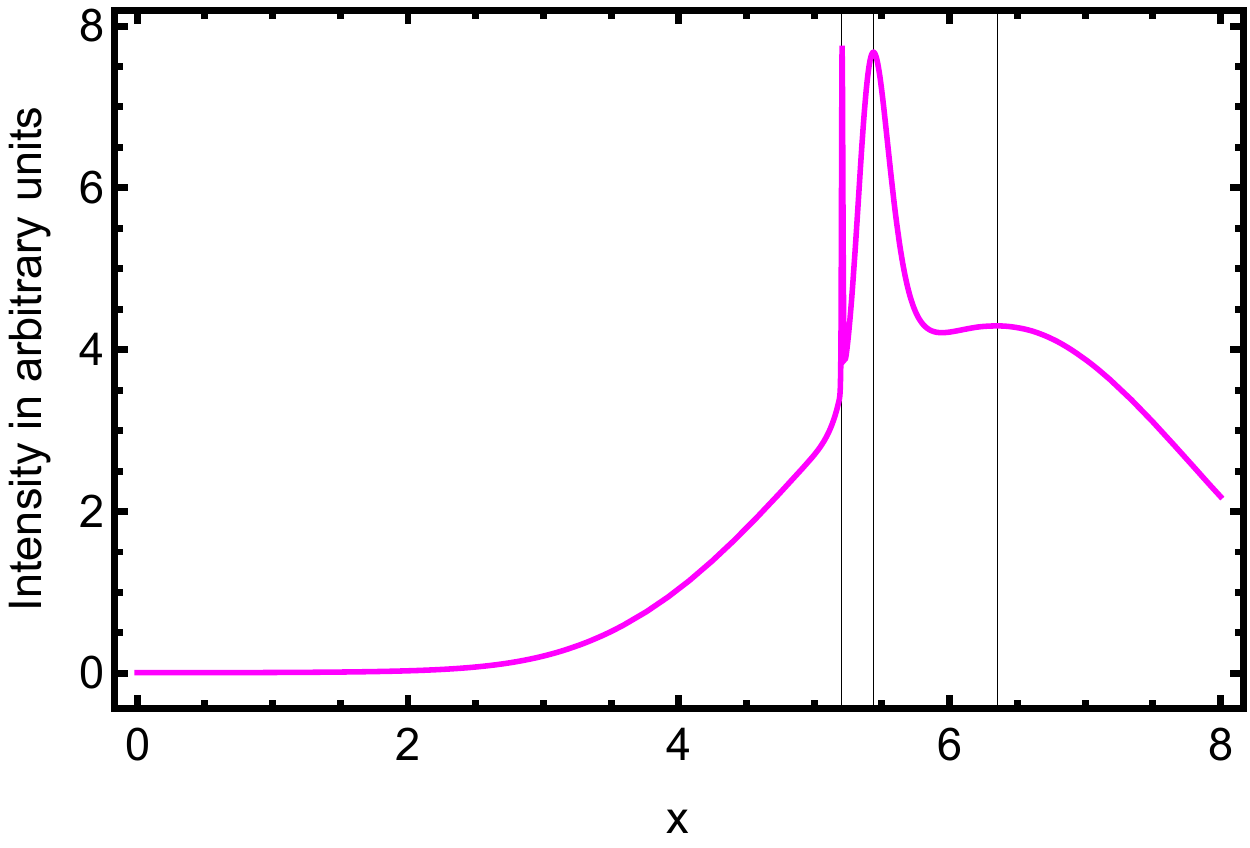}
	\\
	\includegraphics[height=0.3\linewidth]{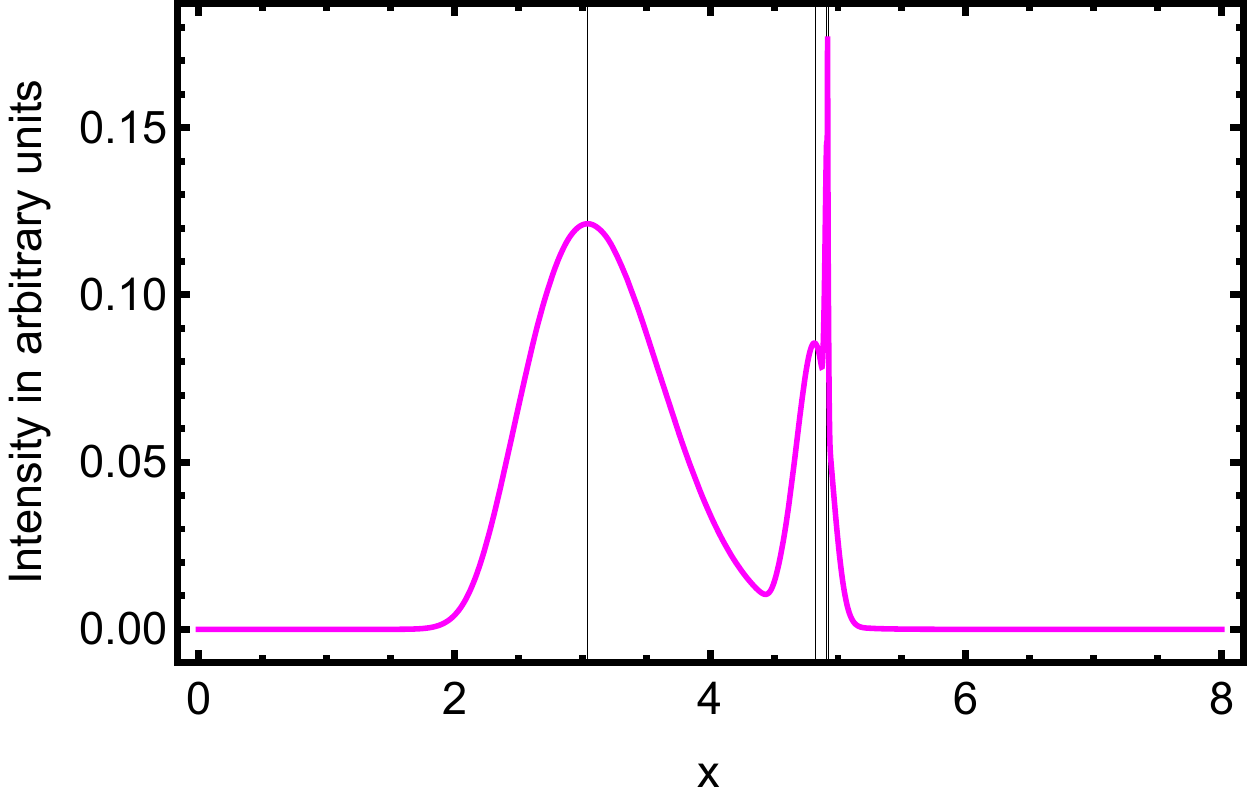}
	\hspace*{1em}
	\includegraphics[height=0.3\linewidth]{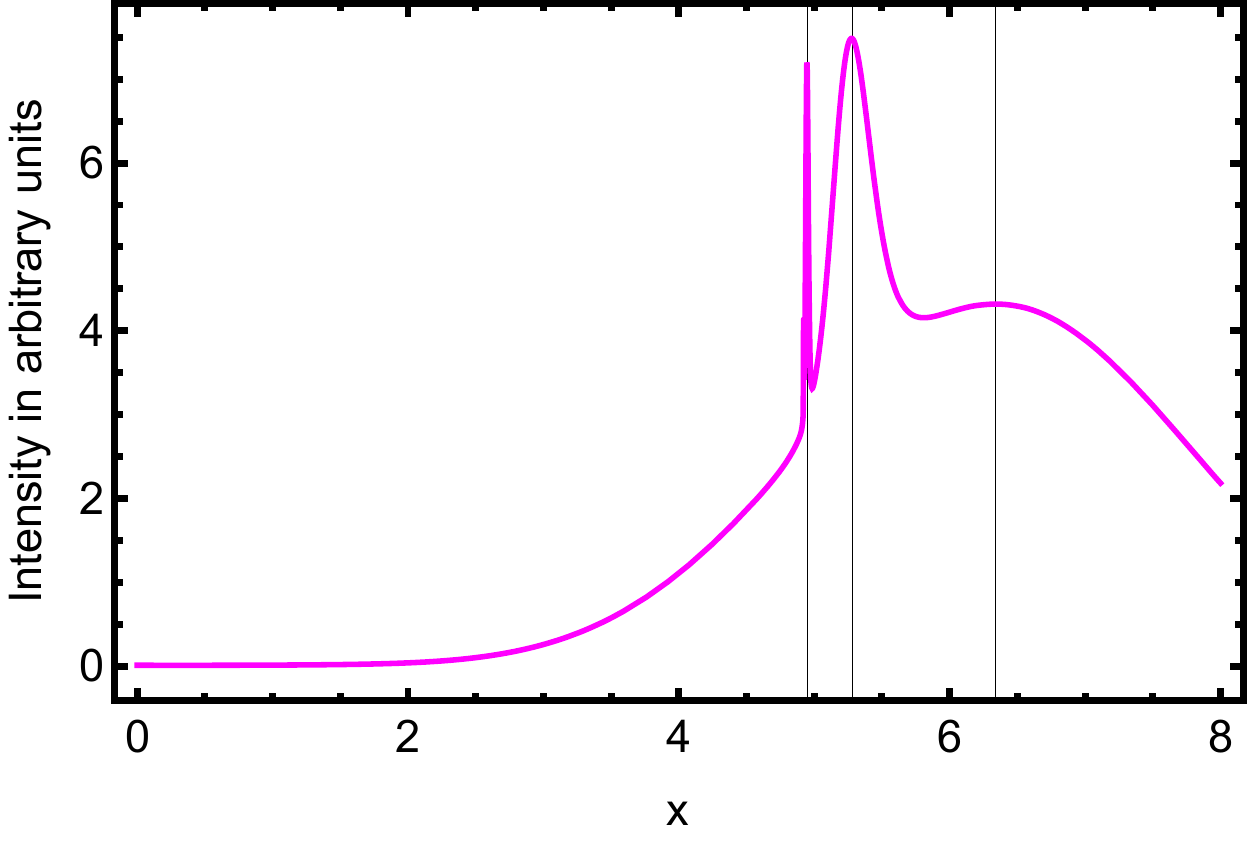}
	\\
	\includegraphics[height=0.3\linewidth]{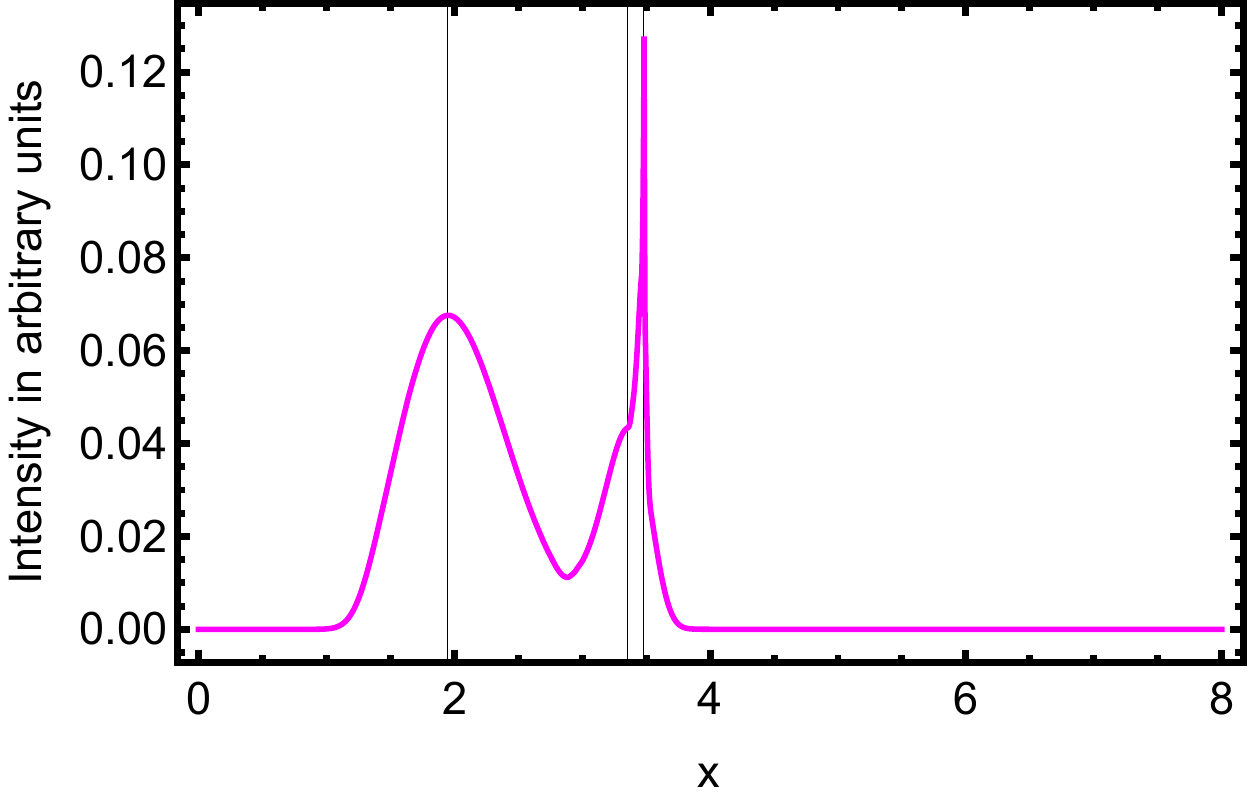}
	\hspace*{1em}
	\includegraphics[height=0.3\linewidth]{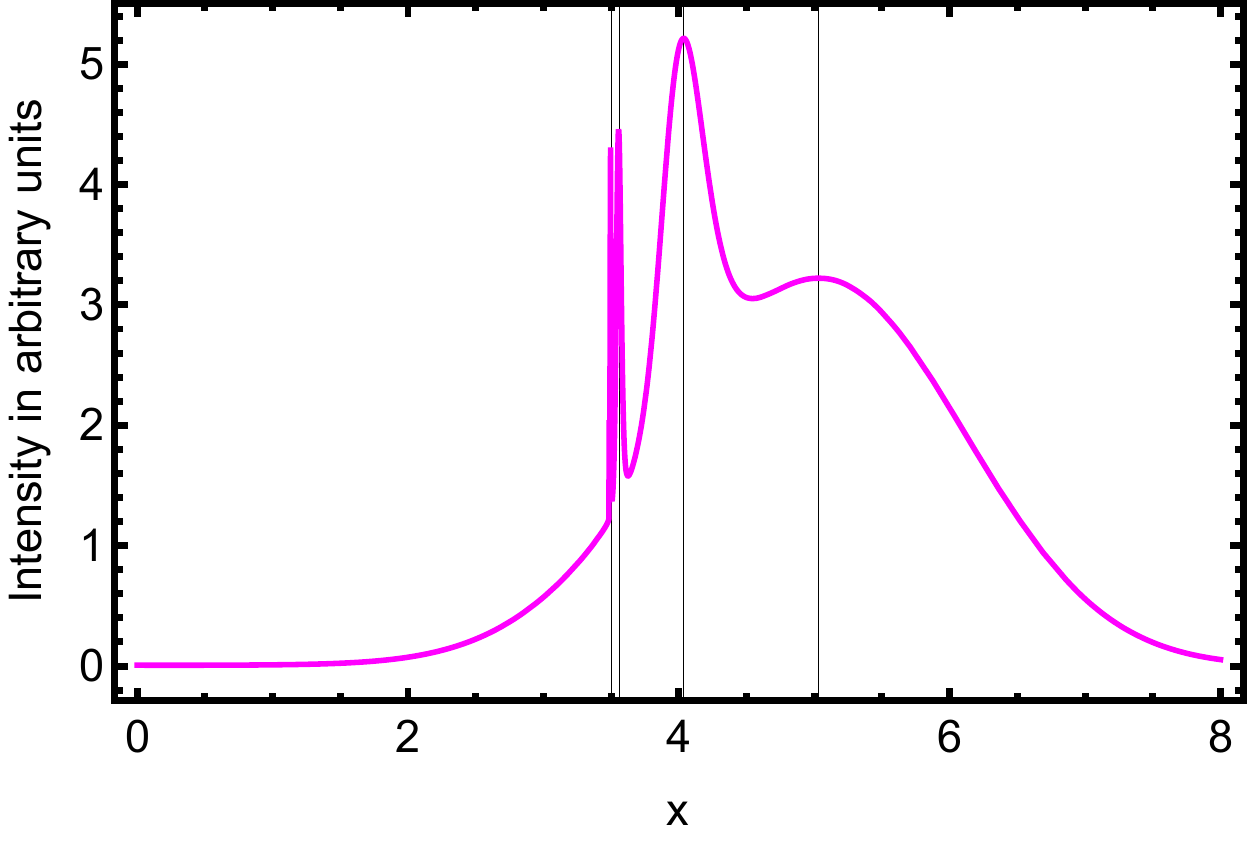}
\caption{\label{fig:ringflip}  Schwarzschild (upper panels), Hayward (central panels) and Simpson-Visser (lower panels). In the three cases, the location of $n=0$ and $n=1$ varies significantly between the left and right panels, but is approximately constant for $n=2$. In the left panels, a thin ring is located very close to the horizon, in the lower right panel the emission region is located further away from the horizon.
	}
\end{figure}

We explore whether both results also hold in simple, regular spacetimes. 
We work with an emission profile of the form
\be
n_{\rm ring}(r, \theta)= r^{-\alpha}\, \exp[-\frac{\cos(\theta)^2}{2h^2}]  \exp[- \frac{(r-r_{\rm cut})^2}{w^2}], 
\ee
with $\alpha=1/8$, $h=1/5$ and $w$ similar to $r_{\rm cut}$. In contrast to the extended accretion disk in Eq.~\eqref{eq:disk-model}, $n_{\rm ring}(r,\theta)$ models a narrow ring, i.e., a radially localized emission profile.

For the regular, simple and local spacetimes that we explore here, the spacetime geometry counteracts the possibility of a “ring flip”: The new-physics effects increase the distance between the rings. If this effect was strong enough, it might not be possible to move the emission region to small enough regions to achieve a “ring flip” of the $n \geq 1$ rings, before the emission region disappears behind the horizon.

As shown in Fig.~\ref{fig:ringflip}, placing the emission region close enough to the horizon of the simple, regular black hole also reverses the ordering of the rings, at least for $n=0,1,2$ that we focus on here.

The reversed ordering of the $n=0$, $n=1$, and $n=2$ light rings provides a particularly striking demonstration of how astrophysics can impact the appearance of low-order photon rings. 
At the same time, the image location of the $n=2$ ring stays essentially constant, even while the other image features, and in particular the image location of the diffuse emission ($n=0$) change dramatically. This strengthens the case that, despite astrophysical uncertainties, meaningful tests of the Kerr paradigm are possible by resolving the $n=2$ photon ring.

\subsection{Inhomogeneous disk models and localized emission}
\label{sec:localized-emission}

Inhomogeneous emission regions, e.g., hotspots \cite{Broderick:2005my,Broderick:2005jj} can provide a more informative map of the spacetime than homogeneous emission and make spacetime tomography possible \cite{Tiede:2020jgo}. From several flaring events at different locations of the disk, a spacetime map can, in principle, be reconstructed based on the lensed images of the flares.
To take a modest first step into the direction of spacetime tomography for simple, regular black holes, we model a localized emission region as a Gaussian in $r, \theta$ and $\phi$, which we center close to the horizon.
In realistic settings, emission from close to the horizon is strongly redshifted. Thus, it depends on the emitted frequencies whether there will be detectable intensity in the EHT band. We leave this question to future work, and thus do not account for redshift here.\\
 For our simple study we consider the emitting matter density to be localized in
a Gaussian patch centred at $(r_0,\theta_0,\phi_0)$, described by the following matter density
\be 
 n_{\rm patch}(r,\theta,\phi) = A \exp\left[-\frac{1}{2}\left(\frac{(r-r_0)^2}{\sigma_r}+ \frac{(\theta-\theta_0)^2}{\sigma_\theta} + \frac{(\phi-\phi_0)^2}{\sigma_\phi}\right)\right],
\ee
with $A$ setting the amplitude, and $\sigma_r$, $\sigma_\theta$, and $\sigma_\phi$ setting the  width of the  patch in $r,\theta,\phi$.
In our conventions, the choice $\phi_0=\pi$ corresponds to a patch located centrally behind the black hole; for $\phi_0=\pi/2$ it is located on the right edge of the black hole. 
\\
\begin{figure*}[!t]
\begin{center}
	\begin{tabular}{m{0.01\linewidth}m{0.3\linewidth}m{0.3\linewidth}m{0.3\linewidth}}
		$\phi_0$ & \hfill$M_{\rm SV} = 1$\hfill${}$ & \hfill$M_{\rm Schw} \simeq 0.81$\hfill${}$ & \hfill$M_{\rm Hay} \simeq 0.85$\hfill${}$
		\\
		$\frac{\pi}{2}$ &
		\includegraphics[width=\linewidth]{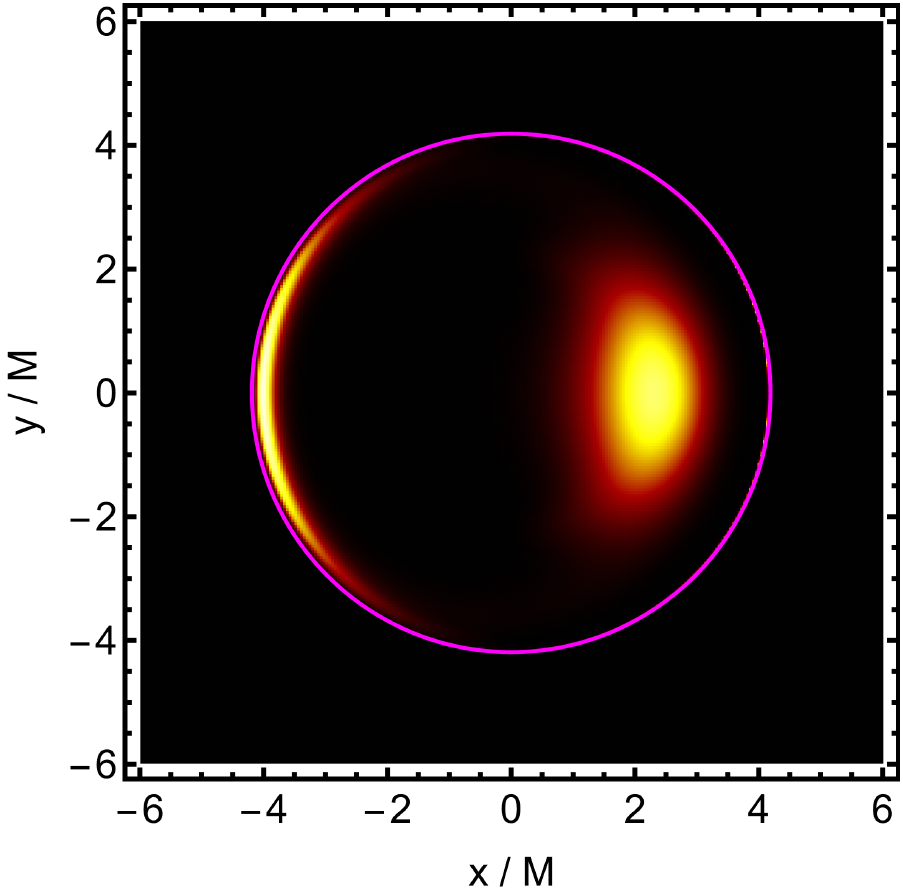} &
		\includegraphics[width=\linewidth]{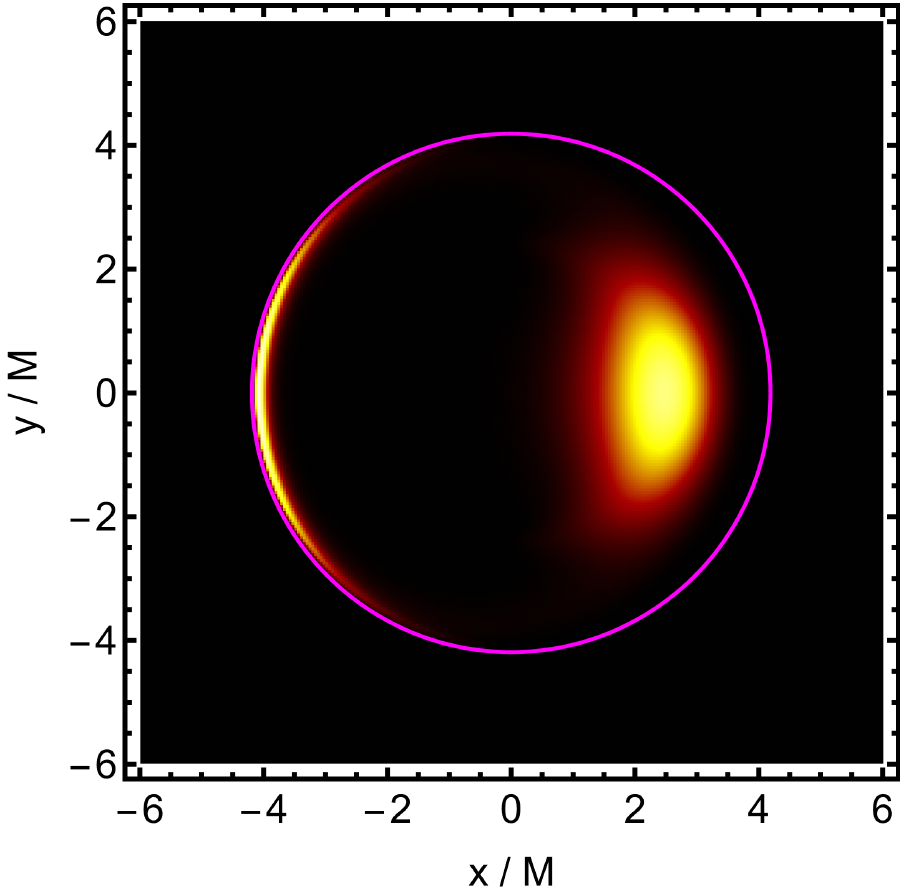} &
		\includegraphics[width=\linewidth]{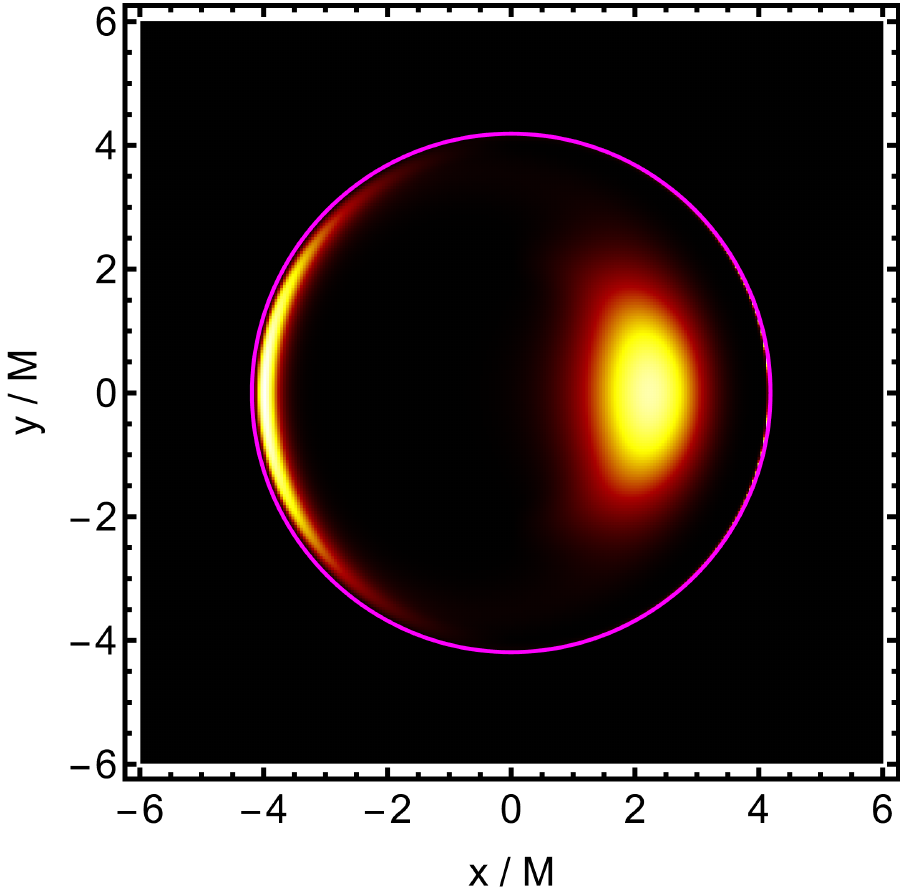}
		\\
		$\frac{3\pi}{4}$ &
		\includegraphics[width=\linewidth]{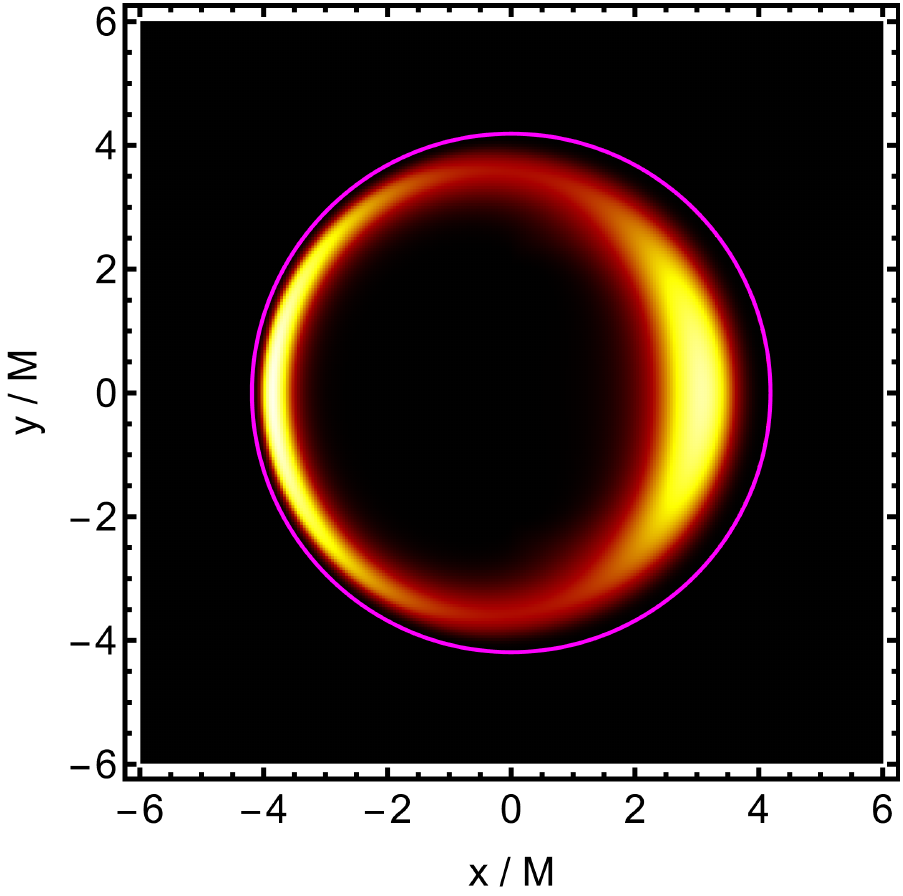} &
		\includegraphics[width=\linewidth]{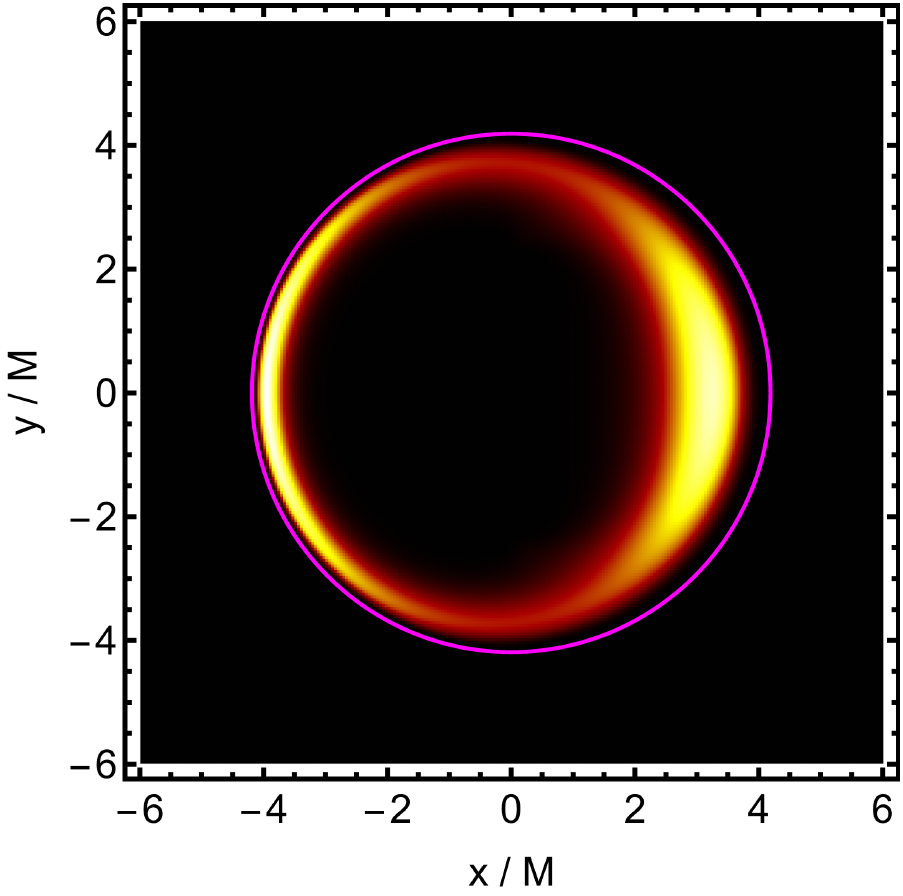} &
		\includegraphics[width=\linewidth]{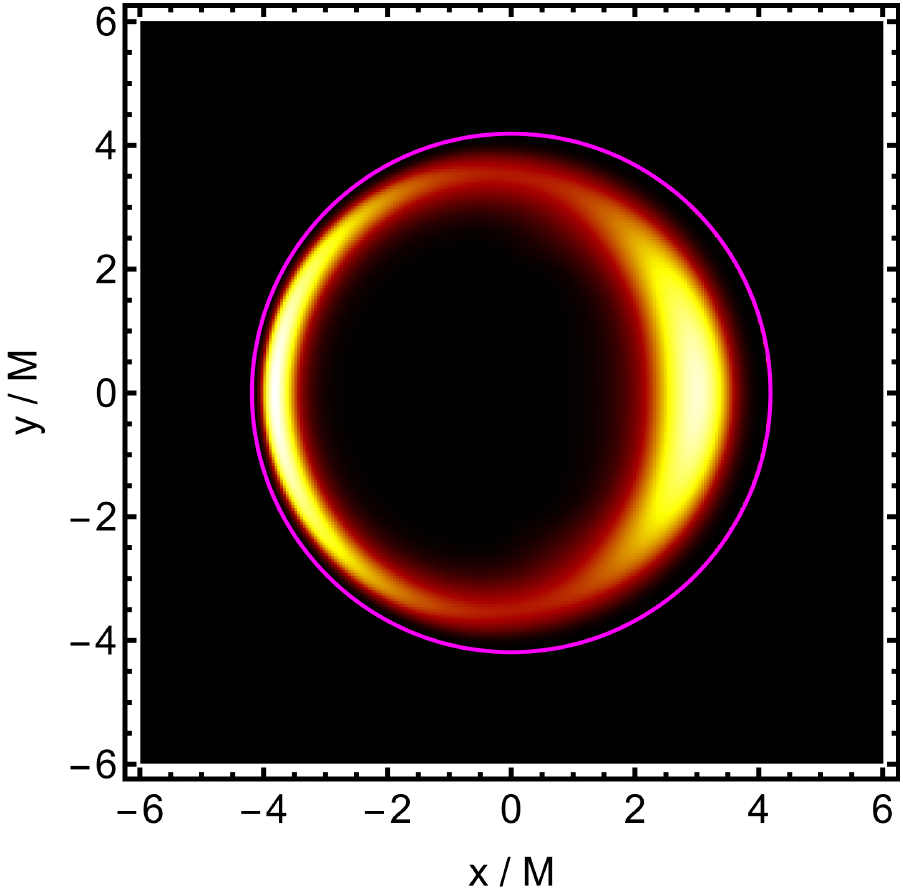}
		\\
		$\pi$ &
		\includegraphics[width=\linewidth]{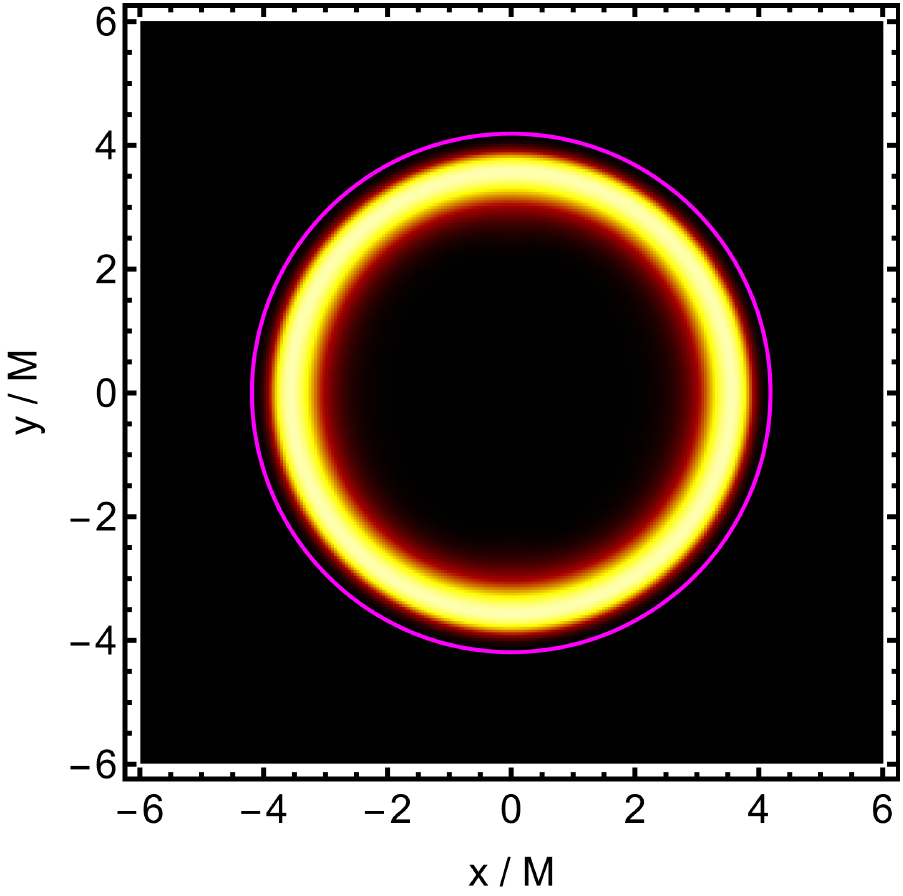} &
		\includegraphics[width=\linewidth]{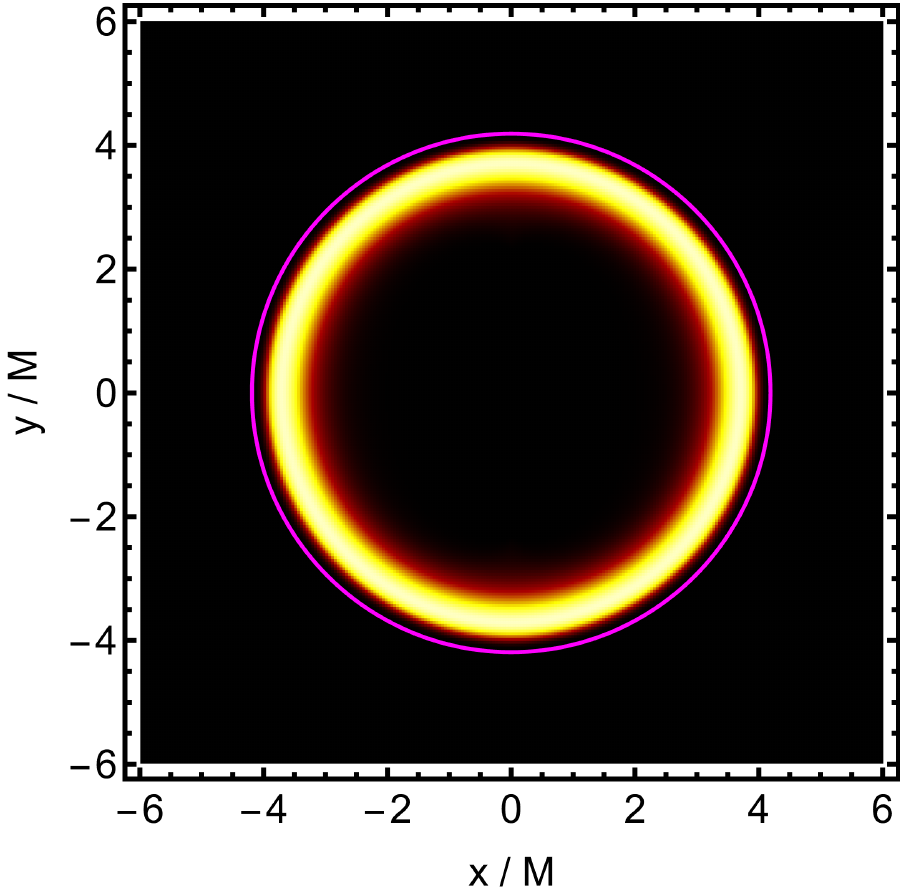} &
		\includegraphics[width=\linewidth]{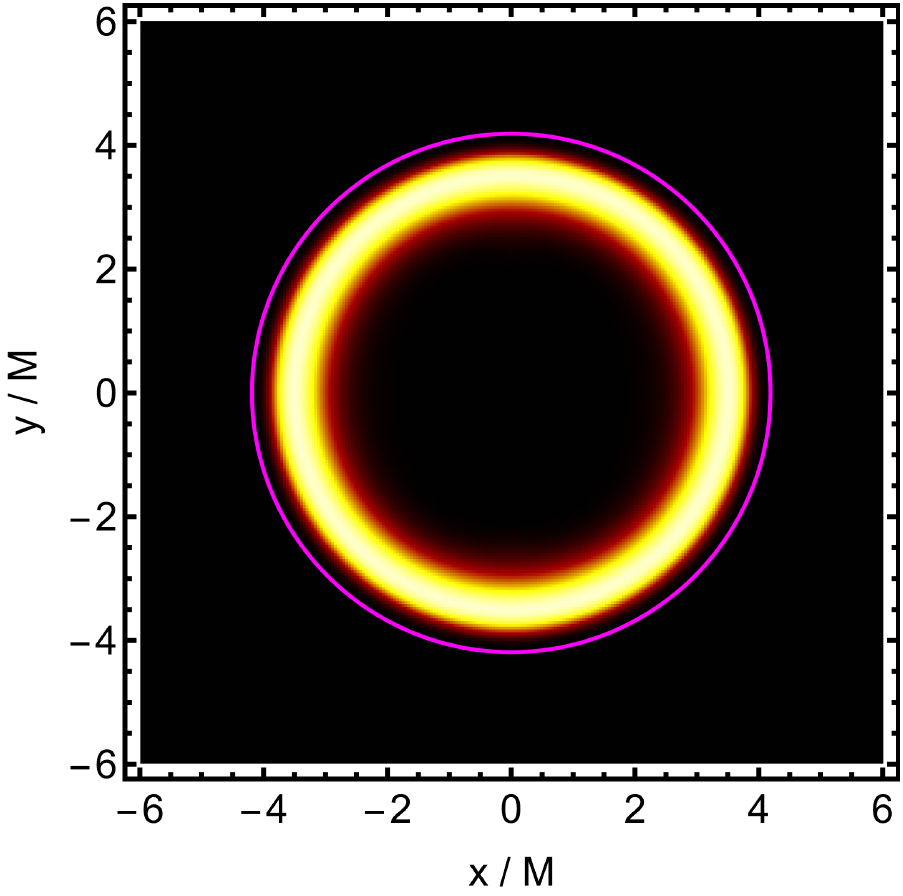}
	\end{tabular}
\end{center}
	\vspace*{-10pt}
	\caption{
	\label{fig:blobintensities}
	We show images for Simpson-Visser (left columns), Schwarzschild (central columns) and Hayward spacetimes (right columns).
	The localized emission regions are centred about $r_0=1.5M_{\rm SV}$ and $\theta=\pi/2$, with $\phi_0= \pi/2$ (top row), $\phi_0=3\pi/4$ (central row) and $\phi_0=\pi$ (bottom row).
	The shadow boundary, which is not visible in the images, is indicated by the magenta circles. Given that the emission regions lie inside the photon sphere, regions of high intensity lie inside the shadow boundary.}
\end{figure*}

\begin{figure*}[!t]
\begin{center}
	\begin{tabular}{m{0.01\linewidth}m{0.33\linewidth}m{0.33\linewidth}}
		$\phi_0$ & \hfill$M_{\rm Schw}$ vs. $M_{\rm SV}$\hfill${}$ & \hfill$M_{\rm Schw}$ vs. $M_{\rm Hay}$\hfill${}$
		\\
		$\frac{\pi}{2}$ &
		\includegraphics[width=\linewidth]{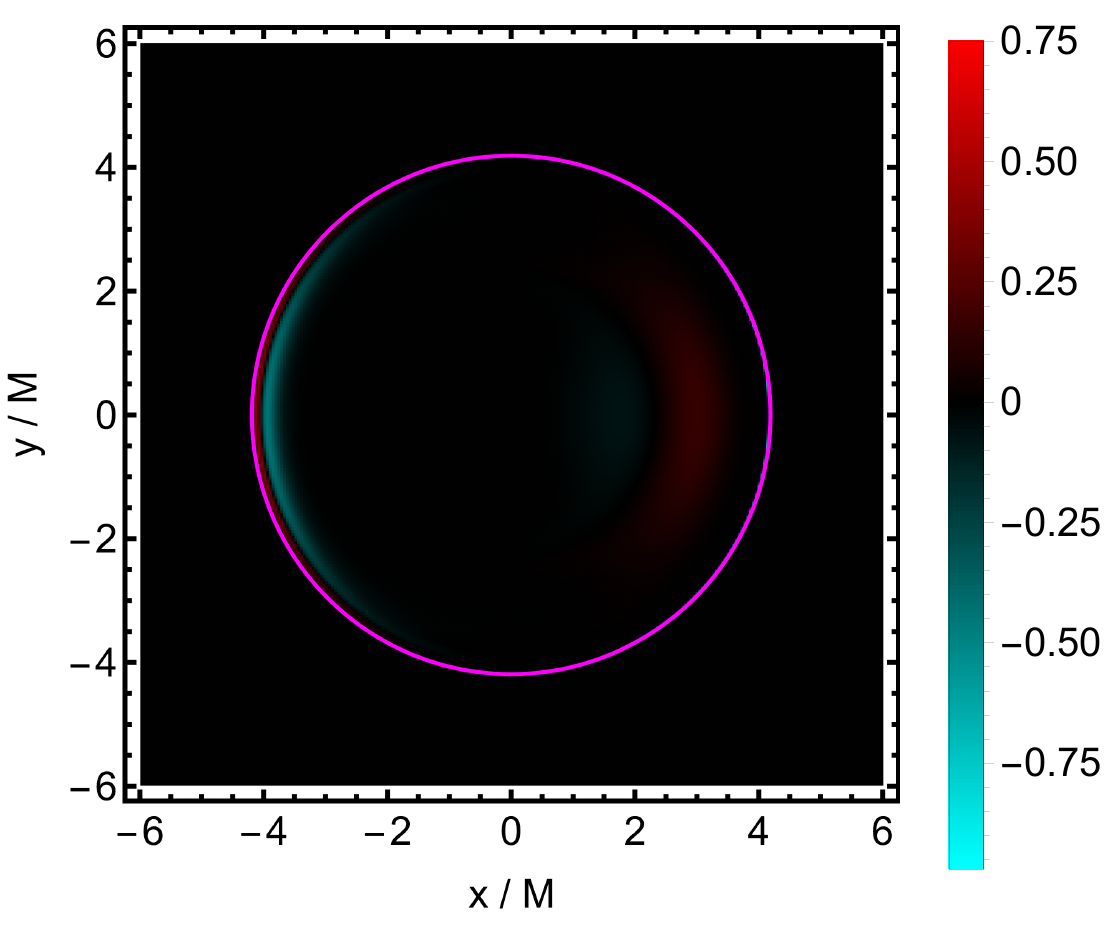} &
		\includegraphics[width=\linewidth]{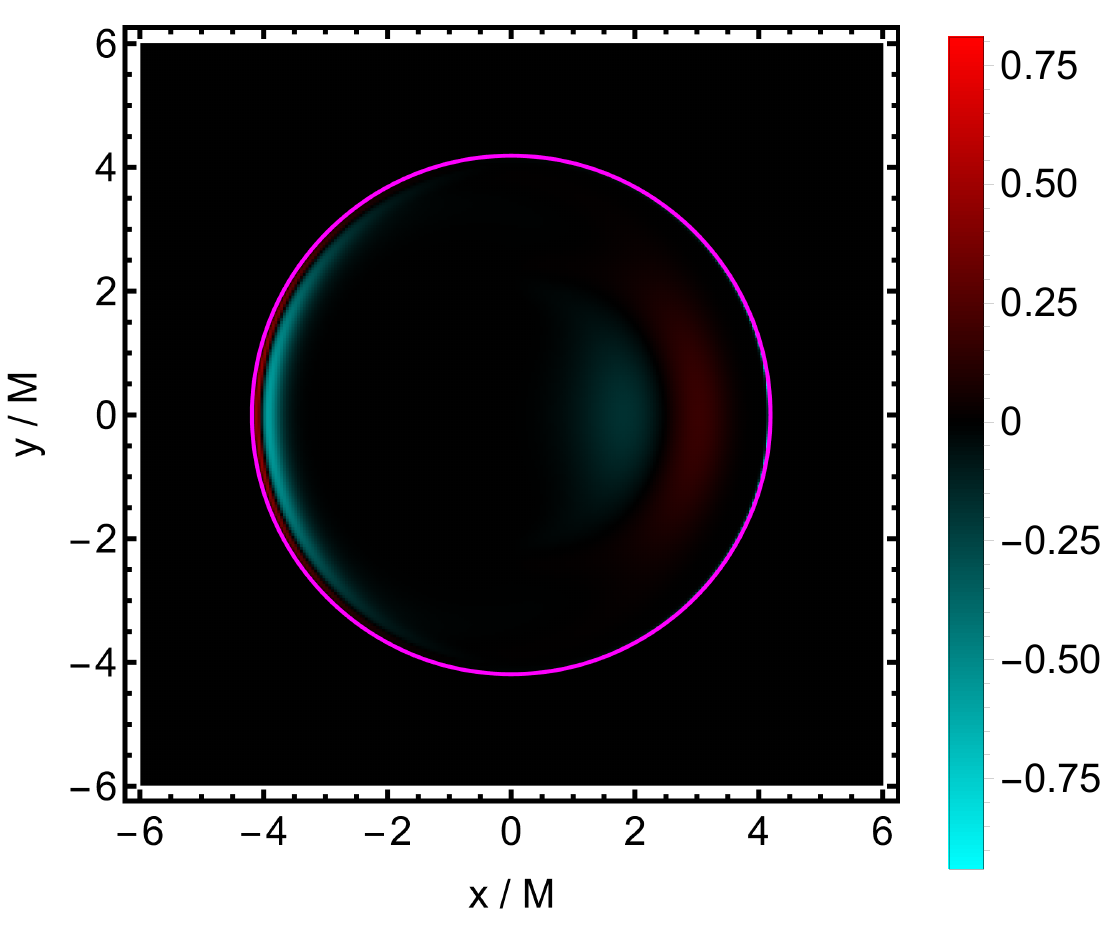}
		\\
		$\frac{3\pi}{4}$ &
		\includegraphics[width=\linewidth]{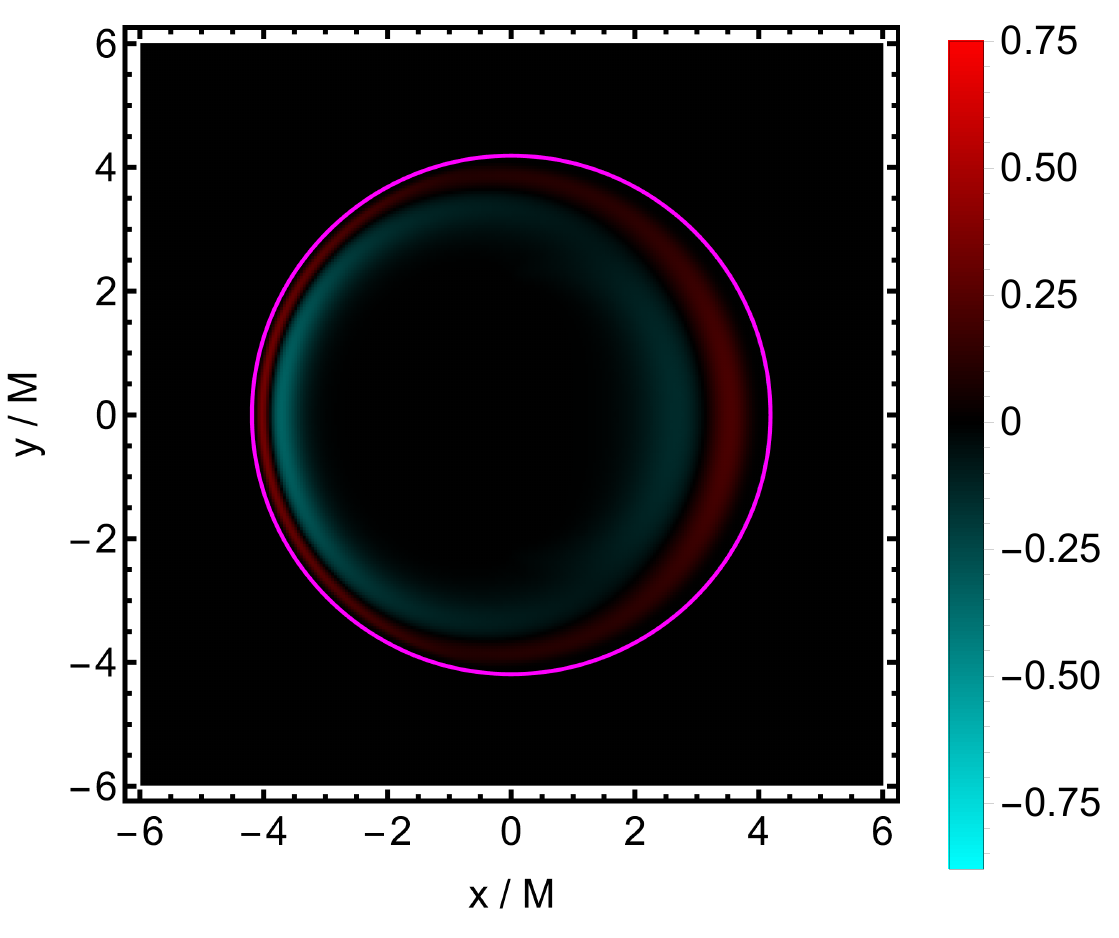} &
		\includegraphics[width=\linewidth]{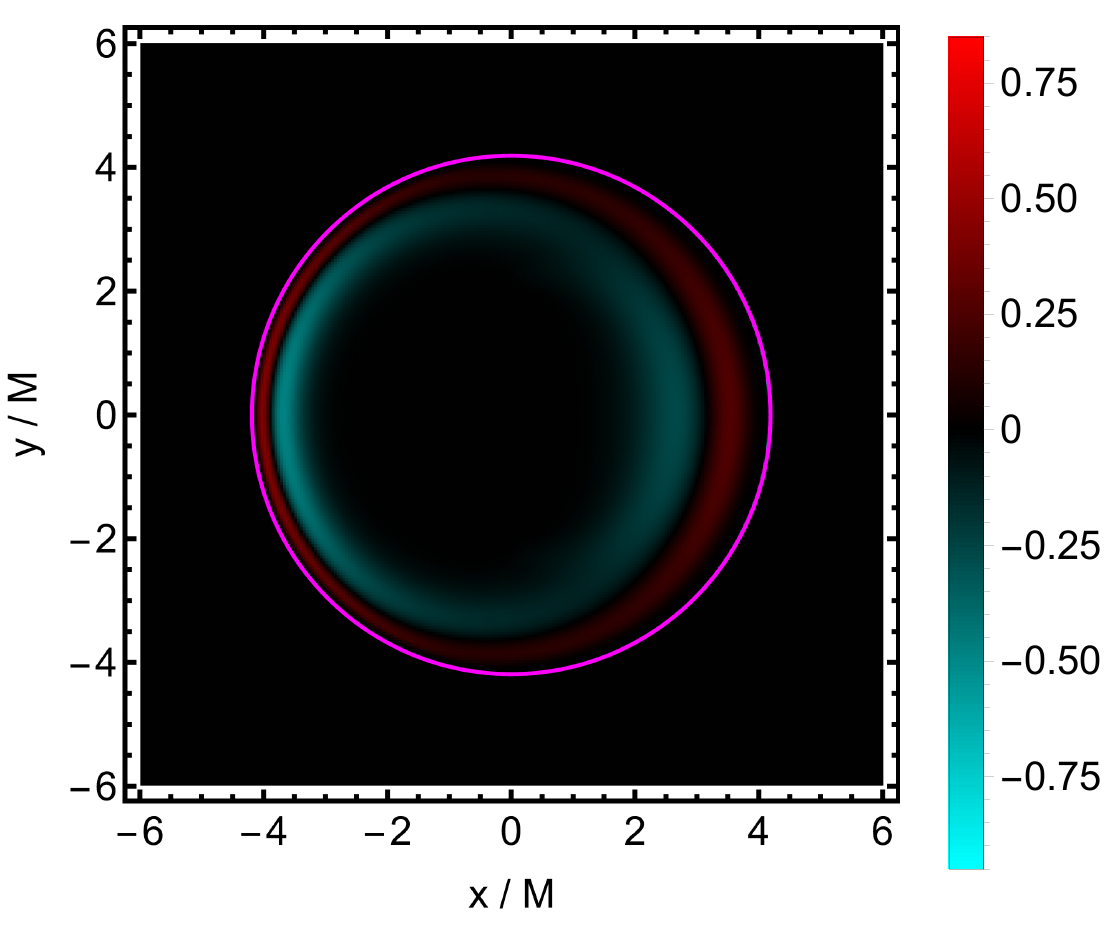}
		\\
		$\pi$ &
		\includegraphics[width=\linewidth]{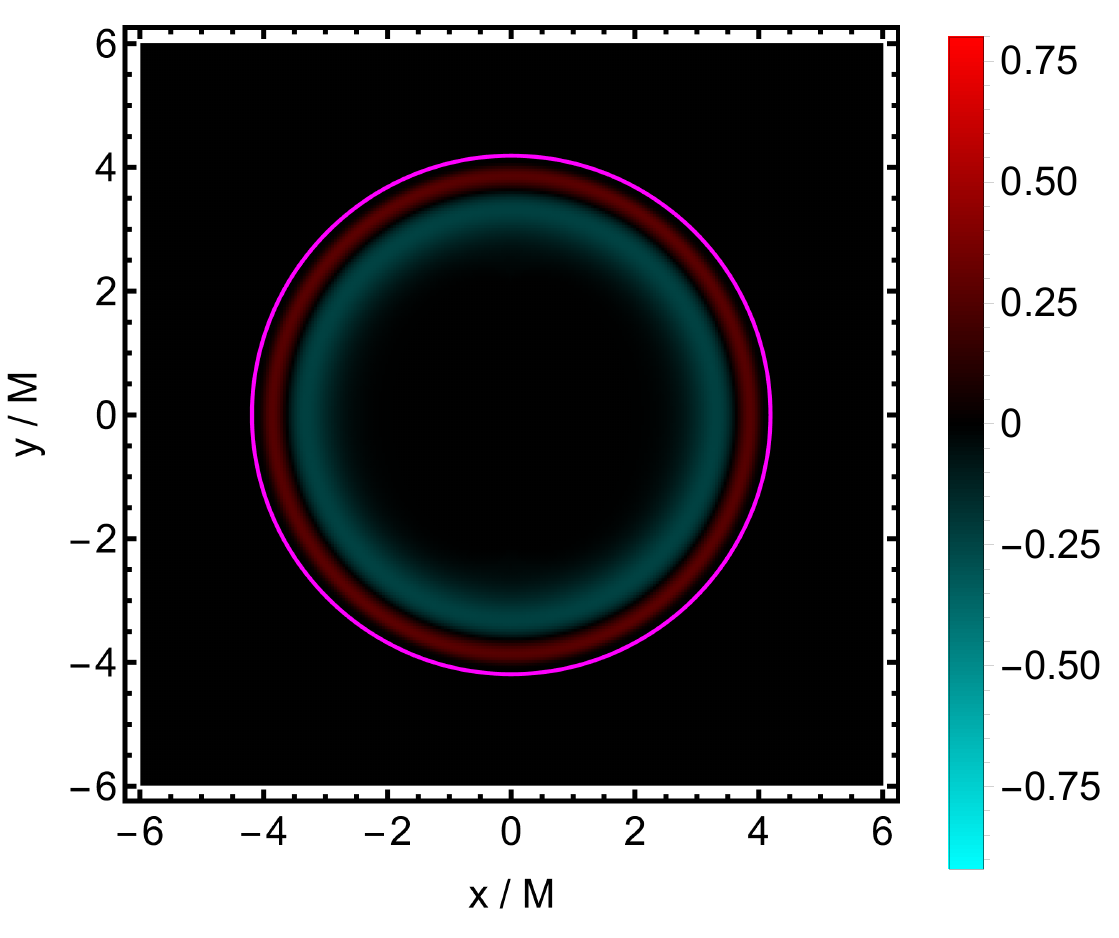} &
		\includegraphics[width=\linewidth]{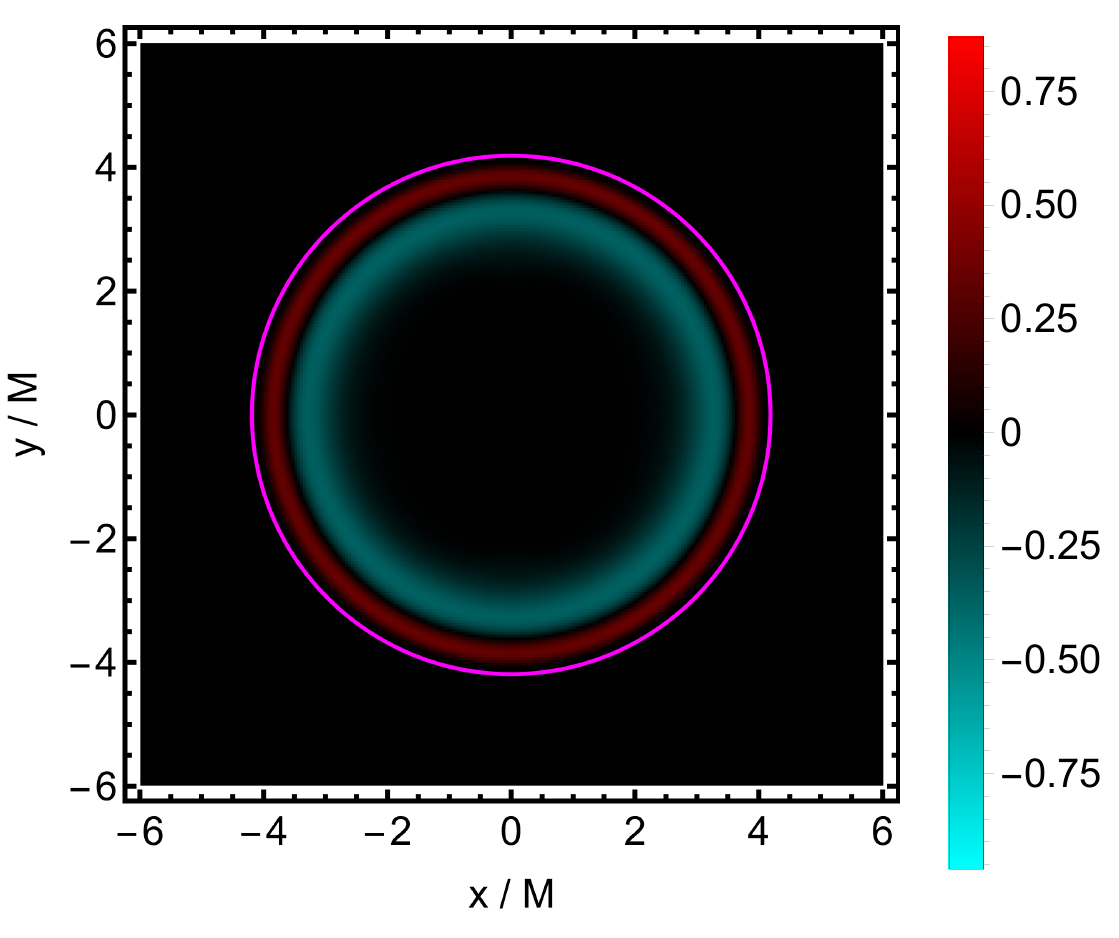}
	\end{tabular}
\end{center}
	\vspace*{-10pt}
	\caption{
	\label{fig:subtractedintensities}
	We show the subtracted, normalized intensities for black-hole spacetimes featuring a localized emission region centred about $r_0=1.5M_{\rm SV}$ and $\theta_0=\pi/2$, shown in Fig.~\ref{fig:blobintensities}. We subtract the image intensity of the Simpson-Visser black hole from the Schwarzschild black hole (left column), which are shown in the left and central column of Fig.~\ref{fig:blobintensities}; and similarly for the Hayward case (right column).
	}
\end{figure*}

\begin{figure*}[!t]
\begin{center}
	\begin{tabular}{m{0.01\linewidth}m{0.3\linewidth}m{0.3\linewidth}m{0.3\linewidth}}
		$\phi_0$ & \hfill$M_{\rm SV} = 1$\hfill${}$ & \hfill$M_{\rm Schw} \simeq 0.81$\hfill${}$ & \hfill$M_{\rm Hay} \simeq 0.85$\hfill${}$
		\\
		$\frac{3\pi}{4}$ &
		\includegraphics[width=\linewidth]{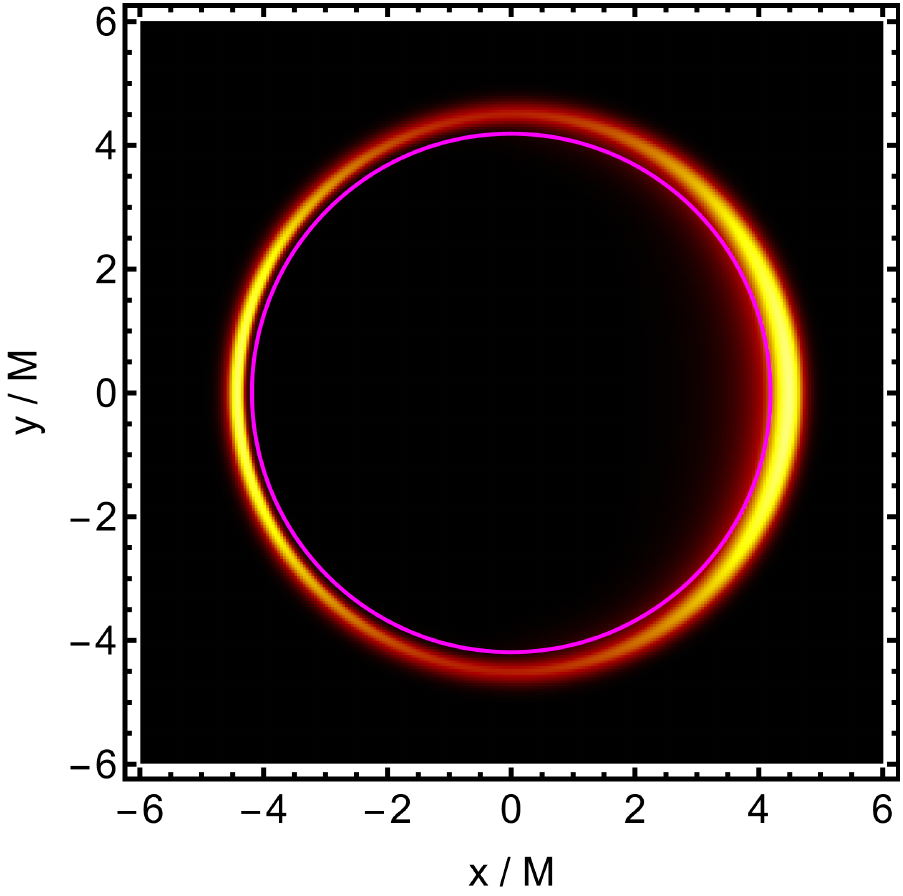} &
		\includegraphics[width=\linewidth]{SimVis_r0_3_phi0_3Pi_4_shadow.pdf} &
		\includegraphics[width=\linewidth]{SimVis_r0_3_phi0_3Pi_4_shadow.pdf}
	\end{tabular}
\end{center}
	\vspace*{-10pt}
	\caption{
	\label{fig:blobintensities_larger_r0}
	We show images for Simpson-Visser (left), Schwarzschild (center) and Hayward spacetimes (right).
	The localized emission regions are centred about $r_0=3M_{\rm SV}$ and $\theta=\pi/2$, with  $\phi_0=3\pi/4$.
	The shadow boundary, which is not visible in the images, is indicated by the magenta circles. Because the emission region is mostly localized outside the photon sphere, the image intensity is concentrated outside of the shadow boundary.
	}
\end{figure*}

\begin{figure*}[!t]
\begin{center}
	\begin{tabular}{m{0.01\linewidth}m{0.33\linewidth}m{0.33\linewidth}}
		$\phi_0$ & \hfill$M_{\rm Schw}$ vs. $M_{\rm SV}$\hfill${}$ & \hfill$M_{\rm Schw}$ vs. $M_{\rm Hay}$\hfill${}$
		\\
		$\frac{3\pi}{4}$ &
		\includegraphics[width=\linewidth]{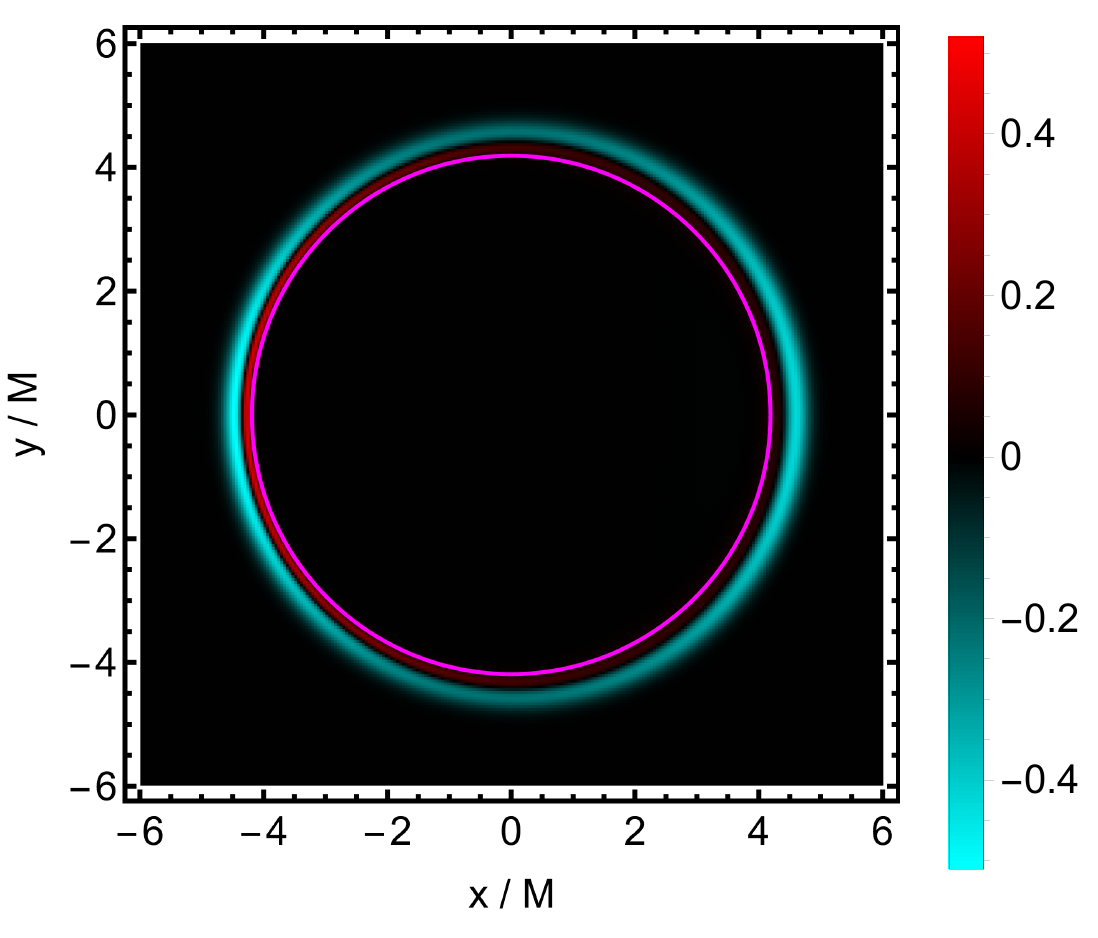} &
		\includegraphics[width=\linewidth]{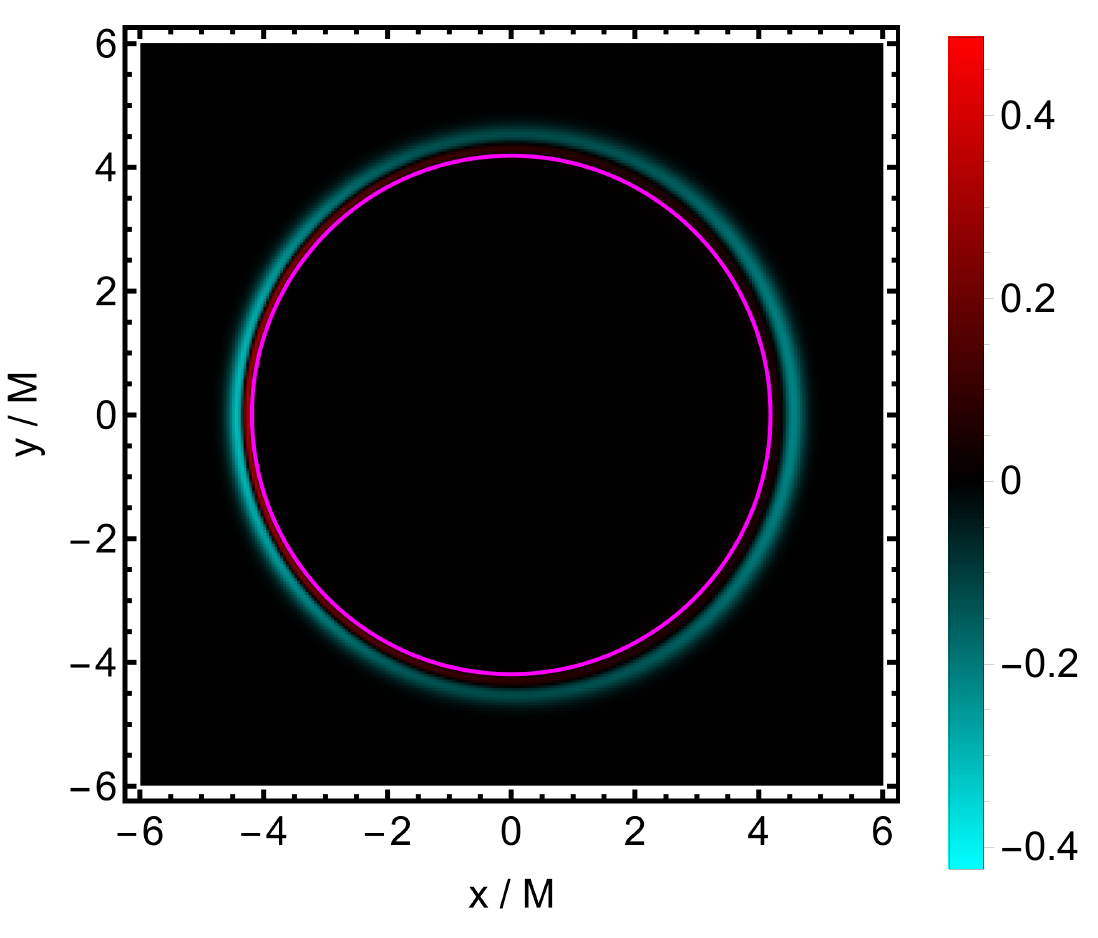}
	\end{tabular}
\end{center}
	\vspace*{-10pt}
	\caption{
	\label{fig:subtractedintensities_larger_r0}
	We show the subtracted, normalized intensities for black-hole spacetimes featuring a localized emission region centred about $r_0=3M_{\rm SV}$ and $\theta_0=\pi/2$ shown in Fig.~\ref{fig:blobintensities_larger_r0}. We subtract the image intensity of the Simpson-Visser black hole from the Schwarzschild black hole (left panel), which are shown in the left and central panel of Fig.~\ref{fig:blobintensities_larger_r0}; and similarly for the Hayward case (right panel).}
\end{figure*}

We choose $r_0$ close to the horizon, where deviations of the spacetime from the Schwarzschild spacetime are maximized. To enable a meaningful comparison of images with the Schwarzschild spacetime, we compare with a Schwarzschild spacetime with rescaled mass parameter, such that the size of the shadow agrees for all spacetimes we study. The rationale behind this choice is the assumption that a previous observation, e.g., with a homogeneous accretion disk, would already have allowed to determine the shadow size. A subsequent observation of an inhomogeneity in the emission could then be based on this calibration. 
The shadow radius $\Lambda_{\rm c}$ can only be given analytically for the Schwarzschild spacetime, where it is
	\be
		\Lambda_{\rm c,\,\rm Schw} = 3\sqrt{3}M_{\rm Schw},
	\ee
whereas for the Hayward and Simpson-Visser spacetime the corresponding equations need to be solved numerically and can be fit by the polynomials in Eq.~\eqref{eq:shadowHayward} and \eqref{eq:shadowSV}.

We use a Simpson-Visser black hole to set the scale, i.e., we set $M_{\rm SV}=1$ for the Simpson-Visser metric, and choose the new-physics parameter $r_{\rm NP}/M_{\rm SV} = 0.5^{3/2}$ in that metric. For a Hayward black hole, which would have a larger shadow size at the same mass, we choose a new-physics parameter $r_{\rm NP}/M_{\rm SV}=\sqrt{5}/3$ and rescale the mass to $M_{\rm Hay}\approx 0.85$ such that the shadow boundary agrees with the Simpson-Visser case.
Finally, the Schwarzschild black hole would have the largest shadow diameter at $M=1$ and thus requires the largest rescaling to $M_{\rm Schw}\approx 0.81$ to match to the same shadow boundary.

In Fig.~\ref{fig:blobintensities} and Fig.~\ref{fig:blobintensities_larger_r0}, we show the resulting images at different choices of the azimuthal angle $\phi_0$ for the placement of the emission region. We observe a commonality between all images in Fig.~\ref{fig:blobintensities}: Because the emission region is located inside the photon sphere, the visible intensity in the images is located inside the shadow boundary. The shadow boundary itself is not visible in the images (but its location is indicated in the plots for clarity). In the case of Fig.~\ref{fig:blobintensities_larger_r0}, where the emission region is centred outside the photon sphere, the image intensity lies outside of the shadow boundary.

We also observe differences between the Schwarzschild case and the two simple, regular black holes. For both of the regular black holes, the lensed image of the emission region (which appears on the left-hand side of the images) is slightly broader than for the image of Schwarzschild spacetime. To make this comparison clearer, we show the subtracted intensities in Fig.~\ref{fig:subtractedintensities}, where we normalize each individual image to peak intensity 1 in arbitrary reference units.
We see differences in the locations of image intensity.

The clear differences visible in Fig.~\ref{fig:subtractedintensities} support the case that a spacetime tomography \cite{2020ApJ...892..132T} for simple, regular black holes could be promising. In line with our previous results that higher-order photon rings lie further inwards, also our study of localized emission patches shows that the image intensities for simple, regular black holes are located at smaller image radii than in the Schwarzschild case. These lensing effects are again qualitatively (though not quantitatively) universal for all regular, simple black holes.
\\

\section{Horizonless objects with and without photon spheres}\label{sec:horless}

\begin{figure}
\begin{center}
\includegraphics[width=0.7\linewidth]{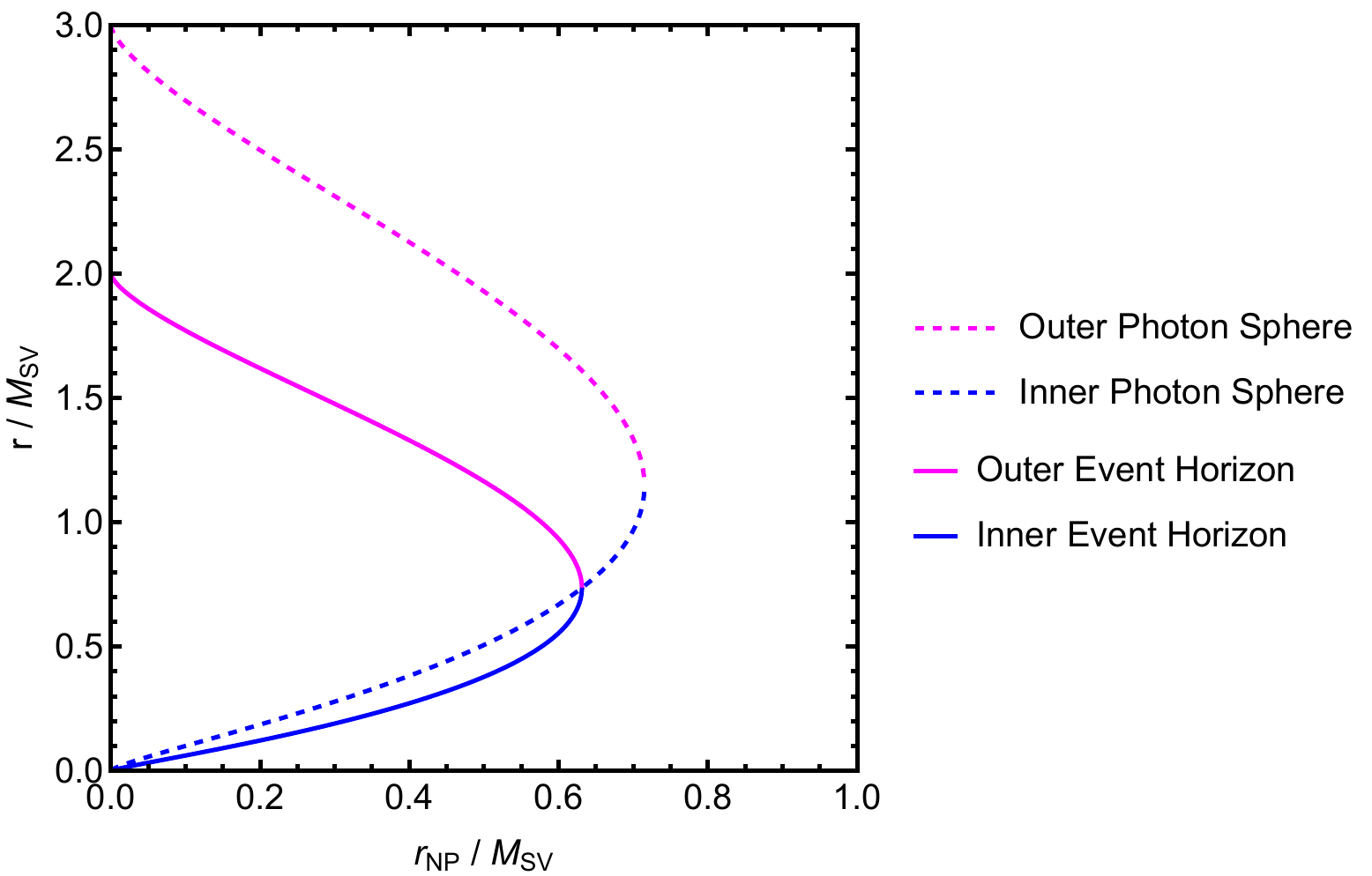}
\end{center}
\caption{\label{fig:photon sphere} We show the radial coordinate of the inner and outer event horizon and photon sphere for a Simpson-Visser spacetime, cf.~\cite{Berry:2020ntz}. For a range of $r_{\rm NP}$, a horizonless object with two intact photon spheres exists, before the photon spheres annihilate for even larger $r_{\rm NP}$.}
\end{figure}

Increasing $r_{\rm NP}$ beyond $r_{\rm NP,\, crit}$ results in a loss of the event horizon. In contrast to GR, where a Schwarzschild spacetime without horizon contains a naked singularity, making it non-viable from a theoretical point of view, simple regular spacetimes do not contain a curvature singularity.
Such a geometry may or may not be an astrophysically viable alternative to a black hole\footnote{One may expect that the accreted matter forms a compact object which emits radiation. A priori, one may expect this radiation to be detectable -- possibly outside the EHT band, depending on the emission frequency and redshift. In \cite{Lu:2017vdx} it has been shown that strong lensing implies that just a fraction of radiation escapes to infinity, see also \cite{Carballo-Rubio:2018jzw}; thus a compact massive object formed out of accreted matter cannot be excluded.}. We leave this question aside and  explore the image features of such spacetimes with and without photon sphere from a more theoretical point of view.\\
Due to the absence of an event horizon, images at $r_{\rm NP}> r_{\rm NP, \, crit}$ show new image features inside the shadow boundary. 
These new features arise because the light emitted by the accretion disk can pass through the spacetime region formerly inside the horizon. The absence of the horizon does not imply the absence of strong lensing; thus the new image features include additional photon rings. They are associated to trajectories which approach the photon sphere from the inside (and would have to cross the event horizon for $r_{\rm NP}< r_{\rm NP, \, crit}$). 
\\

The four principles of regularity, locality, simplicity, and Newtonian limit result not only in two horizons but also in two photon spheres. Just like the two horizons annihilate at $r_{\rm NP}=r_{\rm NP,\,crit}$, also the two photon spheres annihilate at $r_{\rm NP}=r_{\rm NP,\,crit,\,2}$. The simplicity principle implies that the smaller $r$, the larger the increase in compactness. Therefore, the photon spheres annihilate only after the horizons annihilate, i.e., $r_{\rm NP,\,crit,\,2}>r_{\rm NP,\,crit}$. While these properties are universal for any simple, regular black-hole (or horizonless) spacetime, cf.~Sec.~\ref{sec:morecompactgamma}, we focus on the Simpson-Visser spacetime as a concrete and representative example. Fig.~\ref{fig:photon sphere} shows the behavior of the two horizons and the two photon spheres as a function of $r_{\rm NP}$.
\\

\begin{figure}
\begin{center}
\includegraphics[width=\linewidth]{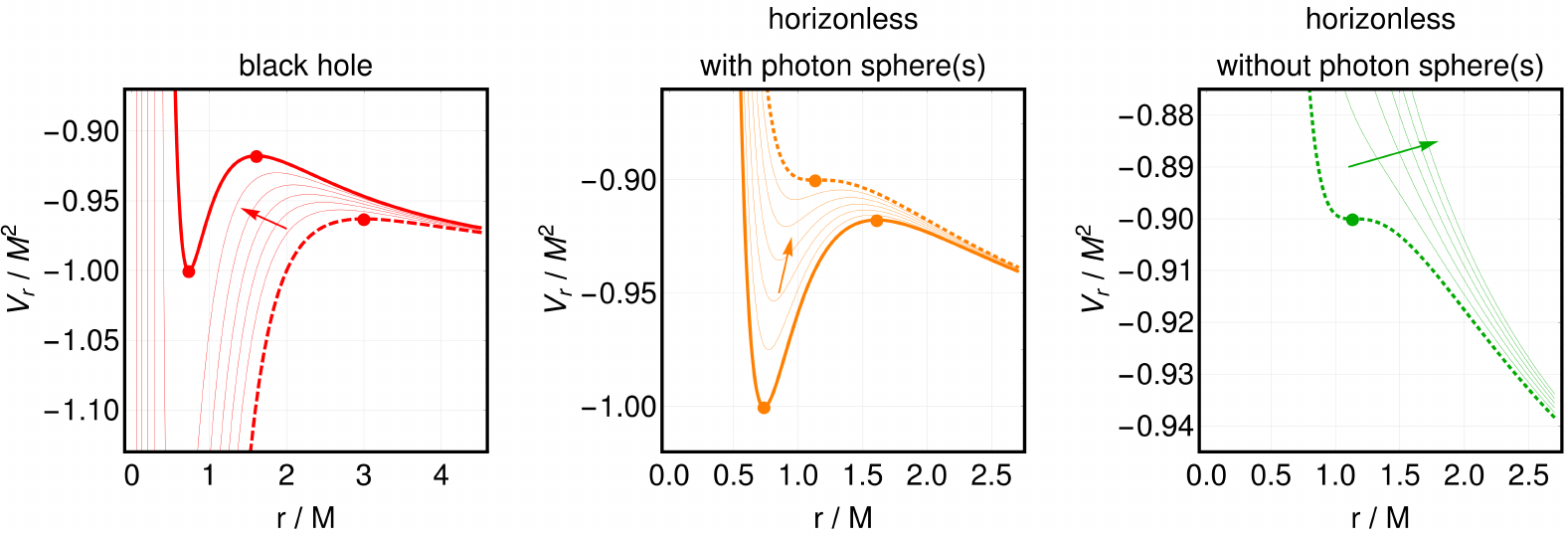}
\end{center}
\caption{\label{fig:SV_Vr}
We show the behavior of the radial potential $V_r(r)$ (cf.~Eq.~\eqref{eq:Vr}) for the Simpson-Visser spacetime with growing $r_\text{NP}$ and for $L=E=M$. From left to right, we show the black-hole case ($0<r_{\rm NP}<r_{\rm NP,\, crit}$), the horizonless case with photon spheres ($r_{\rm NP,\, crit}<r_{\rm NP}<r_{\rm NP,\, crit,\, 2}$), and the horizonless case without photon sphere ($r_{\rm NP,\, crit,\, 2}<r_{\rm NP}$). 
The critical cases are highlighted with thick lines. Points on these thick lines indicate the position of the photon spheres. At $r_{\rm NP}=0$ (thick dashed), there is only a single extremum corresponding to an unstable photon sphere (Schwarzschild). At $r_{\rm NP}=r_{\rm NP,\, crit}$ (thick continuous), there are two extrema, corresponding to an unstable and a stable photon sphere. At $r_{\rm NP}=r_{\rm NP,\, crit,\,2}$ (thick dotted), the two photon spheres annihilate and no extremum is left. We also show intermediate $r_{\rm NP}$ (thin lines) and indicate growing $r_{\rm NP}$ with arrows.
	}
\end{figure}

\begin{figure}
\begin{center}
\includegraphics[width=\linewidth]{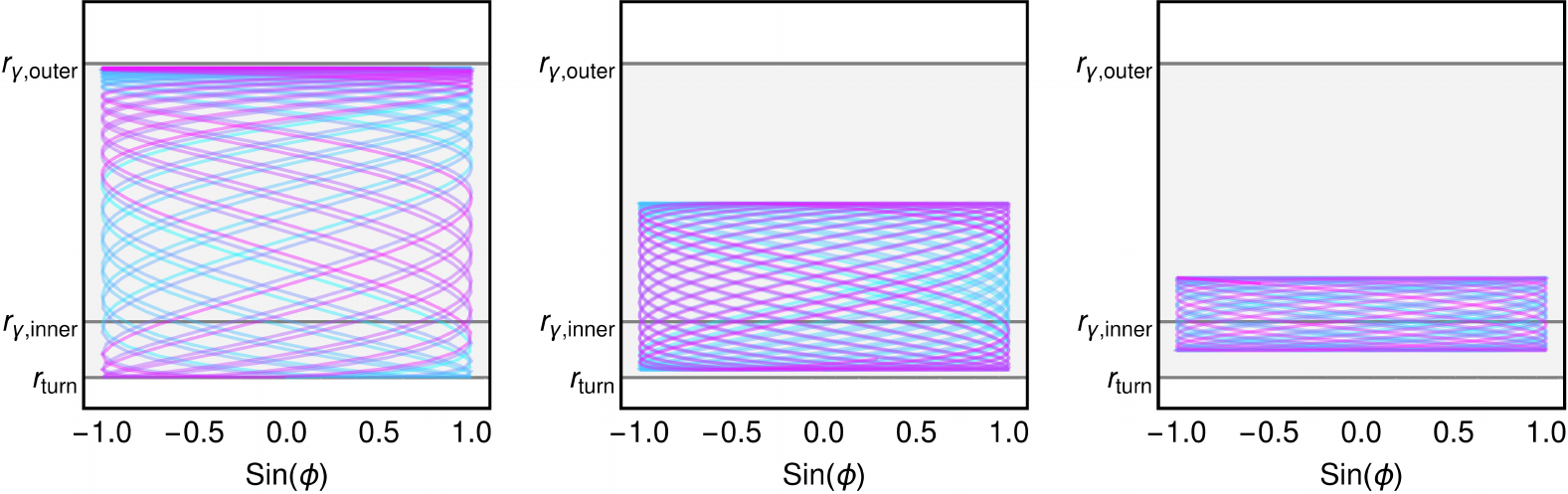}
\end{center}
\caption{\label{fig:SV_boundOrbits} 
We show three examples of bound orbits in the Simpson-Visser spacetime (for $r_{\rm NP,\, crit}<r_{\rm NP}<r_{\rm NP,\, crit,\, 2}$) as parametric curves in $\phi$ and $r$, oscillating around the stable inner photon sphere at $r_{\gamma\rm,inner}$. The critical orbit (left panel) is bound to the region $r\in[r_\text{turn},\,r_{\gamma\rm,outer}]$. The colour scale indicates growing affine parameter (from cyan to magenta). Without loss of generality, $\theta=\pi/2$.
	}
\end{figure}

\begin{figure}[t!]
	\centering	\includegraphics[width=0.49\linewidth]{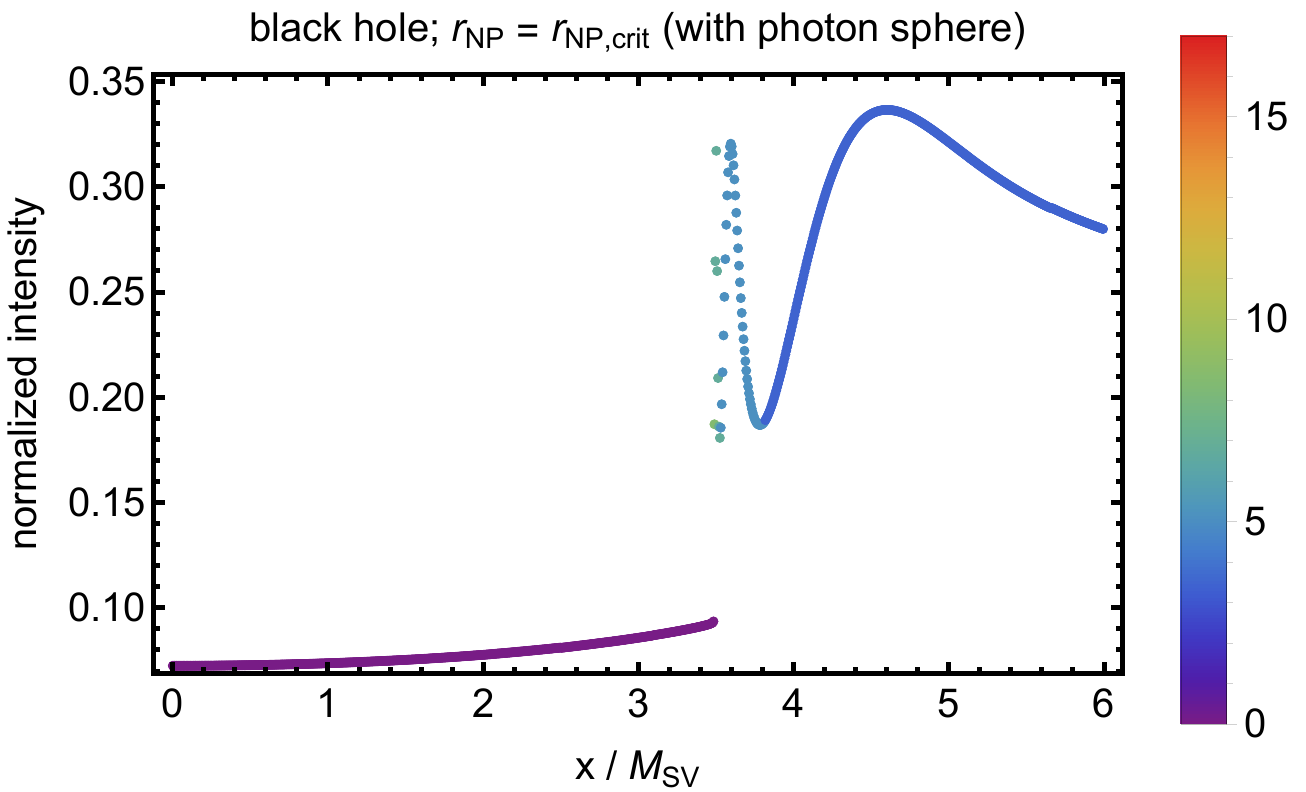}
	\includegraphics[width=0.49\linewidth]{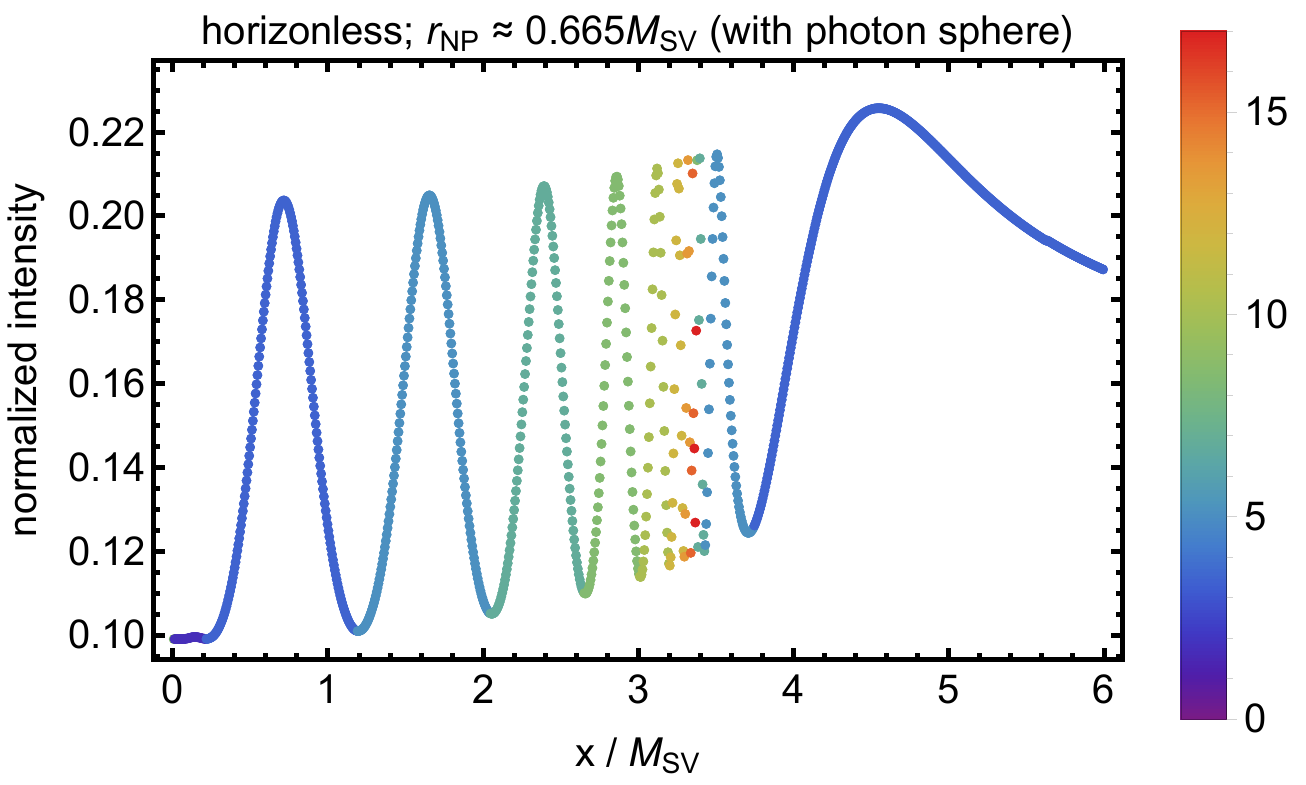}
	\includegraphics[width=0.49\linewidth]{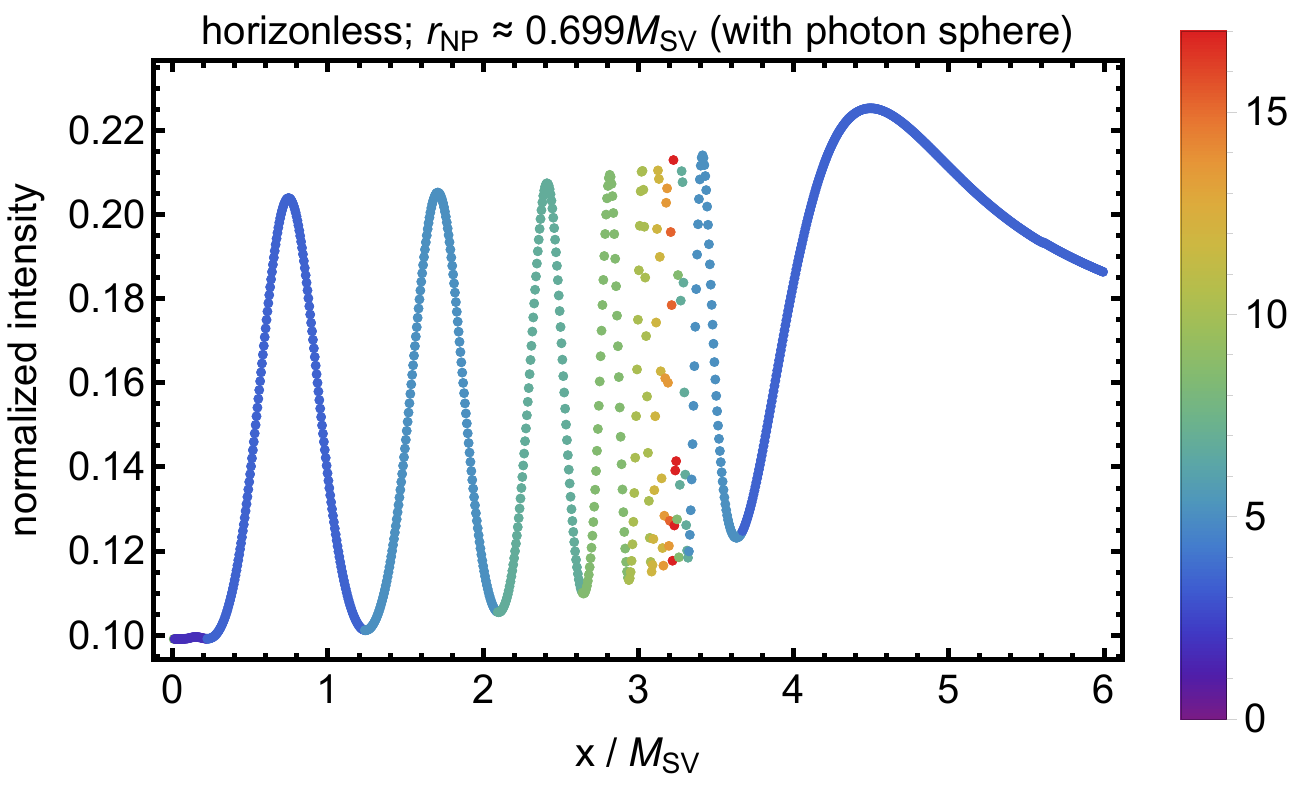}
	\includegraphics[width=0.49\linewidth]{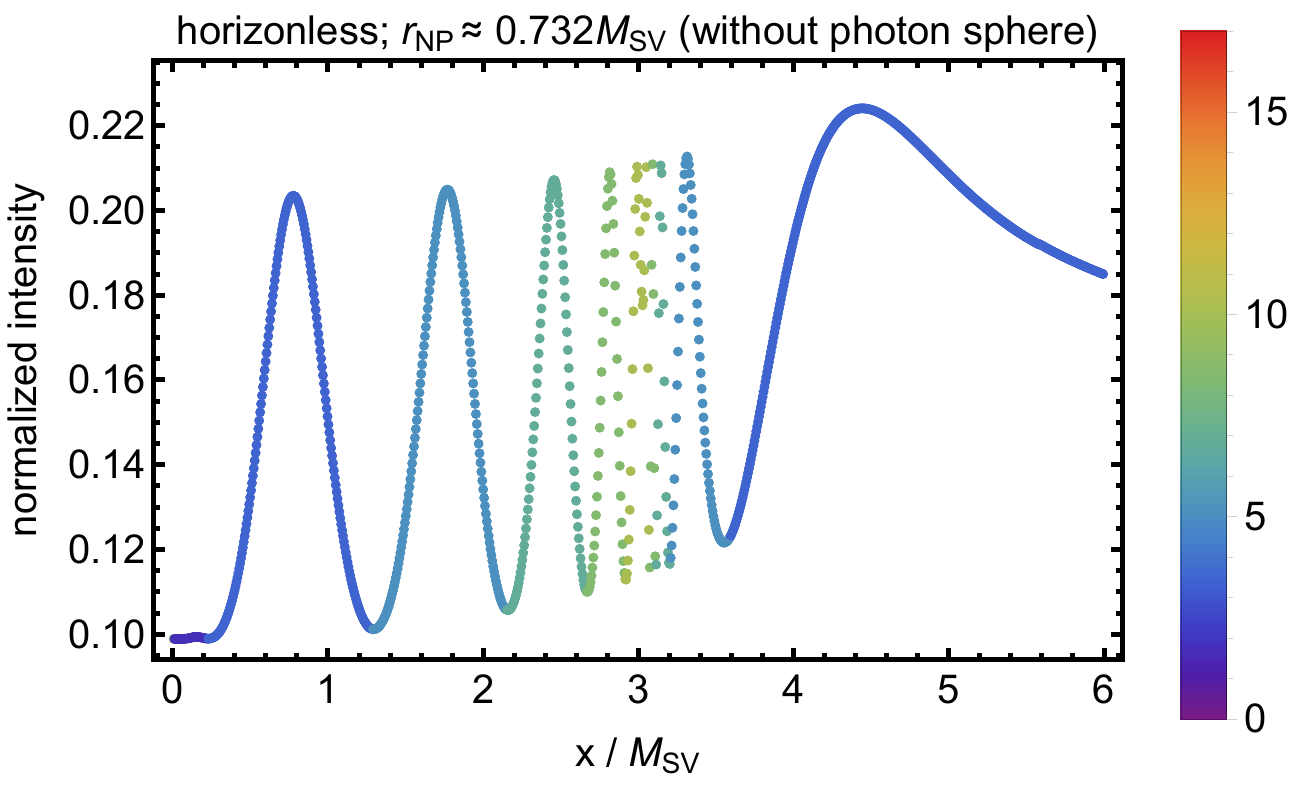}
	\includegraphics[width=0.49\linewidth]{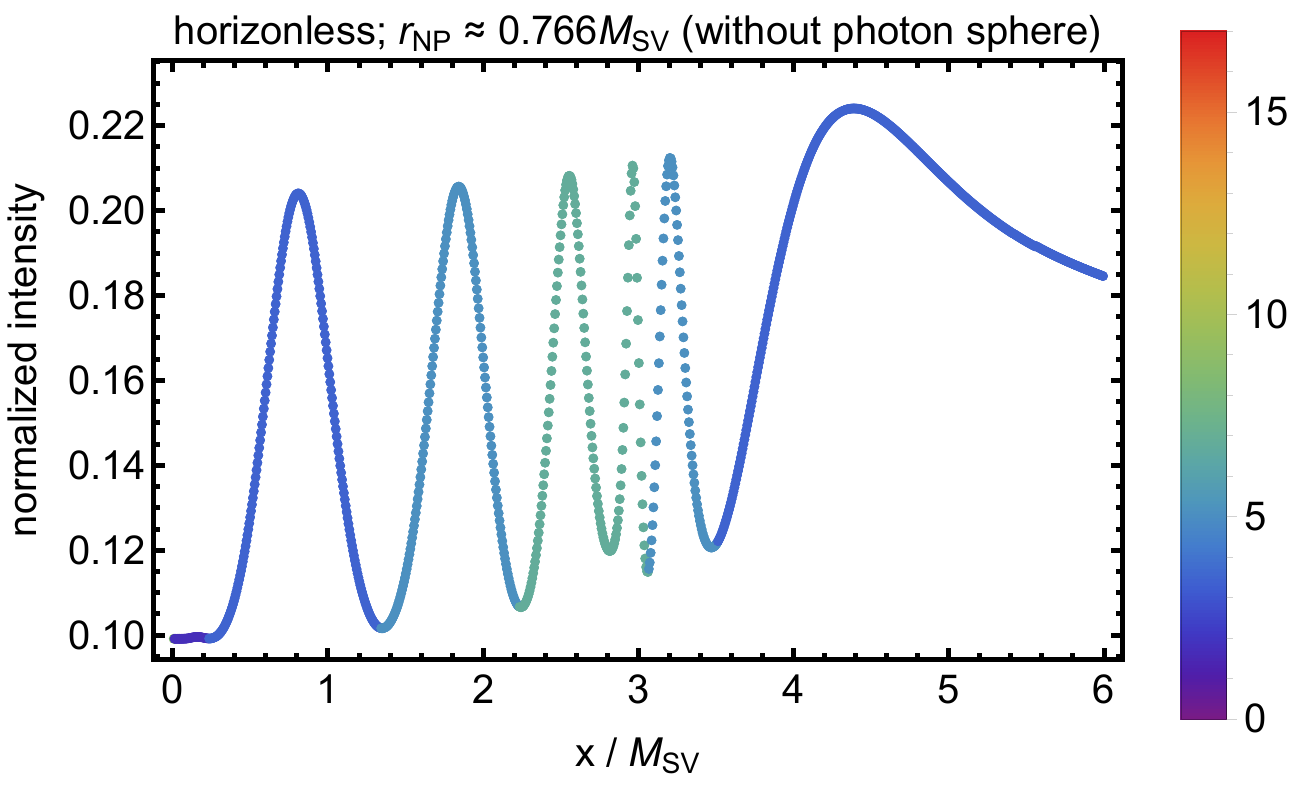}
	\includegraphics[width=0.49\linewidth]{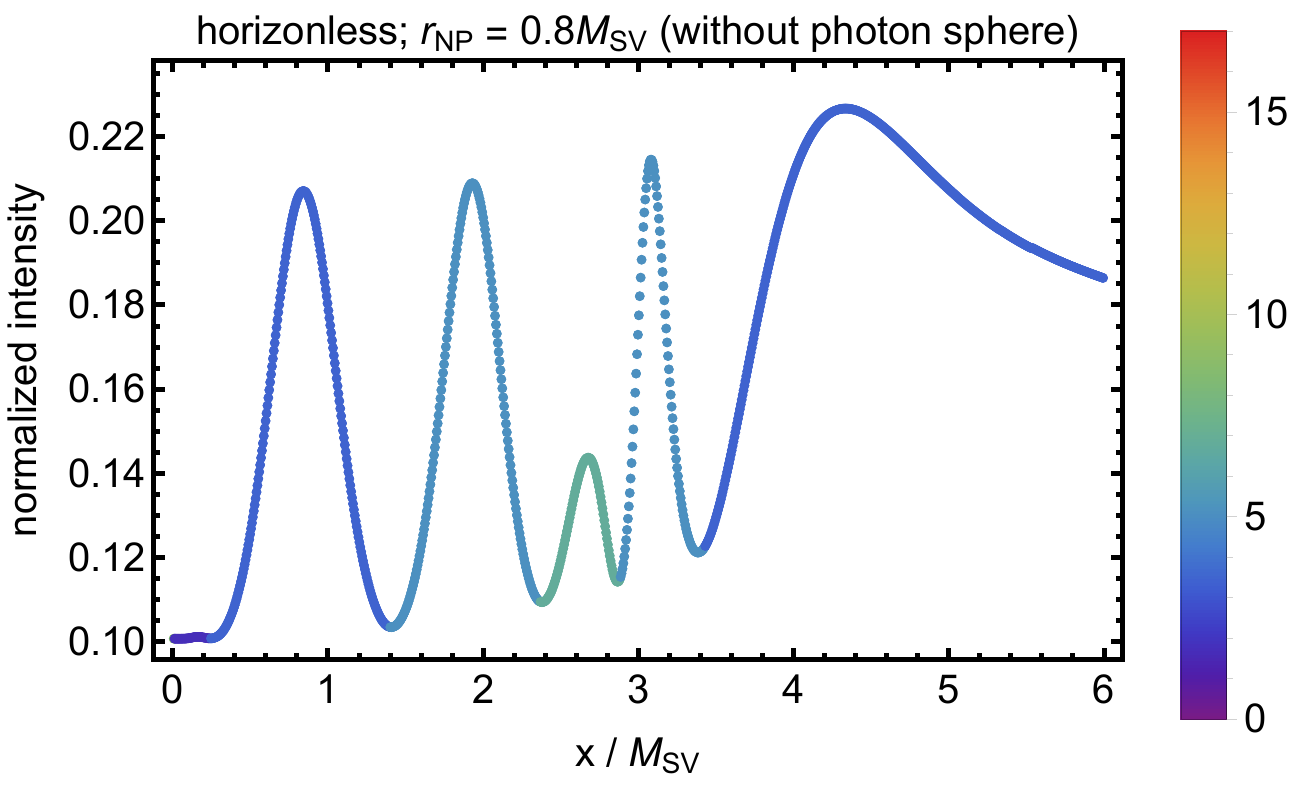}
	\caption{We show cross-sections of the normalised intensity for five equally distributed values of $r_{\rm NP}$ between $r_{\rm NP, \, crit}\leq r_{\rm NP}\leq0.8$. The value of $r_{\rm NP}$ increases from left to right and from top to bottom. The color scale indicates the associated winding number, i.e., how many times the respective geodesic has pierced the equatorial plane.
	The accretion disk is modelled by Eq.~\eqref{eq:disk-model} with parameters $\alpha=1$, $h=1/3$, $r_{\rm cut}=6M_{\rm SV}$, and $w=2$, with the distant observer being located at an inclination $\theta_{\rm cam}=\pi/3$.}
	\label{fig:photon_rings_nohorizon}
\end{figure}

While the outer photon sphere at $r_{\gamma\rm,outer}$ is unstable (i.e., radial perturbations of null geodesics at $r_{\gamma\rm, outer}$ grow), the inner photon sphere at $r_{\gamma\rm, inner}$ is stable (i.e., radial perturbations of null geodesics at $r_{\gamma\rm, inner}\pm\epsilon$ return to and oscillate around $r_{\gamma\rm, inner}$).
This can be understood in terms of the radial potential $V_r(r)$ for null geodesics, cf.~Eq.~\eqref{eq:Vr}. We show the radial potential for the Simpson-Visser spacetime in Fig.~\ref{fig:SV_Vr}. Photon spheres correspond to extrema of the radial potential. A minimum (maximum) corresponds to a stable (unstable) photon sphere. By inspecting the radial potential, we can classify the structure of photon rings in the image plane into three qualitatively different regimes of $r_\text{NP}$. Examples of the respective image cross sections in these three regimes are shown in Fig.~\ref{fig:photon_rings_nohorizon}.

For $0<r_{\rm NP}<r_{\rm NP,\, crit}$, there are two horizons and two photon spheres. Just like for the Schwarzschild spacetime, there is a central brightness depression and one exponentially stacked set of photon rings, cf.~left~upper panel in Fig.~\ref{fig:photon_rings_nohorizon}, because the outer horizon separates both the inner horizon and the inner photon sphere from an observer at asymptotic infinity.

For $r_{\rm NP,\, crit}<r_{\rm NP}<r_{\rm NP,\, crit,\, 2}$, there are no longer any horizons in the spacetime, but the two photon spheres at $r_{\gamma\rm,inner}$ and $r_{\gamma\rm,outer}$ remain intact. Geodesics from asymptotic infinity can approach the outer photon sphere in two distinct ways: they can either approach $r_{\gamma\rm,outer}$ from \emph{above} ($r= r_{\gamma\rm,outer}+\epsilon$), resulting in the usual exponentially stacked set of photon rings, or they can approach $r_{\gamma\rm,outer}$ from \emph{below}, ($r= r_{\gamma\rm,outer}-\epsilon$), resulting in a new exponentially stacked set of photon rings, cf.~Fig.~\ref{fig:photon_rings_nohorizon}.
The inner photon sphere is stable, corresponding to a minimum of the radial potential, such that there is a region of $r$ for which $V_r(r)< V_R(r_{\gamma\rm, outer})$. Therefore, bound orbits occur between the outer photon sphere and the turning point in the radial potential, $r_{\rm turn}$, which is the innermost point at which $V_r(r)< V_R(r_{\gamma\rm, outer})$ holds, cf.~Fig.~\ref{fig:SV_boundOrbits}.
These bound orbits cannot be accessed from asymptotic infinity. The bound orbit which oscillates between $r_{\rm turn}$ and $r_{\gamma\rm, outer}$ can be approached by null geodesics from asymptotic infinity; those are the ones that make up the inner exponentially stacked set of photon rings.

Finally, for $r_{\rm NP}>r_{\rm NP,\, crit,\, 2}$, there are neither horizons nor photon spheres left. As a result, there remain only a finite number of inner and outer photon rings. As $r_{\rm NP}$ increases, more and more inner and outer photon rings annihilate, cf.~lower panels in Fig.~\ref{fig:photon_rings_nohorizon}.
Horizonless objects at $r_{\rm NP}\gg r_{\rm NP,\, crit,\, 2}$ only show a small number of photon rings, cf.~Fig.~\ref{fig:photon_rings_nohorizon}. The higher-order photon rings annihilate first, and the lower-order photon rings remain behind the longest. In this sense, gravitational lensing is weakened as $r_{\rm NP}$ is increased. This is in line with the observation that curvature invariants, such as the Kretschmann scalar, go to zero in the limit $r_{\rm NP} \rightarrow \infty$. Similarly, it can be seen that the metric approaches Minkowski form in this limit.
At such large values of $r_{\rm NP}$, the “repulsive” character of the new-physics effect completely overwhelms the attractive character of the original Schwarzschild solution, such that all gravitational effects cancel.

\section{Conclusions}\label{sec:conclusions}
In order to use radio-VLBI imaging to test GR, the alternatives to black holes in GR as well as their observational features must be mapped out. Here, we take a step in this direction by working in the principled-parameterized approach to black holes and horizonless objects. The principled-parameterized approach is situated inbetween parameterized approache (which attempts to provide a comprehensive characterization of metrics beyond GR, see, e.g., \cite{1979GReGr..10...79B,Cardoso:2014rha,Johannsen:2015pca,Konoplya:2016jvv,Delaporte:2022acp}) and the principled approach (which derives black-hole solutions in specific theories beyond~GR).

We work in spherical symmetry both for its technical simplicity as well as its phenomenological relevance for existing shadow images: Due to its near-face-on inclination and due to finite EHT resolution, the shadow of M87* does not show any presently detectable spin effects \cite{paper5}. Instead, constraints on the spin of M87* arise from combining EHT data with the requirement that the source produces the observed jet (see, e.g., \cite{2018ApJ...855..128W} and references therein).

We investigate a set of principles \cite{Eichhorn:2021etc, Eichhorn:2021iwq} in the principled-parametrized approach: In spherical symmetry, we establish that the four principles of (i) locality, (ii) regularity, (iii) simplicity, and (iv) a Newtonian limit, lead to a more compact event horizon, photon sphere, and shadow boundary (critical curve). This sharpens previous results \cite{Eichhorn:2021etc, Eichhorn:2021iwq}, where the simplicity principle was implicitly assumed, but not spelled out.
The locality principle states that modifications of the metric from the Schwarzschild spacetime are tied to the local curvature of the Schwarzschild spacetime itself, i.e., must be a function of the Kretschmann scalar. 
The regularity principle states that all (non-derivative) curvature invariants must be regular.
The simplicity principle states that modifications of the metric from the Schwarzschild spacetime must be characterized by a single, not several, new-physics scales.
The Newtonian limit states that the spacetime is asymptotically flat and that the Newtonian limit is recovered when approaching asymptotic infinity.
In spherical symmetry, these principles are sufficient to conclude that the horizon, photon sphere, and thus the shadow boundary (critical curve) must shrink when the new-physics scale $r_{\rm NP}$ increases.

We recast the modifications into the form of a radially dependent mass function and focus on three examples from the literature, one polynomial (Hayward \cite{Hayward:2005gi}) and two exponential examples (Dymnikova \cite{Dymnikova:1992ux} and Simpson-Visser \cite{Simpson:2019mud}). We highlight that image features observed for these examples are qualitatively universal across all regular, simple black holes. Only quantitative differences arise for different choices of the mass function.
\\

Turning to observations of these simple, regular black holes, it is not the idealized shadow boundary but rather the low-order photon rings which are observables. The first photon ring is constrained by current EHT measurements. The second photon ring will potentially become accessible with the next-generation EHT or future space-based very-long baseline interferometry, potentially with the support of super-resolution techniques \cite{Broderick:2020wda}. Working with a sample set of disk models, we determine whether shadow observations can constrain the new-physics scale of such spacetimes.
 
Within our sample set of disk models, access to the first photon ring is sufficient to constrain the new-physics parameter of Simpson-Visser-type modifications to black-hole spacetimes, but insufficient to constrain Hayward-type modifications. Such a constraint relies on an independent mass measurement in the Newtonian limit, e.g., from stellar dynamics \cite{Gebhardt:2011yw}.

Additional access to the second photon ring is sufficient to significantly constrain both types of modifications. With access to two photon rings, no independent mass measurement is required. Intriguingly, the new-physics modifications increase the separation of the photon rings. For the largest separation achievable in a Simpson-Visser black hole, the separation for the $n=1$ and $n=2$ photon ring is close to $10\, \mu \rm as$ (for an $n=2$ diameter of $42\, \mu\rm as$), which is the nominal resolution achievable with ground-based VLBI at 345 GHz. The increased relative separation of photon rings is a universal property of simple, regular black holes. The role of simplicity is critical in this context: with simplicity, there is an increase in compactness in all distinct surfaces of a regular black hole. The increase is largest for surfaces at small radii, resulting in an increased relative separation of photon rings. In contrast, without simplicity, an increase in compactness still occurs at small radii, but a second new-physics effect can even lead to a decrease in compactness at larger radii, see App.~\ref{app:simplicity-expansions}. Thus, regularity only impacts image features (such as the relative separation of photon rings) when simplicity holds. Simplicity thus is the bridge that connects regularity -- a key property expected from quantum gravity -- to image features. 

We also take a first step to explore the possibility of removing astrophysical uncertainties with time-dependent emission such as hotspots in accretion flows. We start exploring whether the resulting image intensity provides an independent constraint on the spacetime geometry, at least within the idealized assumptions of (i) an independent measurement of the shadow diameter and (ii) perfect knowledge about the emission profile of the hot-spot model. Here, we only take a first step in demonstrating the constraining power of spacetime tomography approach that is based on inhomogeneous and time-dependent emission. More extensive studies are needed to discover, e.g., degeneracies between modifications of the spacetime and modifications of the properties of the emission region.

Finally, we investigate the image features of horizonless regular spacetimes. If these regular spacetimes are based on the four fundamental principles, they become horizonless once the ratio of new-physics scale and asymptotic mass reaches a critical value. We emphasize that a photon sphere can persist even if the horizon is lost. However, the absence of the horizon qualitatively changes the image features resulting from said photon sphere. In presence of a horizon there is one exponentially stacked and infinite set of photon rings. In absence of a horizon there are two such exponentially stacked and infinite sets of photon rings, pairing one outer and one inner photon ring. We delineate the intricate structure of these image features.
\\
 
In summary, we work within the framework of the principled-parameterized approach and study spherically symmetric black-hole spacetimes. We demand three principles that encode fundamental physics, namely (i) locality (in an EFT sense), (ii) regularity (to address the main shortcoming of GR), (iii) simplicity (encoding the number of scales in a fundamental theory), and also demand the observationally motivated (iv) Newtonian limit. These four principles lead to increased compactness of the event horizon and photon sphere of the resulting simple, regular black holes in comparison to the Schwarzschild spacetime. This is a universal finding and holds for all choices of the free functions and free parameters of simple, regular black holes.
We analyze the universal structure of photon rings surrounding the shadow image of simple, regular spacetimes, with and without horizon. In the black-hole case, observational constraints on the associated new-physics scale $r_\text{NP}$ require two independent measurements, that is, either (i) a measurement of the $n=1$ photon ring and a constraint from the post-Newtonian regime (e.g., stellar dynamics) or (ii) a measurement of the $n=1$ and the $n=2$ photon ring. A first constraint on $r_{\rm NP}$ can, for particular choices of the free function, already be obtained with current data. Most intriguingly, a universal feature of simple, regular black holes is that they feature an increased photon-ring separation compared to the Schwarzschild case, implying that a resolution of the $n=1$ and $n=2$ photon ring might potentially even be within reach of the ngEHT.
\\


\emph{Acknowledgments}\\
We thank R.~Gold for discussions.
A.~E.~is supported by a grant from Villum Fonden, grant no.~29405. The work leading to this publication was supported by the PRIME programme of the German Academic Exchange Service (DAAD) with funds from the German Federal Ministry of Education and Research (BMBF).
During parts of this project, A.~H. was supported by a Royal Society International Newton Fellowship under the grant no. NIF\textbackslash R1\textbackslash 191008. 

\newpage
\appendix
\section{Simplicity criterion and radius of convergence for series expansions}\label{sec:series}
\label{app:simplicity-expansions}
\begin{figure}
\centering
\includegraphics[height=0.32\linewidth]{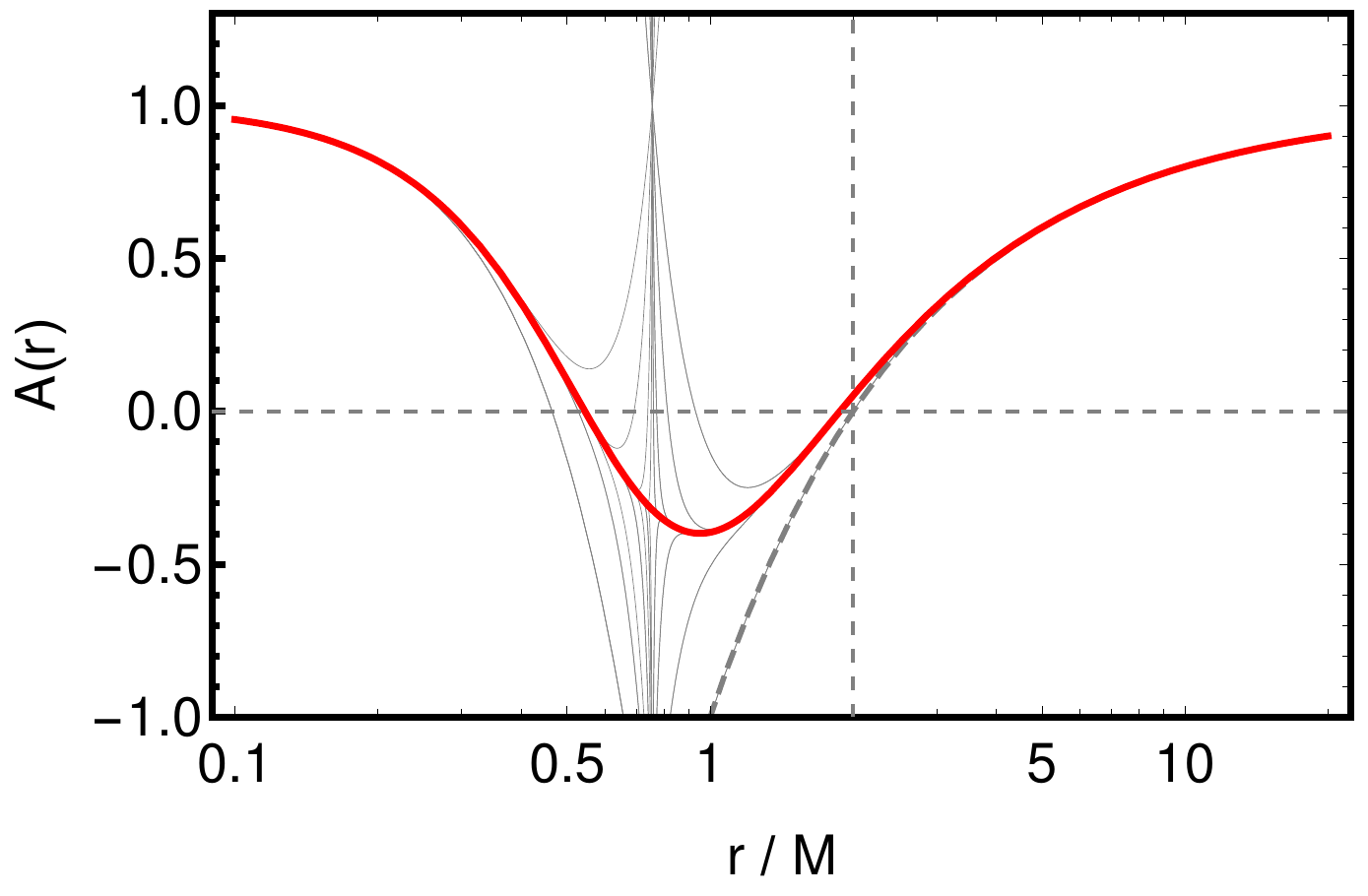}
\hfill
\includegraphics[height=0.32\linewidth]{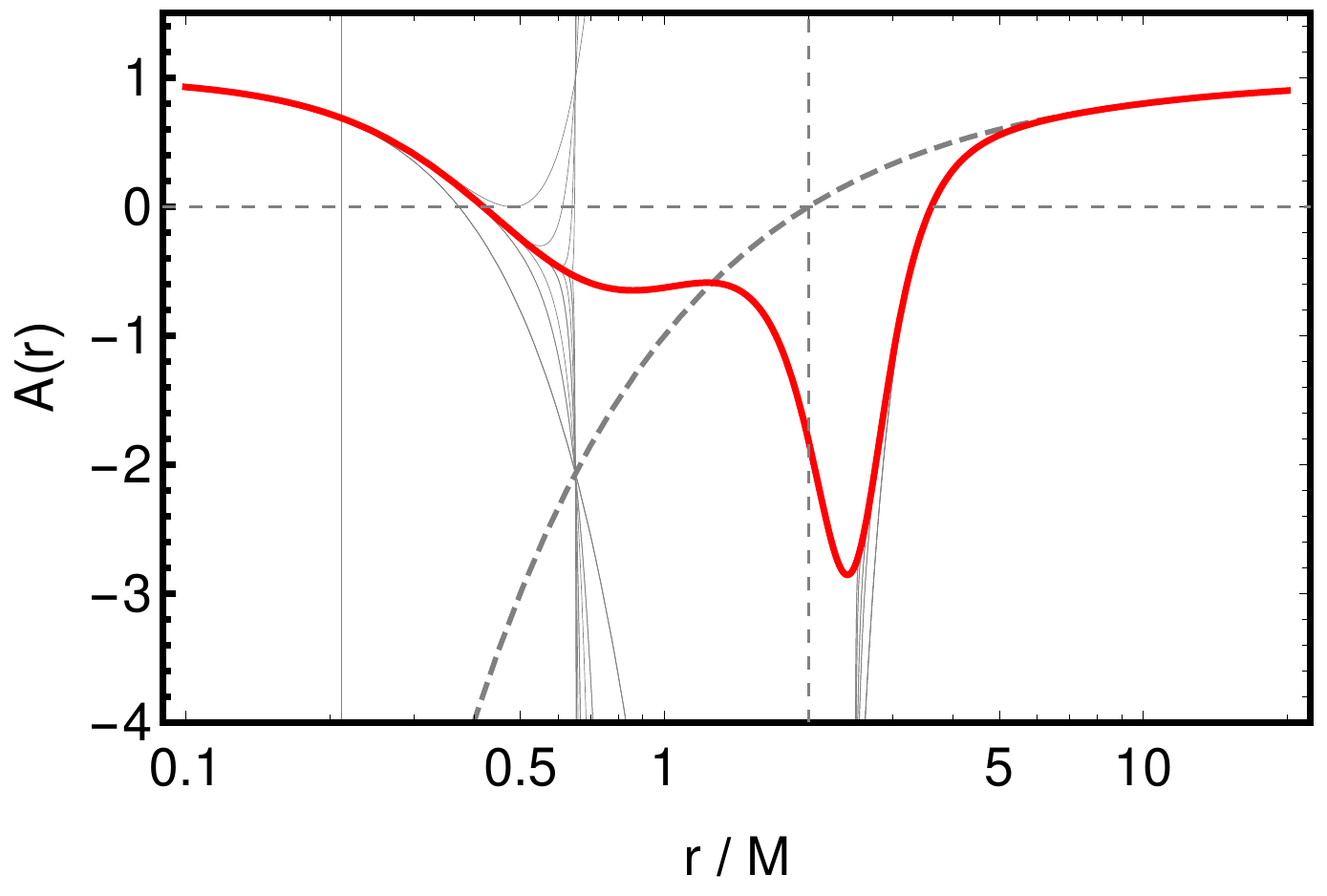}
\caption{\label{fig:expansion-matching} 
Different functions $A(r)$ (thick red line) which interpolate between a regular black-hole core and a Newtonian far-field limit. Series expansions around $r=0$ and $r=\infty$ at order $2^N$ and $2^N+1$ with $N=3,\dots,8$ (continuous thin-gray lines) indicate a finite radius of convergence for both expansions. The left-hand panel shows a `simple' choice of $A(r)$, cf.~Eq.~\eqref{eq:Asimple} (with $\alpha=3$ and $r_\text{NP}=M/4$), for which the two radii of convergence coincide. The right-hand panel shows a `non-simple' choice of $A(r)$, cf.~Eq.~\eqref{eq:Anonsimple} (with $\beta=1$, $\beta_2=2$, $r_\text{NP}=M/5$, and $r_2 = 3M/2$), for which the two radii of convergence do not overlap. For reference, we also show the function $A_\text{Schw}(r) = 1-2M/r$ (gray-dashed).
}
\end{figure}
If $A(r)$ and $B(r)$ admit a series expansion, the simplicity principle can be rephrased as a requirement on these series expansions. 
To establish this, we focus on the case $B= A$ below. 

The single-scale assumption underlying the simplicity principle implies that at most two Taylor expansions are required to encode $A(r)$ at all radii.
Specifically, we consider the Taylor expansions of $A(r)$ around $r=0$ (where the regularity principle provides a constraint) and around $r = \infty$ (where the Newtonian limit principle provides a constraint). 
The simplicity principle implies that the radius of convergence of these two Taylor expansions is such that the two expansion can be glued at a point $r_c$.\\
We start from a power series expansion for $A(r)$ around $r=0$,
\be
A(r) = \sum_{n=-\infty}^\infty a_n r^n.\label{eq:seriessmallr}
\ee
The (dimensionfull) expansion coefficients $a_n$ are constrained to $a_{n<0}=0$, if we demand the Kretschmann scalar \eqref{eq:KretschmannAequalB} to be finite at $r=0$. Additionally, singularity resolution constrains $a_0=1$ and $a_1=0$. \\
 The next step is to consider the far-field regime, which is governed by classical Newtonian gravity.
 We consider a power series for $A(x\equiv 1/r)$,
 \be
A(x) = \sum_{n=0}^{\infty} \bar{a}_n x^n.\label{eq:serieslarger}
\ee
The dimensionful expansion coefficients $\bar{a}_n$ are constrained by the Newtonian limit:	
the Newtonian potential determines the temporal component of the metric, i.e.,
\be
g_{00} = -(1+2\Phi(r)) = -A(r),
\ee
with $\Phi(r)= \frac{M}{r}$ the Newtonian potential. This results in $\bar{a}_0=1$ and $\bar{a}_1=-2M$, whereas all higher-order expansion coefficients remain unconstrained in the Newtonian limit; constraints arise at post-Newtonian order, see App.~\ref{sec:PN}.
\\
Upon comparing the two expansions Eq.~\eqref{eq:seriessmallr} and \eqref{eq:serieslarger}, we note that the expansion coefficients do not match: Eq.~\eqref{eq:seriessmallr} contains only positive powers of $r$, whereas Eq.~\eqref{eq:serieslarger} must contain a term $~r^{-1}$. This is distinct from the classical case, where $A(\bar{x}) = 1 - 2 M\, \bar{x}$ holds at all values of $\bar{x}$.
As a consequence, the radii of convergence of Eq.~\eqref{eq:seriessmallr} and \eqref{eq:serieslarger} cannot overlap.  Thus, there must be (at least) one distinct new scale, corresponding to the radius of convergence.
In the absence of the simplicity criterion, there may be several distinct scales, resulting in more than two distinct expansions.
\\

To exemplify our discussion, we consider a simple, regular black hole and contrast it with a non-simple one.
In the presence of a single new-physics scale, a single dimensionless ratio of scales, $\gamma =\beta_1 (r_{\rm NP}/ M)^{\beta_2}$, with dimensionless constants $\beta_1\in\mathbb{R}$ and $\beta_2\in\mathbb{R}$, is allowed to appear in the line-element. Therefore, the deviation of $A(r)$ from the flat-space limit $A(r)=1$ can, for instance, be an algebraic function, i.e., the ratio of two simple polynomials, where only one of the polynomials can have more than a single term, e.g.,
\begin{align}
	A(r) = 1- \frac{2M}{r}\frac{(r/M)^{\alpha}}{\gamma + (r/M)^\alpha} = 1-\frac{2M}{r}\frac{1}{(r_{\rm NP}^4 K_\text{Schw})^{\frac{\alpha}{6}}+1},\label{eq:Asimple}
\end{align}
with $\alpha\geqslant3$ required by singularity resolution and $\beta_1=48^{\alpha/6}$, $\beta_2=\frac{2}{3}\alpha$ in the last step.

In contrast, a set of functions $A(r)$ which violates the simplicity criterion is given by
\bea
\label{eq:Anonsimple}
	A(r) &=&
1- \frac{2M}{r} \left(\frac{1}{1+\left(K_\text{Schw}\,r_{\rm NP}^4  \right)^{\frac{1}{2}}} + \left(\frac{r_2}{r_{\rm NP}} \right)^{\beta_1} \frac{r_2^4 K_\text{Schw}}{1+ \left(r_2^4 K_\text{Schw} \right)^{\beta_2}}\right),
\eea
for which $r_2$ enters as an additional scale and $\beta_2 \geqslant 3/2$ is required for singularity resolution. 
Both functions as well as their series expansions are shown in Fig.~\ref{fig:expansion-matching} for exemplary values of $r_{\rm NP}$ (and $r_2$). 
\\

\begin{figure}
	\centering
	\includegraphics[width=0.48\linewidth]{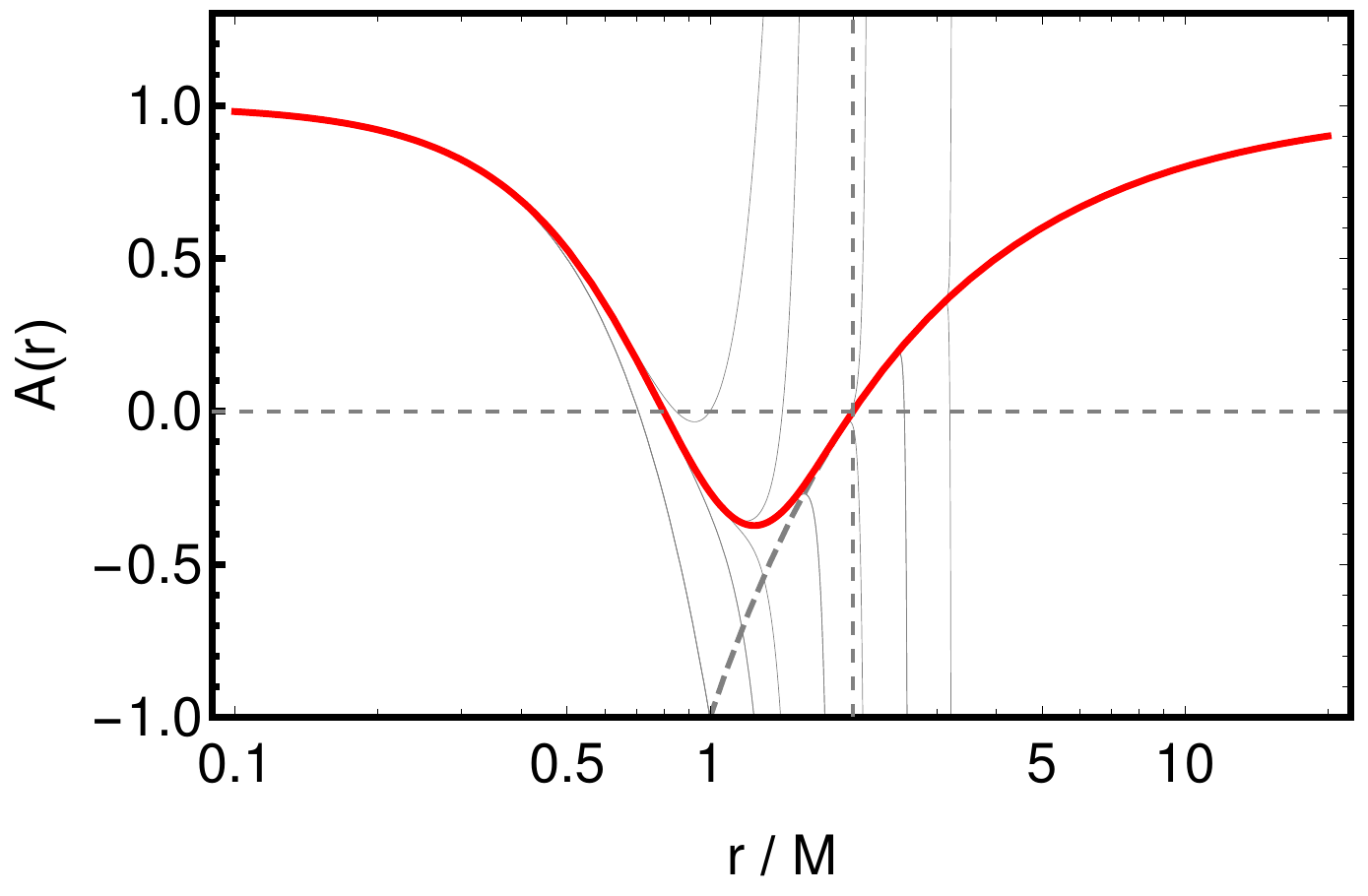}
	\hfill
	\includegraphics[width=0.48\linewidth]{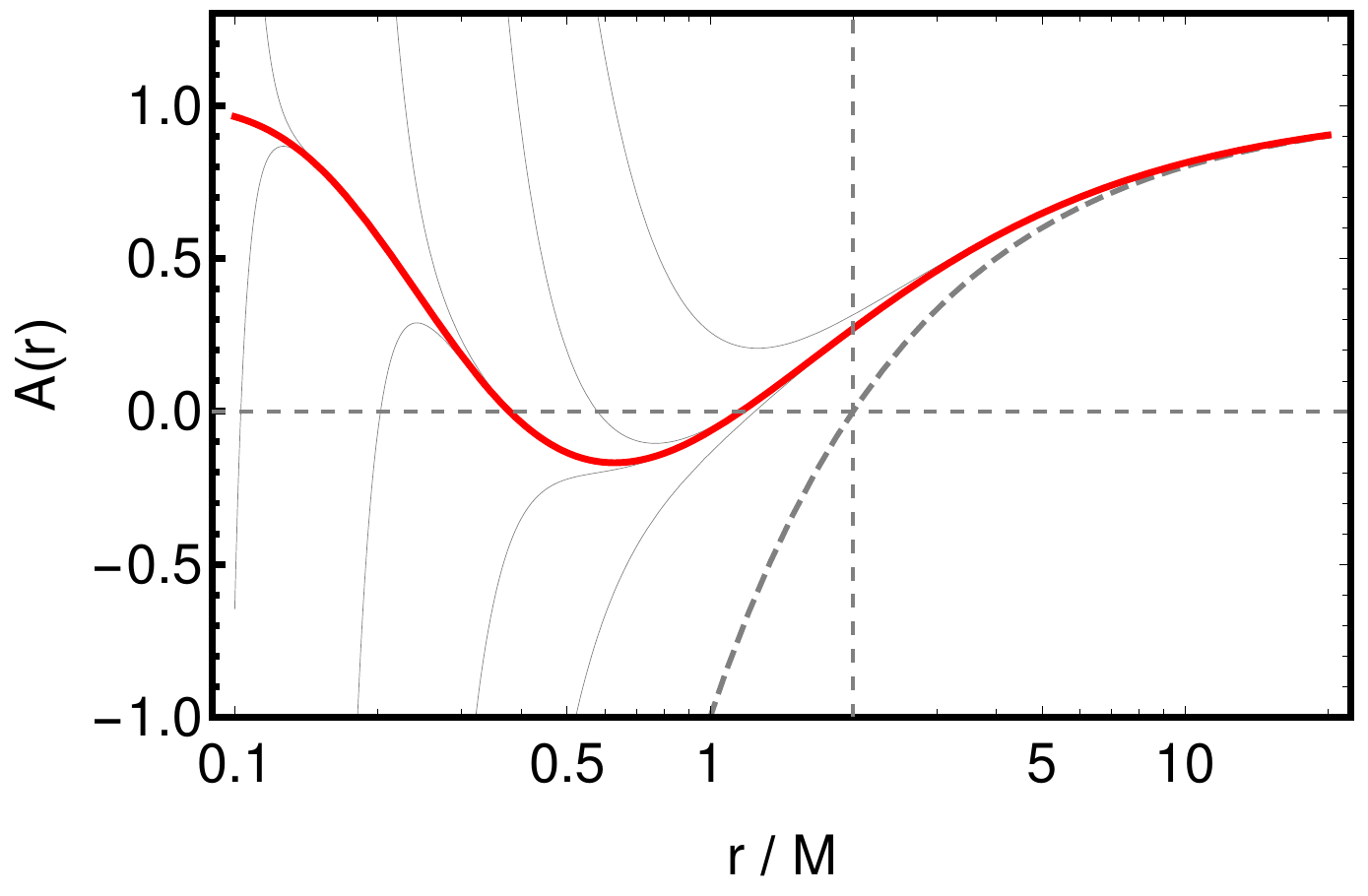}
	\caption{\label{fig:model-expansions} 
	We show metric functions $A(r)$ (thick red line) and one-sided series expansions around $r=0$ and $r=\infty$ at order $2^N$ and $2^N+1$ with $N=3,\dots,8$ (continuous thin-gray lines) of Dymnikova (left-hand panel with $r_\text{NP}=M$) and Simpson-Visser (right-hand panel with $r_\text{NP}=M/2$) spacetime. For reference, we also show the function $A_\text{Schw}(r) = 1-2M/r$ (gray-dashed).
	}
\end{figure}
The three mass functions of Dymnikova-, Hayward- and Simpson-Visser spacetime, cf.~Eqs.~\eqref{eq:fDym}-\eqref{eq:fSV} all fulfill the simplicity criterion:
The Taylor expansions for a Hayward spacetime around $r=0$ and $r=\infty$ have finite radius of convergence and match at finite, nonzero $r$. This case is shown in the left-hand panel of Fig.~\ref{fig:expansion-matching} and is equivalent to Eq.~\eqref{eq:Asimple} with $\alpha=3$. The Taylor expansions for a Dymnikova spacetime around $r=0$ and $r=\infty$ have infinite and vanishing radius of convergence, respectively and match at $r=\infty$. This case is shown in the left-hand panel of Fig.~\ref{fig:model-expansions}. The Taylor expansions for a Simpson-Visser spacetime around $r=0$ and $r=\infty$ have vanishing and infinite radius of convergence, respectively and match at $r=0$. This case is shown in the right-hand panel of Fig.~\ref{fig:model-expansions}.
\\

The non-simple, regular black holes in \cite{Bonanno:2000ep}, cf.~Eq.~\eqref{eq:fBR}, provide an example for how multiple new-physics scales, here $\tilde{\omega}$ and $\gamma$, break the simplicity principle and can thereby alter the conclusions about a more compact horizon and photon sphere. 

The quantum contributions calculated in \cite{Bonanno:2000ep, Donoghue:1993eb} result in $\gamma>0$ and $\tilde{\omega}>0$ such that $f_\text{BR}(r)$ remains a monotonically increasing function of $r$. Hence, the increase in compactness discussed for simple black holes in the main text persists. 

\begin{figure}
	\centering
	\includegraphics[width=0.95\linewidth]{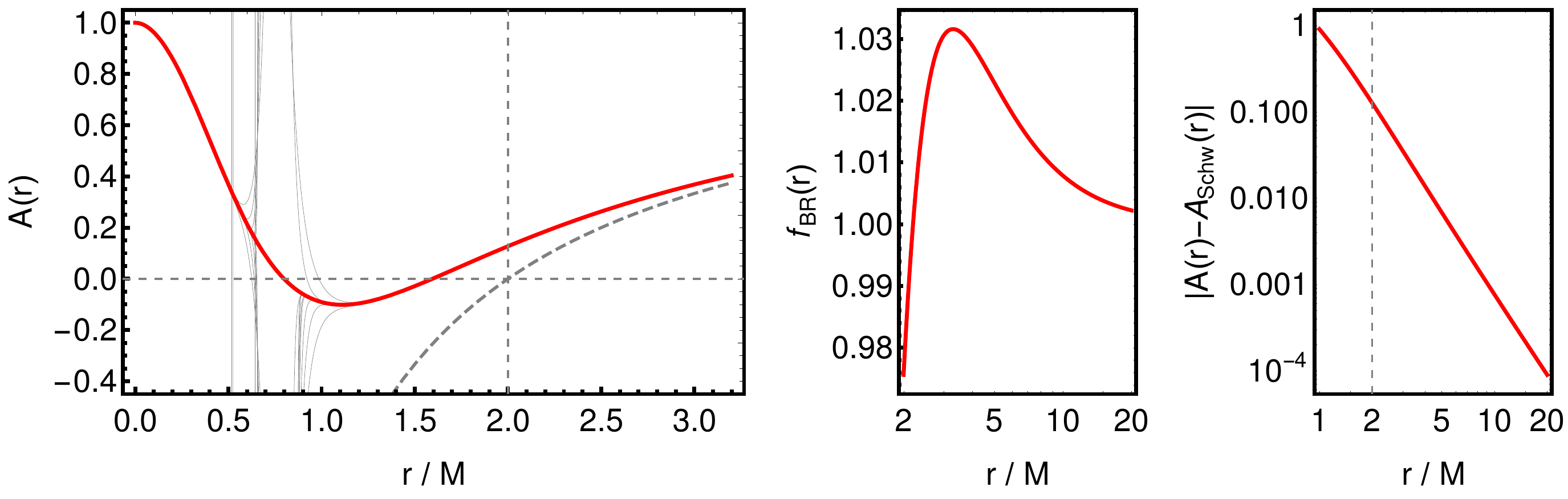}
	\caption{
	\label{fig:simplicity_BR}
	In the left-hand panel, we show metric functions $A(r)$ (thick red line) and series expansions around $r=0$ and $r=\infty$ at order $2^N$ and $2^N+1$ with $N=4,\dots,8$ (continuous thin-gray lines) for the non-simple regular spacetime with $f_\text{BR}(r)$ but with opposite signs (compared to \cite{Bonanno:2000ep}) of the new-physics parameters $\tilde{\omega}/M^2=-1$ and $\gamma=-2.2$. We also show the function $A_\text{Schw}(r) = 1-2M/r$ (gray-dashed). In the middle panel, we show the non-monotonic behavior of $f_\text{BR}(r)$. In the right-hand panel, we show the difference $|A(r)-A_\text{Schw}(r)|$ which takes positive values (continuous) at the horizon and negative values (dashed) at the location of the photon sphere.
	}
\end{figure}

To demonstrate how a violation of the simplicity principle can impact the properties of the spacetime, we may formally flip the signs and choose $\gamma<0$ and $\tilde{\omega}<0$. In this case there exist parameter choices that lead to a non-monotonic $f_\text{BR}(r)$, cf.~middle panel in Fig.~\ref{fig:simplicity_BR}. In specific cases, compactness at horizon scales may increase while compactness at the photon sphere decreases, cf.~right-hand panel in Fig.~\ref{fig:simplicity_BR}. For instance, $\tilde{\omega}/M^2=-1$ and $\gamma=-2.2$ leads to a shrinking horizon $r_{\rm H} \approx 1.93M < 2M$ but a growing photon sphere $r_\gamma \approx 3.08M > 3M$. Such behavior is only possible due to a violation of the simplicity principle, as can be seen from the expansions in the left-hand panel of Fig.~\ref{fig:simplicity_BR}.

\section{Post-Newtonian constraints}\label{sec:PN}
\label{app:PN}
We can make the, arguably strong, assumption that a uniqueness theorem analogous to Birkhoff's theorem holds beyond GR. Under that assumption we can use post-Newtonian constraints arising from solar-system observations to constrain the spherically symmetric metrics.

The observational constraint on the leading coefficient from the parameterized post-Newtonian expansion affects the Simpson-Visser metric; the corresponding coefficient vanishes for the other metrics. In the case of Simpson-Visser spacetime, weak-field observations constrain the new-physics scale $r_{\rm NP}$ as follows: 
We expand $f_{\rm SV}$ in $M/r$ to obtain
\bea
f_{\rm SV} &=& \exp\left(-\left(\frac{r_\text{NP}^4 K_\text{Schw}}{48}\right)^{1/6}\right)= \exp\left(-r_\text{NP}^{\frac{2}{3}} \frac{M^{\frac{1}{3}}}{r}\right) = \exp\left(-\left(\frac{r_\text{NP}}{M}\right)^{\frac{2}{3}} \frac{M}{r}\right)\nonumber\\
&=& 1- \left(\frac{r_{\rm NP}}{M}\right)^{\frac{2}{3}} \frac{M}{r}
+ \mathcal{O} \left(\frac{M}{r} \right)^2.
\eea
In weak-field tests in the solar system, the coefficient of the $\frac{M^2}{r^2}$ term in $A(r)$ (i.e., the coefficient of $\frac{M}{r}$ in $f(r)$) is constrained to be less than $10^{-5}$ \cite{Will:2014kxa}. Given that these tests use the sun as the gravitating mass, this constraint can be converted into a constraint on the new-physics scale:
\be
\frac{r_{\rm NP}}{M_\odot} < 3.2 \cdot 10^{-8}.
\ee
We can reinstate $G$ and $c$ to convert this into a length-scale, to obtain
\be
r_{\rm NP}<3.2\cdot10^{-8} \frac{G\, M_\odot}{c^2} \approx 4.7 \cdot 10^{-5}\, \rm{m}.
\ee

If we now assume that angular momentum can be neglected to leading order, a spherically symmetric line element can be used to describe the spacetime of spinning black holes.
This implies that for a supermassive black hole with a mass roughly like ${\rm M}87^\ast$, the critical value 
$(r_{\rm NP}/M)^{\frac{2}{3}} \approx 0.735$ (at which the compact object becomes horizonless) cannot be reached without violating solar-system constraints. Instead, for a mass of $M_{\rm M87^\ast}=6.5\cdot 10^9 \, M_{\odot}$ \cite{paper6}, we obtain $(r_{\rm NP}/M)^{\frac{2}{3}} = 2.9\cdot 10^{-12}$.
Thereby the weak-field measurements exclude that ${\rm M}87^\ast$ is a horizonless object under the assumption that it is described by the Simpson-Visser metric and that a uniqueness theorem holds. Both are very strong assumptions. Therefore, while we report the constraint on $r_{\rm NP}$, we also work with larger values of $r_{\rm NP}$ for the Simpson-Visser metric in the body of this paper.

\bibliography{References}

\begin{thebibliography}{10}

\bibitem{Abbott:2016blz}
B.~P. Abbott et~al.
\newblock {Observation of Gravitational Waves from a Binary Black Hole Merger}.
\newblock {\em Phys. Rev. Lett.}, 116(6):061102, 2016.

\bibitem{paper1}
Event Horizon~Telescope Collaboration.
\newblock {First M87 Event Horizon Telescope Results. I. The Shadow of the
  Supermassive Black Hole}.
\newblock {\em Astrophys. J.}, 875(1):L1, 2019.

\bibitem{paper2}
Event Horizon~Telescope Collaboration.
\newblock {First M87 Event Horizon Telescope Results. II. Array and
  Instrumentation}.
\newblock {\em Astrophys. J.}, 875(1):L2, 2019.

\bibitem{paper3}
Event Horizon~Telescope Collaboration.
\newblock {First M87 Event Horizon Telescope Results. III. Data Processing and
  Calibration}.
\newblock {\em Astrophys. J.}, 875(1):L3, 2019.

\bibitem{paper4}
Event Horizon~Telescope Collaboration.
\newblock {First M87 Event Horizon Telescope Results. IV. Imaging the Central
  Supermassive Black Hole}.
\newblock {\em Astrophys. J.}, 875(1):L4, 2019.

\bibitem{paper5}
Event Horizon~Telescope Collaboration.
\newblock {First M87 Event Horizon Telescope Results. V. Physical Origin of the
  Asymmetric Ring}.
\newblock {\em Astrophys. J.}, 875(1):L5, 2019.

\bibitem{paper6}
Event Horizon~Telescope Collaboration.
\newblock {First M87 Event Horizon Telescope Results. VI. The Shadow and Mass
  of the Central Black Hole}.
\newblock {\em Astrophys. J.}, 875(1):L6, 2019.

\bibitem{paper7}
Event Horizon~Telescope Collaboration.
\newblock {First M87 Event Horizon Telescope Results. VII. Polarization of the
  Ring}.
\newblock {\em Astrophys. J. Lett.}, 910(1), 2021.

\bibitem{TheLIGOScientific:2016src}
B.~P. Abbott et~al.
\newblock {Tests of general relativity with GW150914}.
\newblock {\em Phys. Rev. Lett.}, 116(22):221101, 2016.
\newblock [Erratum: Phys.Rev.Lett. 121, 129902 (2018)].

\bibitem{Do:2019txf}
Tuan Do et~al.
\newblock {Relativistic redshift of the star S0-2 orbiting the Galactic center
  supermassive black hole}.
\newblock {\em Science}, 365(6454):664--668, 2019.

\bibitem{LIGOScientific:2019fpa}
B.~P. Abbott et~al.
\newblock {Tests of General Relativity with the Binary Black Hole Signals from
  the LIGO-Virgo Catalog GWTC-1}.
\newblock {\em Phys. Rev. D}, 100(10):104036, 2019.

\bibitem{Cardoso:2019rvt}
Vitor Cardoso and Paolo Pani.
\newblock {Testing the nature of dark compact objects: a status report}.
\newblock {\em Living Rev. Rel.}, 22(1):4, 2019.

\bibitem{LIGOScientific:2020tif}
R.~Abbott et~al.
\newblock {Tests of general relativity with binary black holes from the second
  LIGO-Virgo gravitational-wave transient catalog}.
\newblock {\em Phys. Rev. D}, 103(12):122002, 2021.

\bibitem{Volkel:2020xlc}
Sebastian~H. V\"olkel, Enrico Barausse, Nicola Franchini, and Avery~E.
  Broderick.
\newblock {EHT tests of the strong-field regime of General Relativity}.
\newblock {\em Class. Quant. Grav.}, 38:21, 11 2020.

\bibitem{LIGOScientific:2021sio}
R.~Abbott et~al.
\newblock {Tests of General Relativity with GWTC-3}.
\newblock 12 2021.

\bibitem{Penrose:1964wq}
Roger Penrose.
\newblock {Gravitational collapse and space-time singularities}.
\newblock {\em Phys. Rev. Lett.}, 14:57--59, 1965.

\bibitem{Starobinsky:1980te}
Alexei~A. Starobinsky.
\newblock {A New Type of Isotropic Cosmological Models Without Singularity}.
\newblock {\em Adv. Ser. Astrophys. Cosmol.}, 3:130--133, 1987.

\bibitem{Almheiri:2012rt}
Ahmed Almheiri, Donald Marolf, Joseph Polchinski, and James Sully.
\newblock {Black Holes: Complementarity or Firewalls?}
\newblock {\em JHEP}, 02:062, 2013.

\bibitem{Rovelli:2014cta}
Carlo Rovelli and Francesca Vidotto.
\newblock {Planck stars}.
\newblock {\em Int. J. Mod. Phys.}, D23(12):1442026, 2014.

\bibitem{Haggard:2016ibp}
Hal~M. Haggard and Carlo Rovelli.
\newblock {Quantum Gravity Effects around Sagittarius A*}.
\newblock {\em Int. J. Mod. Phys.}, D25(12):1644021, 2016.

\bibitem{Giddings:2016btb}
Steven~B. Giddings and Dimitrios Psaltis.
\newblock {Event Horizon Telescope Observations as Probes for Quantum Structure
  of Astrophysical Black Holes}.
\newblock {\em Phys. Rev. D}, 97(8):084035, 2018.

\bibitem{Giddings:2019jwy}
Steven~B. Giddings.
\newblock {Searching for quantum black hole structure with the Event Horizon
  Telescope}.
\newblock {\em Universe}, 5(9):201, 2019.

\bibitem{Held:2019xde}
Aaron Held, Roman Gold, and Astrid Eichhorn.
\newblock {Asymptotic safety casts its shadow}.
\newblock {\em JCAP}, 1906:029, 2019.

\bibitem{Bacchini:2021fig}
Fabio Bacchini, Daniel~R. Mayerson, Bart Ripperda, Jordy Davelaar, H\'ector
  Olivares, Thomas Hertog, and Bert Vercnocke.
\newblock {Fuzzball Shadows: Emergent Horizons from Microstructure}.
\newblock {\em Phys. Rev. Lett.}, 127(17):171601, 2021.

\bibitem{Carballo-Rubio:2019nel}
Ra\'ul Carballo-Rubio, Francesco Di~Filippo, Stefano Liberati, and Matt Visser.
\newblock {Opening the Pandora's box at the core of black holes}.
\newblock 8 2019.

\bibitem{Carballo-Rubio:2021ayp}
Ra\'ul Carballo-Rubio, Francesco Di~Filippo, and Stefano Liberati.
\newblock {Hearts of Darkness: the inside out probing of black holes}.
\newblock 6 2021.

\bibitem{Bonanno:2020fgp}
Alfio Bonanno, Amir-Pouyan Khosravi, and Frank Saueressig.
\newblock {Regular black holes with stable cores}.
\newblock {\em Phys. Rev. D}, 103(12):124027, 2021.

\bibitem{Barcelo:2020mjw}
Carlos Barcel\'o, Valentin Boyanov, Ra\'ul Carballo-Rubio, and Luis~J. Garay.
\newblock {Black hole inner horizon evaporation in semiclassical gravity}.
\newblock {\em Class. Quant. Grav.}, 38(12):125003, 2021.

\bibitem{Lu:2015psa}
H.~Lü, A.~Perkins, C.~N. Pope, and K.~S. Stelle.
\newblock {Spherically Symmetric Solutions in Higher-Derivative Gravity}.
\newblock {\em Phys. Rev.}, D92(12):124019, 2015.

\bibitem{Lu:2017kzi}
H.~Lü, A.~Perkins, C.~N. Pope, and K.~S. Stelle.
\newblock {Lichnerowicz Modes and Black Hole Families in Ricci Quadratic
  Gravity}.
\newblock {\em Phys. Rev.}, D96(4):046006, 2017.

\bibitem{Pravda:2016fue}
Vojtech Pravda, Alena Pravdov\'a, Jiri Podolsk\'y, and Robert \v{S}varc.
\newblock {Exact solutions to quadratic gravity}.
\newblock {\em Phys. Rev. D}, 95(8):084025, 2017.

\bibitem{Podolsky:2019gro}
Jiri Podolsk\'y, Robert \v{S}varc, Vojtech Pravda, and Alena Pravdov\'a.
\newblock {Black holes and other exact spherical solutions in Quadratic
  Gravity}.
\newblock {\em Phys. Rev. D}, 101(2):024027, 2020.

\bibitem{Brito:2013wya}
Richard Brito, Vitor Cardoso, and Paolo Pani.
\newblock {Massive spin-2 fields on black hole spacetimes: Instability of the
  Schwarzschild and Kerr solutions and bounds on the graviton mass}.
\newblock {\em Phys. Rev. D}, 88(2):023514, 2013.

\bibitem{Cayuso:2020lca}
Ramiro Cayuso and Luis Lehner.
\newblock {Nonlinear, noniterative treatment of EFT-motivated gravity}.
\newblock {\em Phys. Rev. D}, 102(8):084008, 2020.

\bibitem{Held:2021pht}
Aaron Held and Hyun Lim.
\newblock {Nonlinear dynamics of quadratic gravity in spherical symmetry}.
\newblock {\em Phys. Rev. D}, 104(8):084075, 2021.

\bibitem{Eichhorn:2021etc}
Astrid Eichhorn and Aaron Held.
\newblock {Image features of spinning regular black holes based on a locality
  principle}.
\newblock {\em Eur. Phys. J. C}, 81(10):933, 2021.

\bibitem{Eichhorn:2021iwq}
Astrid Eichhorn and Aaron Held.
\newblock {From a locality-principle for new physics to image features of
  regular spinning black holes with disks}.
\newblock {\em JCAP}, 05:073, 2021.

\bibitem{1979GReGr..10...79B}
S.~{Benenti} and M.~{Francaviglia}.
\newblock {Remarks on certain separability structures and their applications to
  general relativity}.
\newblock {\em General Relativity and Gravitation}, 10(1):79--92, January 1979.

\bibitem{Cardoso:2014rha}
Vitor Cardoso, Paolo Pani, and Jo\~{a}o Rico.
\newblock {On generic parametrizations of spinning black-hole geometries}.
\newblock {\em Phys. Rev. D}, 89:064007, 2014.

\bibitem{Johannsen:2015pca}
Tim Johannsen.
\newblock {Regular Black Hole Metric with Three Constants of Motion}.
\newblock {\em Phys. Rev.}, D88(4):044002, 2013.

\bibitem{Konoplya:2016jvv}
Roman Konoplya, Luciano Rezzolla, and Alexander Zhidenko.
\newblock {General parametrization of axisymmetric black holes in metric
  theories of gravity}.
\newblock {\em Phys. Rev.}, D93(6):064015, 2016.

\bibitem{Delaporte:2022acp}
H\'elo\"\i{}se Delaporte, Astrid Eichhorn, and Aaron Held.
\newblock {Parameterizations of black-hole spacetimes beyond circularity}.
\newblock 2 2022.

\bibitem{Cardenas-Avendano:2019zxd}
Alejandro C\'ardenas-Avenda\~{n}o, Sourabh Nampalliwar, and Nicol\'as Yunes.
\newblock {Gravitational-wave versus X-ray tests of strong-field gravity}.
\newblock {\em Class. Quant. Grav.}, 37(13):135008, 2020.

\bibitem{Shashank:2021giy}
Swarnim Shashank and Cosimo Bambi.
\newblock {Constraining the Konoplya-Rezzolla-Zhidenko deformation parameters
  III: limits from stellar-mass black holes using gravitational-wave
  observations}.
\newblock 12 2021.

\bibitem{Hayward:2005gi}
Sean~A. Hayward.
\newblock {Formation and evaporation of regular black holes}.
\newblock {\em Phys. Rev. Lett.}, 96:031103, 2006.

\bibitem{Dymnikova:1992ux}
I.~Dymnikova.
\newblock {Vacuum nonsingular black hole}.
\newblock {\em Gen. Rel. Grav.}, 24:235--242, 1992.

\bibitem{Simpson:2019mud}
Alex Simpson and Matt Visser.
\newblock {Regular black holes with asymptotically Minkowski cores}.
\newblock {\em Universe}, 6(1):8, 2019.

\bibitem{Tiede:2020jgo}
Paul Tiede, Hung-Yi Pu, Avery~E. Broderick, Roman Gold, Mansour Karami, and
  Jorge~A. Preciado-L\'opez.
\newblock {Spacetime Tomography Using The Event Horizon Telescope}.
\newblock 2 2020.

\bibitem{Broderick:2005my}
Avery~E. Broderick and Abraham Loeb.
\newblock {Imaging bright spots in the accretion flow near the black hole
  horizon of Sgr A*}.
\newblock {\em Mon. Not. Roy. Astron. Soc.}, 363:353--362, 2005.

\bibitem{Broderick:2005jj}
Avery~E. Broderick and Abraham Loeb.
\newblock {Imaging optically-thin hot spots near the black hole horizon of sgr
  a* at radio and near-infrared wavelengths}.
\newblock {\em Mon. Not. Roy. Astron. Soc.}, 367:905--916, 2006.

\bibitem{Pesce:2021adg}
Dominic~W. Pesce, Daniel C.~M. Palumbo, Ramesh Narayan, Lindy Blackburn,
  Sheperd~S. Doeleman, Michael~D. Johnson, Chung-Pei Ma, Neil~M. Nagar,
  Priyamvada Natarajan, and Angelo Ricarte.
\newblock {Toward Determining the Number of Observable Supermassive Black Hole
  Shadows}.
\newblock {\em Astrophys. J.}, 923(2):260, 2021.

\bibitem{Carballo-Rubio:2019fnb}
Ra\'ul Carballo-Rubio, Francesco Di~Filippo, Stefano Liberati, and Matt Visser.
\newblock {Geodesically complete black holes}.
\newblock {\em Phys. Rev. D}, 101:084047, 2020.

\bibitem{1991JMP....32.3135C}
J.~{Carminati} and R.~G. {McLenaghan}.
\newblock {Algebraic invariants of the Riemann tensor in a four-dimensional
  Lorentzian space}.
\newblock {\em Journal of Mathematical Physics}, 32(11):3135--3140, November
  1991.

\bibitem{1997GReGr..29..539Z}
E.~{Zakhary} and C.~B.~G. {McIntosh}.
\newblock {A Complete Set of Riemann Invariants}.
\newblock {\em General Relativity and Gravitation}, 29(5):539--581, May 1997.

\bibitem{Bardeen}
J.~M. Bardeen.
\newblock {}.
\newblock {\em Conference Proceedings of GR5, Tbilisi, USSR}, page 174, 1968.

\bibitem{Bonanno:2000ep}
Alfio Bonanno and Martin Reuter.
\newblock {Renormalization group improved black hole space-times}.
\newblock {\em Phys. Rev.}, D62:043008, 2000.

\bibitem{Nicolini:2008aj}
Piero Nicolini.
\newblock {Noncommutative Black Holes, The Final Appeal To Quantum Gravity: A
  Review}.
\newblock {\em Int. J. Mod. Phys.}, A24:1229--1308, 2009.

\bibitem{Platania:2019kyx}
Alessia Platania.
\newblock {Dynamical renormalization of black-hole spacetimes}.
\newblock {\em Eur. Phys. J. C}, 79(6):470, 2019.

\bibitem{Nicolini:2019irw}
Piero Nicolini, Euro Spallucci, and Michael~F. Wondrak.
\newblock {Quantum Corrected Black Holes from String T-Duality}.
\newblock {\em arXiv preprint}, 2019.

\bibitem{Zhou:2020eth}
Biao Zhou, Askar~B. Abdikamalov, Dimitry Ayzenberg, Cosimo Bambi, Sourabh
  Nampalliwar, and Ashutosh Tripathi.
\newblock {Shining X-rays on asymptotically safe quantum gravity}.
\newblock {\em JCAP}, 01:047, 2021.

\bibitem{Donoghue:1993eb}
John~F. Donoghue.
\newblock {Leading quantum correction to the Newtonian potential}.
\newblock {\em Phys. Rev. Lett.}, 72:2996--2999, 1994.

\bibitem{Cardoso:2008bp}
Vitor Cardoso, Alex~S. Miranda, Emanuele Berti, Helvi Witek, and Vilson~T.
  Zanchin.
\newblock {Geodesic stability, Lyapunov exponents and quasinormal modes}.
\newblock {\em Phys. Rev. D}, 79(6):064016, 2009.

\bibitem{Bardeen:1972fi}
James~M. Bardeen, William~H. Press, and Saul~A. Teukolsky.
\newblock {Rotating black holes: Locally nonrotating frames, energy extraction,
  and scalar synchrotron radiation}.
\newblock {\em Astrophys. J.}, 178:347, 1972.

\bibitem{Ghez:2008ms}
A.~M. Ghez et~al.
\newblock {Measuring Distance and Properties of the Milky Way's Central
  Supermassive Black Hole with Stellar Orbits}.
\newblock {\em Astrophys. J.}, 689:1044--1062, 2008.

\bibitem{2015MNRAS.447..948C}
S.~{Chatzopoulos}, T.~K. {Fritz}, O.~{Gerhard}, S.~{Gillessen}, C.~{Wegg},
  R.~{Genzel}, and O.~{Pfuhl}.
\newblock {The old nuclear star cluster in the Milky Way: dynamics, mass,
  statistical parallax, and black hole mass}.
\newblock {\em \mnras}, 447:948--968, February 2015.

\bibitem{Gebhardt:2011yw}
Karl Gebhardt, Joshua Adams, Douglas Richstone, Tod~R. Lauer, S.~M. Faber,
  Kayhan Gültekin, Jeremy Murphy, and Scott Tremaine.
\newblock {The Black-Hole Mass in M87 from Gemini/NIFS Adaptive Optics
  Observations}.
\newblock {\em Astrophys. J.}, 729:119, 2011.

\bibitem{Berti:2015itd}
Emanuele Berti et~al.
\newblock {Testing General Relativity with Present and Future Astrophysical
  Observations}.
\newblock {\em Class. Quant. Grav.}, 32:243001, 2015.

\bibitem{Abdujabbarov:2016hnw}
Ahmadjon Abdujabbarov, Muhammed Amir, Bobomurat Ahmedov, and Sushant~G. Ghosh.
\newblock {Shadow of rotating regular black holes}.
\newblock {\em Phys. Rev.}, D93(10):104004, 2016.

\bibitem{Kumar:2019ohr}
Rahul Kumar, Balendra~Pratap Singh, and Sushant~G. Ghosh.
\newblock {Shadow and deflection angle of rotating black hole in asymptotically
  safe gravity}.
\newblock {\em Annals Phys.}, 420:168252, 2020.

\bibitem{Liu:2020ola}
Cheng Liu, Tao Zhu, Qiang Wu, Kimet Jusufi, Mubasher Jamil, Mustapha
  Azreg-A\"\i{}nou, and Anzhong Wang.
\newblock {Shadow and Quasinormal Modes of a Rotating Loop Quantum Black Hole}.
\newblock {\em Phys. Rev. D}, 101(8):084001, 2020.

\bibitem{Stuchlik:2019uvf}
Zdenek Stuchl\'\i{}k and Jan Schee.
\newblock {Shadow of the regular Bardeen black holes and comparison of the
  motion of photons and neutrinos}.
\newblock {\em Eur. Phys. J. C}, 79(1):44, 2019.

\bibitem{Wielgus:2021peu}
Maciek Wielgus.
\newblock {Photon rings of spherically symmetric black holes and robust tests
  of non-Kerr metrics}.
\newblock 9 2021.

\bibitem{Johnson:2019ljv}
Michael~D. Johnson et~al.
\newblock {Universal Interferometric Signatures of a Black Hole's Photon Ring}.
\newblock {\em Sci. Adv.}, 6(12):eaaz1310, 2020.

\bibitem{EventHorizonTelescope:2021dqv}
Prashant Kocherlakota et~al.
\newblock {Constraints on black-hole charges with the 2017 EHT observations of
  M87*}.
\newblock {\em Phys. Rev. D}, 103(10):104047, 2021.

\bibitem{2010A&A...524A..71B}
S.~{Bird}, W.~E. {Harris}, J.~P. {Blakeslee}, and C.~{Flynn}.
\newblock {The inner halo of M87: a first direct view of the red-giant
  population}.
\newblock {\em \aap}, 524:A71, December 2010.

\bibitem{2009ApJ...694..556B}
John~P. {Blakeslee}, Andr{\'e}s {Jord{\'a}n}, Simona {Mei}, Patrick
  {C{\^o}t{\'e}}, Laura {Ferrarese}, Leopoldo {Infante}, Eric~W. {Peng},
  John~L. {Tonry}, and Michael~J. {West}.
\newblock {The ACS Fornax Cluster Survey. V. Measurement and Recalibration of
  Surface Brightness Fluctuations and a Precise Value of the Fornax-Virgo
  Relative Distance}.
\newblock {\em \apj}, 694(1):556--572, March 2009.

\bibitem{2018ApJ...856..126C}
Michele {Cantiello}, John~P. {Blakeslee}, Laura {Ferrarese}, Patrick
  {C{\^o}t{\'e}}, Joel~C. {Roediger}, Gabriella {Raimondo}, Eric~W. {Peng},
  Stephen {Gwyn}, Patrick~R. {Durrell}, and Jean-Charles {Cuillandre}.
\newblock {The Next Generation Virgo Cluster Survey (NGVS). XVIII. Measurement
  and Calibration of Surface Brightness Fluctuation Distances for Bright
  Galaxies in Virgo (and Beyond)}.
\newblock {\em \apj}, 856(2):126, April 2018.

\bibitem{Lima:2021las}
Haroldo C.~D. Lima, Junior., Lu\'\i{}s C.~B. Crispino, Pedro V.~P. Cunha, and
  Carlos A.~R. Herdeiro.
\newblock {Can different black holes cast the same shadow?}
\newblock {\em Phys. Rev. D}, 103(8):084040, 2021.

\bibitem{Blackburn:2019bly}
Lindy Blackburn et~al.
\newblock {Studying Black Holes on Horizon Scales with VLBI Ground Arrays}.
\newblock 9 2019.

\bibitem{Broderick:2021ohx}
Avery~E. Broderick, Paul Tiede, Dominic~W. Pesce, and Roman Gold.
\newblock {Measuring Spin from Relative Photon Ring Sizes}.
\newblock 5 2021.

\bibitem{Ayzenberg:2022twz}
Dimitry Ayzenberg.
\newblock {Testing Gravity with Black Hole Shadow Subrings}.
\newblock 2 2022.

\bibitem{Johnson:2015iwg}
Michael~D. Johnson et~al.
\newblock {Resolved Magnetic-Field Structure and Variability Near the Event
  Horizon of Sagittarius A*}.
\newblock {\em Science}, 350(6265):1242--1245, 2015.

\bibitem{Gralla:2020srx}
Samuel~E. Gralla, Alexandru Lupsasca, and Daniel~P. Marrone.
\newblock {The shape of the black hole photon ring: A precise test of
  strong-field general relativity}.
\newblock {\em Phys. Rev. D}, 102(12):124004, 2020.

\bibitem{2020ApJ...892..132T}
Paul {Tiede}, Hung-Yi {Pu}, Avery~E. {Broderick}, Roman {Gold}, Mansour
  {Karami}, and Jorge~A. {Preciado-L{\'o}pez}.
\newblock {Spacetime Tomography Using the Event Horizon Telescope}.
\newblock {\em \apj}, 892(2):132, April 2020.

\bibitem{Berry:2020ntz}
Thomas Berry, Alex Simpson, and Matt Visser.
\newblock {Photon spheres, ISCOs, and OSCOs: Astrophysical observables for
  regular black holes with asymptotically Minkowski cores}.
\newblock {\em Universe}, 7(1):2, 2020.

\bibitem{Lu:2017vdx}
Wenbin Lu, Pawan Kumar, and Ramesh Narayan.
\newblock {Stellar disruption events support the existence of the black hole
  event horizon}.
\newblock {\em Mon. Not. Roy. Astron. Soc.}, 468(1):910--919, 2017.

\bibitem{Carballo-Rubio:2018jzw}
Ra\'ul Carballo-Rubio, Francesco Di~Filippo, Stefano Liberati, and Matt Visser.
\newblock {Phenomenological aspects of black holes beyond general relativity}.
\newblock {\em Phys. Rev. D}, 98(12):124009, 2018.

\bibitem{2018ApJ...855..128W}
R.~Craig {Walker}, Philip~E. {Hardee}, Frederick~B. {Davies}, Chun {Ly}, and
  William {Junor}.
\newblock {The Structure and Dynamics of the Subparsec Jet in M87 Based on 50
  VLBA Observations over 17 Years at 43 GHz}.
\newblock {\em \apj}, 855(2):128, March 2018.

\bibitem{Broderick:2020wda}
Avery~E. Broderick, Dominic~W. Pesce, Paul Tiede, Hung-Yi Pu, and Roman Gold.
\newblock {Hybrid Very Long Baseline Interferometry Imaging and Modeling with
  themis}.
\newblock {\em Astrophys. J.}, 898(1):9, 2020.

\bibitem{Will:2014kxa}
Clifford~M. Will.
\newblock {The Confrontation between General Relativity and Experiment}.
\newblock {\em Living Rev. Rel.}, 17:4, 2014.

\end{thebibliography}
    
\end{document}